\documentclass[12pt]{spieman}  
\usepackage{amsmath,amsfonts,amssymb}
\usepackage{graphicx}
\usepackage{setspace}
\usepackage{tocloft}
\usepackage{lineno}

\usepackage{natbib}
\usepackage{etoolbox}
\newcounter{lsaref}
\makeatletter
\apptocmd{\@lbibitem}{\stepcounter{lsaref}[\arabic{lsaref}]}{}{}
\makeatother

\usepackage{authblk}

\usepackage{xcolor}

\setcitestyle{square}

\usepackage{setspace}
\title{GAMERA: A three-dimensional finite-volume MHD solver for non-orthogonal curvilinear geometries}

\author[a]{Binzheng Zhang}
\author[b]{Kareem A. Sorathia}
\author[c]{John G. Lyon}
\author[b]{Viacheslav G. Merkin}
\author[b]{Jeffrey S. Garretson}
\author[d]{Michael Wiltberger}

\affil[a]{Department of Earth Sciences, the University of Hong Kong}
\affil[b]{Applied Physics Laboratory, Johns Hopkins University}
\affil[c]{Department of Physics and Astronomy, Dartmouth College }
\affil[d]{National Center for Atmospheric Research }

\cftpagenumbersoff{figure}
\cftpagenumbersoff{table} 
\begin{document} 
\maketitle

\begin{abstract}
Efficient simulation of plasmas in various contexts often involves the use of meshes that conform to the intrinsic geometry of the system under consideration. We present here a description of a new magnetohydrodynamic code, Gamera (Grid Agnostic MHD for Extended Research Applications), designed to combine geometric flexibility with high-order spatial reconstruction and constrained transport to maintain the divergence-free magnetic field. Gamera carries on the legacy of its predecessor, the LFM (Lyon-Fedder-Mobarry), a research code whose use in space physics has spanned three decades. At the time of its initial development the LFM code had a number of novel features: eighth-order centered spatial differencing, the Partial Donor Cell Method limiter for shock capturing, a non-orthogonal staggered mesh with constrained transport, and conservative averaging-reconstruction for axis singularities. A capability to handle multiple ion species was also added later. Gamera preserves the core numerical philosophy of LFM while also incorporating numerous algorithmic and computational improvements. The upgrades in the numerical schemes include accurate grid metric calculations using high-order Gaussian quadrature techniques, high-order upwind reconstruction, non-clipping options for interface values, and improved treatment of axis singularities. The improvements in the code implementation include the use of data structures and memory access patterns conducive to aligned vector operations and the implementation of hybrid parallelism, using MPI and OMP. Gamera is designed to be a portable and easy-to-use code that implements multi-dimensional MHD simulations in arbitrary non-orthogonal curvilinear geometries on modern supercomputer architectures.
\end{abstract}


\begin{spacing}{1.5}   

\tableofcontents


\section{Introduction}

Numerical magnetohydrodynamics (MHD), which solves the equations of ideal or non-ideal MHD, is used extensively in research areas such as basic plasma physics, astrophysics, solar physics, geospace environment modeling, planetary sciences and space weather applications. Therefore the development of accurate, multi-dimensional, general-purposed MHD solvers along with comprehensive descriptions of their implementation is important for building powerful computational tools and also for the advancement of scientific understanding, and education of the new generation of scientists. In the past two decades, many general-purposed MHD codes have been developed by research teams from various scientific research communities including but not limited to, the ZEUS code \citep{Stone1992}, the Athena code \citep{Stone2008}, the BATS-R-US code \citep{Powell1999}, the VAC code \citep{Toth1997}, the Nirvana \citep{Ziegler2005}, the RAMSES \citep{Fromang2006}, the PLUTO code\citep{Mignone2007}, the AstroBEAR \citep{Cunningham2011}, the FLASH code \citep{Dubey2008} and the PENCIL code \citep{Dobler2006}. One of the MHD codes that has been widely used within the space physics community is the Lyon-Fedder-Mobarry (LFM) code \citep{Lyon2004}. 

The LFM MHD code, initially developed at the \textit{Naval Research Laboratory} in the early 80s, is one of the pioneers of solving three-dimensional MHD equations in non-orthogonal curvilinear geometry with high-quality advection schemes \citep{Lyon2004,Lyon1981, Brecht1982}. The MHD Kernel of the LFM code had a number of novel features: an eighth-order centered spatial reconstruction method to reduce numerical diffusion, a universal-type non-linear limiter to capture shocks \citep{Hain1987}, a non-orthogonal spherical mesh with high-order constrained transport algorithm to maintain zero magnetic divergence at round-off, a fix for axis singularities and multi-fluid extensions to handle multiple ion species \citep{Brambles2011}. The LFM code has been well known as a global terrestrial magnetosphere model, which has been one of the workhorses for geospace research and space weather applications \citep{Luhmann2004}. In the past decade, as the backbone of a whole geospace model, the LFM has been coupled to other first-principles codes, e.g. the magnetosphere-ionosphere coupler/solver code for high-latitude ionospheric electrodynamics \citep{Merkin2010}, the Rice Convection Model of the inner magnetosphere\citep{Pembroke2012}, the NCAR thermosphere-ionosphere model for upper atmospheric dynamics \citep{Wiltberger2004} and the ionosphere polar wind model \citep{Varney2016} for collisionless heavy ion transport. Beyond geospace research, the LFM code has also been adapted to be an inner heliosphere model \citep{Merkin2011}, a global heliosphere model \citep{NcNutt1999}, an unmagnetized planetary magnetosphere model for Venus \citep{Kallio1998}, a giant planetary magnetosphere model for Jupiter \citep{Zhang2018} and basic plasma simulations for magnetic reconnection \citep{Merkin2015}. The code is actively used by research teams in the space physics community, and the list of applications developed based on the LFM code is still growing.

For a code to maintain its place in computational science for almost $40$ years is a testament to the quality of its core numerics, but the age of the LFM code makes it difficult to adapt to modern and future architectures.  To address these issues, we completed a rewrite of the LFM code from scratch, and re-envisaging it as \textit{GAMERA} - \textit{Grid Agnostic MHD for Extended Research Applications}.  The goal of the GAMERA project was to rebuild the LFM code with two underlying design principles: retain and improve the high-heritage MHD numerics; and prepare the code for the exascale era.  These improvements include 1) flexible grid specifications; 2) high-order grid metric calculations based on Gaussian quadrature; 3) seventh-order upwind spatial reconstruction; 4) non-clipping options in the limiters; 5) higher-order averaging-reconstruction method for axis singularity. While based on the best elements of the LFM legacy, GAMERA is designed to be easily applicable to two- or three-dimensional MHD simulations using general curvilinear computational grids adapted to specific problems. It is highly efficient; for example usable Earth or planetary magnetospheric simulations can be run on a laptop close to real time.


The goal of this paper is to provide a general description of the GAMERA code, including both the numerical schemes and the implementation details. The intent is that the paper will serve as a reference for future users and developers to use, modify and eventually extend the code to their own research applications. The description contains enough details so that the numerical techniques and equations used in the MHD kernel of the GAMERA code can be easily identified, or a functionally equivalent code can be built based on the details described in this paper. An example MATLAB code with the GAMERA algorithms described in this paper is provided in the supplement material for potential users to further understand the numerical algorithms. With a set of standard test problems as references, the quality of the numerical schemes used in GAMERA can be evaluated and improved when necessary. The paper is organized as follows: Section 2 describes the default MHD equations solved in the GAMERA code. Section 3 describes the detailed numerical schemes used in the GAMERA code, including grid discretization, time stepping, spatial reconstruction, flux functions, constrained transport algorithms in non-orthogonal geometry, and details about the code implementation including vectorization and parallelization. Section 4 shows results from five representative test simulation problems to demonstrate the quality of the numeric algorithms, and summary and conclusions are included in Section 5.

\section{The Basic MHD Equations}

The default equation set solved in the GAMERA code is the single-fluid, normalized ideal MHD in a semi-conservative form with the plasma energy equation:
\begin{equation}
\frac{\partial\rho}{\partial t} = -\nabla\cdot\left( \rho \mathbf{u}\right) \label{eqn:MHD1}
\end{equation}
\begin{equation}
\frac{\partial\rho \mathbf{u}}{\partial t} = -\nabla\cdot\left( \rho\mathbf{u}\mathbf{u} + \bar{\mathbf{I}}P \right) -\nabla\cdot\left( \bar{\mathbf{I}}\frac{B^2}{2} - \mathbf{BB} \right)\label{eqn:MHD2}
\end{equation}
\begin{equation}
\frac{\partial E_P}{\partial t} = -\nabla\cdot\left[ \mathbf{u} \left( E_P + P \right) \right] - \mathbf{u}\cdot\nabla\cdot\left( \frac{B^2}{2}\bar{\mathbf{I}} - \mathbf{BB} \right)\label{eqn:MHD3}
\end{equation}
\begin{equation}
\frac{\partial \mathbf{B}}{\partial t} = -\nabla\times\mathbf{E}, \label{eqn:MHD4}
\end{equation}
where $\rho$ is the plasma mass density, $\mathbf{u}$ is the plasma bulk velocity, $\bar{\mathbf{I}}$ is the unit tensor, $P$ is the plasma thermal pressure, $\mathbf{B}$ is the magnetic field, and $\mathbf{E}=-\mathbf{u}\times\mathbf{B}$ is the electric field based on the ideal Ohm's law. $E_P$ is the plasma energy defined as the sum of kinetic and thermal energy:
\begin{equation}
E_P = \frac{1}{2}\rho u^2 + \frac{P}{\gamma -1},\label{eqn:E_P}
\end{equation}
with $\gamma=\frac{5}{3}$ defined as the ratio of specific heat. The normalization of equations (\ref{eqn:MHD1}) - (\ref{eqn:MHD4}) can be found in Appendix A.

The fluid part of the semi-conservative form of the ideal MHD equation set can be written in a compact vector form:
\begin{equation}
\frac{\partial\mathbf{U^C}}{\partial t} = \nabla\cdot\mathbf{F}\left(\mathbf{U^C}\right) + \mathbf{M}\left(\mathbf{U^C}\right), \label{eqn:FV-vector_form}
\end{equation}
where $\mathbf{U^C}$ is the vector form of the conserved density variables defined as:
\begin{equation}
\mathbf{U^C}=\left(\begin{array}{c} \rho \\ \rho\mathbf{u} \\ E_P \end{array}\right). \label{eqn:U_C}
\end{equation}
$\mathbf{F}\left(\mathbf{U^C}\right)$ is the vector form of the fluid part of the ideal MHD equations (ideal hydrodynamics):
\begin{equation}
\mathbf{F}\left(\mathbf{U^C}\right)=-\left[\begin{array}{c} \rho\mathbf{u} \\ \rho\mathbf{u}\mathbf{u} +\overline{\mathbf{I}}P \\ \mathbf{u}(E_P+P) \end{array}\right],
\end{equation}
and $\mathbf{M}\left(\mathbf{U^C}\right)$ is the vector form of the magnetic terms from the Lorentz force defined as:
\begin{equation}
\mathbf{M}\left(\mathbf{U^C}\right)=-\left[\begin{array}{c} 0 \\ \nabla\cdot\left( \bar{\mathbf{I}}\frac{B^2}{2} - \mathbf{BB} \right)  \\ \mathbf{u}\cdot\nabla\cdot\left( \bar{\mathbf{I}}\frac{B^2}{2} - \mathbf{BB} \right) \end{array}\right].
\end{equation}
The corresponding state vector of the primitive variables $\mathbf{U^P}$ is defined as
\begin{equation}
\mathbf{U^P}=\left(\begin{array}{c} \rho \\ \mathbf{u} \\ P \end{array}\right),
\end{equation}
which is mainly used in the reconstruction module. Another important set of variables used in the MHD solver is the volume integrated conserved densities, with a vector form defined as:
\begin{equation}
\mathbf{U^V}=\int\mathbf{U^C}dV=\int\left(\begin{array}{c} \rho \\ \rho\mathbf{u} \\ E_P \end{array}\right)dV, \label{eqn:vol-int-variable_def}
\end{equation}
which are the actually conserved quantities when the MHD equations are evolved. 

The use of the plasma energy equation has significant advantages in a number of problems, especially in modeling planetary magnetospheres with strong background magnetic fields and cold ambient plasmas (e.g., $\beta<10^{-6}$). However in using the plasma energy equation instead of the total energy equation, the system of equations is no longer in a fully-conservative form. Non-conservative forms of the energy equation can lead to non-physical results if non-ideal processes occur, whether physical or numerical. However, \citet{Lyon2004} showed that the use of plasma energy equation does not change the Rankine-Hugoniot relations to within the numerical truncation error. The result is independent of whether or not the electric field is carried by dissipative processes through the shock. The truncation level of numerical error is not quite as low as the round-off error from the total energy formulation, but it is sufficient to ensure that the behavior of shocks is correct to a good approximation. 

\section{Numerical Schemes}

\subsection{Finite-Volume Spatial Discretization}\label{sec:fv_def}

In this section we describe the definitions of the curvilinear computational grids, the plasma and magnetic variables, and basic metric calculations such as face area, face normal unit vector and cell volume used in the non-orthogonal, finite-volume MHD solver.

\subsubsection{Grid Definitions}

\begin{figure}[h!] 
	\noindent\includegraphics[width=39pc]{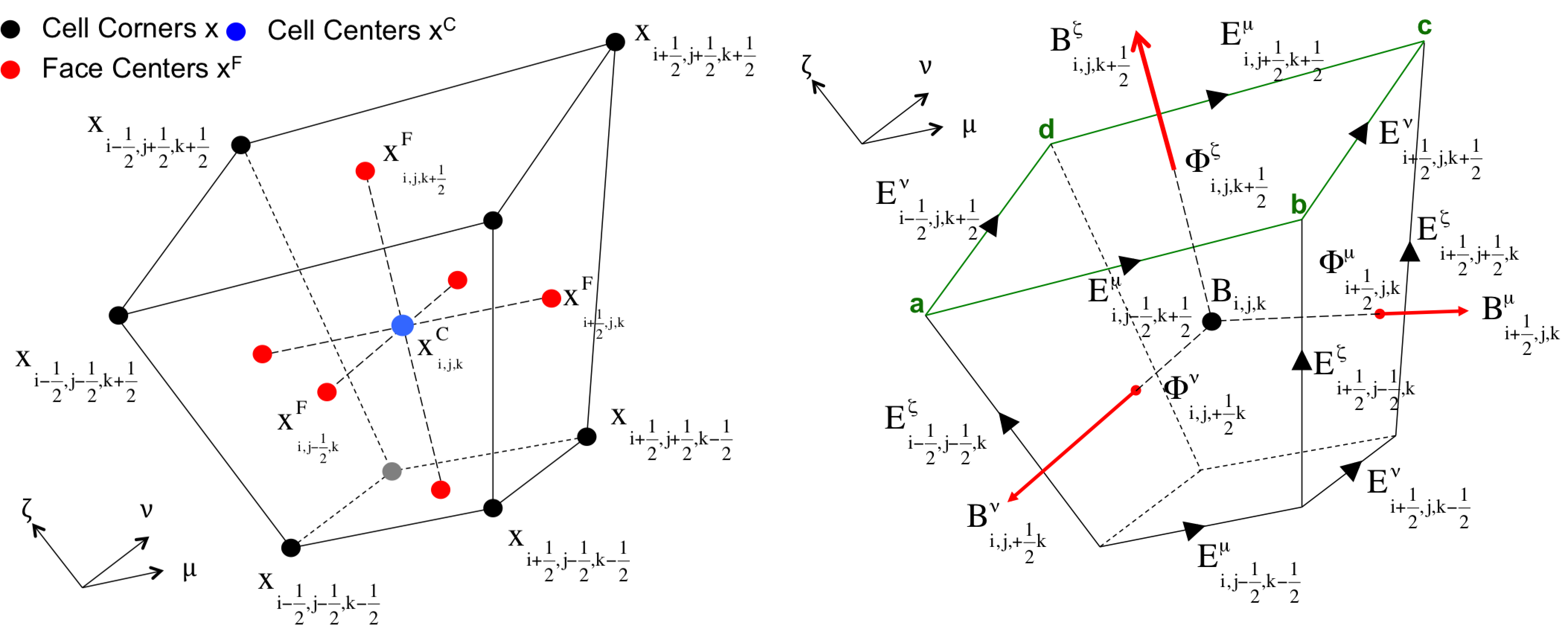}%
	\centering
	\caption{\textbf{Left}) The grid definition for cell center $\mathbf{x}^C$, cell face $\mathbf{x}^F$ and cell corner $\mathbf{x}$ locations in a computational cell. \textbf{Right}) The staggered grid definition for magnetic fluxes/fields and electric fields.}\label{fig:FV-grid-def}
\end{figure}

The basic cell structure defined in the finite-volume solver is that of a general hexahedron without the requirement of coordinate orthogonality as shown in Figure \ref{fig:FV-grid-def}. Although the physical cells used in the solver can be general hexahedra, the grid structure is logically Cartesian in the computational space. In other words, the grid used in the MHD solver is curvilinear. In the finite volume method, the differential form of the MHD equations (\ref{eqn:MHD1}) -- (\ref{eqn:MHD3}) are volume-integrated over individual cells. Thus the quantities then referred to are not the point values, as they would be in a finite difference formulation, but the volume-averaged values within a computational cell. Using Gauss's law, the conservative part of the volume-integrated form of Equation (\ref{eqn:FV-vector_form}) becomes a surface integral:
\begin{equation}
\frac{\partial}{\partial t} \int_V \mathbf{U^C} dV= \frac{\partial}{\partial t}\mathbf{U^V}= \int_V\nabla\cdot\mathbf{F}\left(\mathbf{U^C}\right) dV = \oint\mathbf{F}\left(\mathbf{U^C}\right)\cdot d\mathbf{S},\label{eqn:fv_form}
\end{equation}
where $\mathbf{U^V}$ is the vector form of the volume-integrated conserved variables as defined in (\ref{eqn:vol-int-variable_def}). The Lorentz terms $\mathbf{M}\left(\mathbf{U^C}\right)$ in (\ref{eqn:FV-vector_form}) are treated in a similar way using an operator splitting technique, which is discussed in Section \ref{sec:pim}. The differential form of the Maxwell's equation (\ref{eqn:MHD4}) is integrated over cell interfaces such that the actual magnetic variable evolved in the code is the magnetic flux through cell interfaces, which is discussed in detail in Section \ref{sec:CT}. While it is possible to solve for the non-Cartesian representations of the velocity and magnetic field using a finite volume technique, it generally leads to geometry-related source terms in the integral form of the MHD equations in which cell integrals must be done explicitly in (\ref{eqn:fv_form}). Therefore in the GAMERA code, all the cell-centered vector components are represented as \textit{Cartesian} fields in order to avoid such source terms originating from general curvilinear coordinates. In the GAMERA code, the Cartesian components of the cell-centered velocity $\mathbf{u}$ ($u_x, u_y, u_z$) and magnetic field vectors $\mathbf{B}$ ($B_x,B_y,B_z$) are defined in the ``Base Cartesian Coordinate System''.

In the following sections, the Base Cartesian coordinates are denoted as $(x,y,z)$, while the curvilinear coordinates are denoted as $(\mu,\nu,\zeta)$ as shown in Figure \ref{fig:FV-grid-def}. In the computational space, the cell-centered index of the curvilinear grid is given by integers $i,j$ and $k$ for the $\mu$-, $\nu$- and $\zeta$-direction, respectively. Thus the corresponding cell corner and face index is represented using half integers such as $i\pm\frac{1}{2},j\pm\frac{1}{2},k\pm\frac{1}{2}$, respectively. Following these index notations, the spatial locations of the ``cell centers'' are denoted as $\mathbf{x}_{i,j,k}^C$, the ``cell corners'' are then located at $\mathbf{x}_{i\pm\frac{1}{2},j\pm\frac{1}{2},k\pm\frac{1}{2}}$ given that the $\pm\frac{1}{2}$ index indicates displacement halfway between cell centers. The spatial location of the face centers at the $\mu$-, $\nu$- and $\zeta$-face are denoted as $\mathbf{x}_{i\pm\frac{1}{2},j,k}^\mu$, $\mathbf{x}_{i,j\pm\frac{1}{2},k}^\nu$  and $\mathbf{x}_{i,j,k\pm\frac{1}{2}}^\zeta$, respectively. Note that in the GAMERA code, the \textit{primary} grid is the cell corner coordinates $\mathbf{x}_{i+\frac{1}{2},j+\frac{1}{2},k+\frac{1}{2}}$, while cell centers, face centers and other grid metric calculations are derived from the cell corners. These grid metric calculations include the cell volume, face area and face-normal unit vector. Using the indices defined for the curvilinear grid, the cell volume is denoted as $V_{i,j,k}$, the face areas at the $\mu$-, $\nu$- and $\zeta$-interface are denoted as $A^\mu_{i\pm\frac{1}{2},j,k}$, $A^\nu_{i,j\pm\frac{1}{2},k}$ and $A^\zeta_{i,j,k\pm\frac{1}{2}}$, respectively. Another set of grid metric quantity used in the MHD solver for the calculations of interface fluxes is the unit vectors of the interface normal directions, which is denoted as $\mathbf{n}^\mu_{i,j+\frac{1}{2},k}$, $\mathbf{n}^\nu_{i,j+\frac{1}{2},k}$ and $\mathbf{n}^\zeta_{i,j+\frac{1}{2},k}$ for the $\mu$-, $\nu$- and $\zeta$-interface, respectively. The grid definition and notations used in the following sections are listed in Table 1. The calculations of cell volume ($V$), face-center locations ($\mathbf{x}^F$) and face area ($A$) use the $12^{th}$-order Gaussian quadrature and are described in Appendix B. The face-normal vector and the associated coordinate transforms are used in computing numerical fluxes  through the cell interfaces, which is introduced in Section \ref{sec:flux_func}.

\begin{table}
\centering
\begin{tabular}{llcl}
\hline
Grid Variable & Grid Location & Grid Index      & Notation used in paper                     \\ \hline
$\mathbf{x}$ (primary)  & Cell corners      & $i+\frac{1}{2},j+\frac{1}{2},k+\frac{1}{2}$ & $\mathbf{x}_{i+\frac{1}{2},j+\frac{1}{2},k+\frac{1}{2}}$ \\
$\mathbf{x}^C$          & Cell centers      & $i,j,k$                                     & $\mathbf{x}^C_{i,j,k}$                                   \\
                        & face centers ($\mu$)& $i+\frac{1}{2},j,k$                         & $\mathbf{x}^\mu_{i+\frac{1}{2},j,k}$              \\ 
$\mathbf{x}^F$          & face centers ($\nu$)& $i,j+\frac{1}{2},k$                         & $\mathbf{x}^\nu_{i,j+\frac{1}{2},k}$             \\
                        & face centers ($\zeta$)&  $i,j,k+\frac{1}{2}$                      & $\mathbf{x}^\zeta_{i,j,k+\frac{1}{2}}$    \\ \hline
$Volume$                & Cell volume      & $i,j,k$                                     & $V_{i,j,k}$                                   \\
                        & Face area ($\mu$)& $i+\frac{1}{2},j,k$                         & $A^\mu_{i+\frac{1}{2},j,k}$              \\ 
$A^F$                   & Face area ($\nu$)& $i,j+\frac{1}{2},k$                         & $A^\nu_{i,j+\frac{1}{2},k}$             \\
                        & Face area ($\zeta$)&  $i,j,k+\frac{1}{2}$                      & $A^\zeta_{i,j,k+\frac{1}{2}}$    \\
                        & Face normal unit vector ($\mu$)& $i+\frac{1}{2},j,k$                         & $\mathbf{n}_{\mu,i+\frac{1}{2},j,k}$              \\ 
$\mathbf{n}_F$          & Face normal unit vector ($\nu$)& $i,j+\frac{1}{2},k$                         & $\mathbf{n}_{\nu,i,j+\frac{1}{2},k}$             \\
                        & Face normal unit vector ($\zeta$)&  $i,j,k+\frac{1}{2}$                      & $\mathbf{n}_{\zeta,i,j,k+\frac{1}{2}}$    \\ \hline
 \end{tabular}\label{table:var}
 \caption{The grid definitions and index notations used in the MHD solver.}
\end{table}

Based on the finite-volume discretization shown in Figure \ref{fig:FV-grid-def}, for a single control volume indexed as ($i,j,k$), the volume-integrated form of Equation (\ref{eqn:fv_form}) becomes:
\begin{equation}
\begin{split}
\frac{\partial}{\partial t} \mathbf{U}_{i,j,k}^V & = \left( A_{i-\frac{1}{2},j,k}^\mu\mathbf{n}^\mu_{i-\frac{1}{2},j,k}\cdot\mathbf{F}^\mu_{i-\frac{1}{2},j,k} + A_{i+\frac{1}{2},j,k}^\mu\mathbf{n}^\mu_{i+\frac{1}{2},j,k}\cdot\mathbf{F}^\mu_{i+\frac{1}{2},j,k}\right) \\
& + \left( A^\nu_{i,j-\frac{1}{2},k}\mathbf{n}^\nu_{i,j-\frac{1}{2},k}\cdot\mathbf{F}^\nu_{i,j-\frac{1}{2},k} + A^\nu_{i,j+\frac{1}{2},k}\mathbf{n}^\nu_{i,j+\frac{1}{2},k}\cdot\mathbf{F}^\nu_{i,j+\frac{1}{2},k}\right) \\
& + \left( A^\zeta_{i,j,k-\frac{1}{2}}\mathbf{n}^\zeta_{i,j,k-\frac{1}{2}}\cdot\mathbf{F}^\zeta_{i,j,k-\frac{1}{2}} + A^\zeta_{i,j,k+\frac{1}{2}}\mathbf{n}^\zeta_{i,j,k+\frac{1}{2}}\cdot\mathbf{F}^\zeta_{i,j,k+\frac{1}{2}}\right)
\end{split}\label{eqn:FV_integral_form}
\end{equation}
where $\mathbf{U}_{i,j,k}^V$ is the volume-integrated conservative quantities and $\mathbf{F}$ is the numerical flux through the corresponding cell interfaces. 

\subsubsection{Variable Definitions}

The MHD variables used in the GAMERA code are listed in Table 2. The detailed definitions and calculations about these variables are described in the following sections.

\begin{table}[htb!]
\centering
\begin{tabular}{llclc}
\hline
Variable Type     & Variable Notation & Location defined      & Property  & Coordinates                     \\ \hline
\                  & $\rho$        & cell center              & scalar     & $-$             \\ 
Plasma State ($\mathbf{U}$)  & $\mathbf{u}$ ($u_x,u_y,u_z$)  & cell center              & vector     & Base-$(x,y,z)$  \\
                  & $P$           & cell center              & scalar     & $-$             \\ \hline
Magnetic Flux ($\Phi$)         & $\Phi^{\mu,\nu,\zeta}$ ($\Phi_x,\Phi_y,\Phi_z$)      & face     & scalar     & $-$  \\
Magnetic Field ($\mathbf{B}$)  & $\mathbf{B}^{x,y,z}$ ($B_x,B_y,B_z$)       & cell center     & vector     & Base-$(x,y,z)$  \\
Electric Field ($\mathbf{E}$)  & $\mathbf{E}^{\mu,\nu,\zeta}$ ($E_\mu,E_\nu,E_\zeta$) & edge     & vector     & edge aligned-$(\mu,\nu,\zeta)$  \\ \hline
                        \end{tabular}\label{tab:mhd}
                        \caption{The variable definitions in the MHD solver.}                       
\end{table}

\subsubsection*{Plasma variables}

The fluid state vector $\mathbf{U}_{i,j,k}^V$ of the mass, momentum and energy in the GAMERA code is defined as the volume integral of the conserved densities $\mathbf{U}_{i,j,k}^C$ over the controlled volume $(i,j,k)$:
\begin{equation}
\mathbf{U}_{i,j,k}^V = \int_{k-\frac{1}{2}}^{k+\frac{1}{2}}d\zeta\int_{j-\frac{1}{2}}^{j+\frac{1}{2}}d\nu\int_{i-\frac{1}{2}}^{i+\frac{1}{2}}\mathbf{U}^C\left(x,y,z\right) d\mu = V_{i,j,k}\cdot\mathbf{U}^C_{i,j,k}. \label{Volume_int}
\end{equation}
Here, $\mathbf{U}_{i,j,k}^C$ are the mean densities (mass density, momentum density, energy density) within the cell $(i,j,k)$. As a result, what is actually conserved in the MHD solver is the volume-integrated quantities $\mathbf{U}_{i,j,k}^V$. In the reconstruction algorithm introduced in Section \ref{sec:recon}, the high-order reconstruction method for interface value used in the GAMERA code is based on the volume-integrated conserved variables $\mathbf{U}_{i,j,k}^V$ such that the variation of grid geometry is taken into account in the reconstruction process for non-uniform grids. The limiting step for the left- and right-state at an interface is applied to the primitive state vector $\mathbf{U}_{i,j,k}^P$ without the impact of variation from grid geometry. The details are described in Section \ref{sec:recon}.

\subsubsection*{Magnetic Fluxes and Fields}

The MHD solver adapts the Yee-Grid \citep{Yee1966} to non-orthogonal geometries in order to enforce the conservation of the volume-integrated magnetic divergence to round-off error, which is illustrated in the right panel of Figure \ref{fig:FV-grid-def}. Therefore the primary electric and magnetic variables are defined on a staggered, non-orthogonal grid that are not co-located with the plasma variables at the cell centers. In the GAMERA code, the primary magnetic field variables evolved by the Faraday's law are the \textit{magnetic flux} $\Phi$ through cell faces. Using the $\Phi^\mu$ component shown in the right panel of Figure \ref{fig:FV-grid-def} as an example, the magnetic flux $\Phi_{i+\frac{1}{2},j,k}^\mu$ threading a face in the $\mu$-direction indexed as $(i+\frac{1}{2},j,k)$ is calculated as
\begin{equation}
\Phi^\mu_{i+\frac{1}{2},j,k}=\int_{k-\frac{1}{2}}^{k+\frac{1}{2}}d\zeta\int_{j-\frac{1}{2}}^{j+\frac{1}{2}}d\nu\mathbf{B}\cdot\mathbf{A}^\mu = \mathbf{B}^\mu_{i+\frac{1}{2},j,k}\cdot\mathbf{n}^\mu_{i+\frac{1}{2},j,k}A^\mu_{i+\frac{1}{2},j,k}
\end{equation}
where $\mathbf{B}^\mu_{i+\frac{1}{2},j,k}$ is the mean magnetic field vector at the center of a cell face in the $\mu$-direction. In the GAMERA code, since we track the magnetic flux $\Phi$, the face-centered magnetic field vectors at face centers are nowhere computed directly. Using constrained transport based on the staggered grid, the magnetic fluxes $\Phi^{\mu,\nu,\zeta}$ are evolved using the cell-edge centered electric fields (or electric potential) surrounding the face flux through the integral form of the Faraday's law. For example using the $\Phi_{i,j,k+\frac{1}{2}}^\zeta$ component at the $(i,j,k+\frac{1}{2})$ cell face as an example shown in the right panel of Figure \ref{fig:FV-grid-def}, according to the Stokes' theorem, the integral form of the Faraday's law (\ref{eqn:MHD4}) is expressed by integrating the electric field along the green contour surrounding $\Phi_{i,j,k+\frac{1}{2}}^\zeta$:
\begin{equation}
\frac{\partial}{\partial t}\Phi_{i,j,k+\frac{1}{2}}^\zeta = \oint_{a-b-c-d-a}\mathbf{E}\cdot d\mathbf{l},\label{eqn:faraday}
\end{equation}
where $\mathbf{E}$ is the electric field component along the edge and $d\mathbf{l}$ is the corresponding length of the cell edge vector. The integral path on the RHS of (\ref{eqn:faraday}) is defined as the closed contour $a-b-c-d-a$ as illustrated in the right panel of Figure \ref{fig:FV-grid-def}. In the MHD solver, the contour integral is done explicitly in a piece-wise way, i.e., the quantities $\mathbf{E}\cdot d\mathbf{l}$ are computed at each cell edges with the electric field $\mathbf{E}$ calculated using high-order reconstruction schemes adapted to cell edges.

Using the staggered discretization of the magnetic flux and electric field, it is straightforward to show that the divergence of the magnetic field $\nabla\cdot\mathbf{B}$ is conserved in an integral sense \citep{Evans1988}, as long as $\nabla\cdot\mathbf{B}=0$ is satisfied initially. In other words the action of an electric field along any cell-edge leaves $\nabla\cdot\mathbf{B}$ unchanged during computations. This staggered discretization of the magnetic field allows non-linear limiters which modify the electric field locally to still conserve the solenoidal behavior of the magnetic field, which is essential to the quality of the numerical solutions to MHD equations.

In order to compute the Lorentz force terms $-\nabla\cdot\left( \overline{\overline{\mathbf{I}}}\frac{B^2}{2} - \mathbf{BB} \right) $ on the bulk plasma using magnetic flux functions described in Section \ref{sec:flux_func}, the Cartesian magnetic field vector denoted as $\mathbf{B}_{i,j,k}^{x,y,z}$ (defined in the Base Cartesian coordinates) co-located at cell centers $\mathbf{x}^C_{i,j,k}$ with plasma variables are needed. To compute cell-centered magnetic field vectors $\mathbf{B}_{i,j,k}^{x,y,z}$ using face-centered magnetic fluxes $\Phi^{\mu,\nu,\zeta}$, consider the volume integral of $\nabla\cdot(\mathbf{rB})$ within a controlled volume $V$ given that $\nabla\cdot\mathbf{B}\equiv0$:
\begin{equation}
\begin{split}
\int_V\nabla\cdot\left(\mathbf{r}\mathbf{B}\right) dV & = \int_V\left(\mathbf{B}\cdot\nabla\right)\mathbf{r}dV+\int_V\left(\nabla\cdot\mathbf{B}\right)\mathbf{r}dV \\
& = \int_V \mathbf{B}\cdot dV \\
& = \overline{\mathbf{B}}V,
\end{split}\label{eqn:bflux2field}
\end{equation}
where $\overline{\mathbf{B}}$ is the mean magnetic field vector within a finite-volume cell, and $V$ is the corresponding cell volume. Applying Gauss's law to the LHS of equation (\ref{eqn:bflux2field}), the volume integral is then converted to the following face integral:
\begin{equation}
\int_V\nabla\cdot\left(\mathbf{r}\mathbf{B}\right) dV = \oint_S\left(\mathbf{r}\mathbf{B}\right)\cdot d\mathbf{S} = \oint_S\mathbf{r}\left(\mathbf{B}\cdot d\mathbf{S}\right).\label{eqn:bflux2field_2}
\end{equation}
Using equations (\ref{eqn:bflux2field}) and (\ref{eqn:bflux2field_2}) with $\nabla\cdot\mathbf{B}\equiv0$, the cell centered magnetic fields $\mathbf{B}$ is computed as
\begin{equation}
\overline{\mathbf{B}} = \frac{1}{V}\oint_S\mathbf{r}\left(\mathbf{B}\cdot d\mathbf{S}\right)=\frac{1}{V}\sum_{S=\mu,\nu,\zeta}\mathbf{r}^S\Phi^S.\label{eqn:bflux2field_3}
\end{equation}
The RHS of equation (\ref{eqn:bflux2field_3}) is a face integral using the primary magnetic flux variables $\Phi^{\mu,\nu,\zeta}$. In the GAMERA code, a second order approximation based on the face-centered magnetic flux is used to approximate the face integral on the RHS of equation (\ref{eqn:bflux2field_3}), i.e., the cell-centered magnetic fields $\mathbf{B}_{i,j,k}^{xyz}$ using estimations of magnetic flux at cell centers are computed as
\begin{equation}
\begin{split}
\mathbf{B}_{i,j,k}^{xyz} = \frac{1}{V_{i,j,k}}&\left[\left(\Phi^\mu_{i+\frac{1}{2},j,k}\mathbf{x}^\mu_{i+\frac{1}{2},j,k}-\Phi^\mu_{i-\frac{1}{2},j,k}\mathbf{x}^\mu_{i-\frac{1}{2},j,k}\right)\right.+ \\
&  \left(\Phi^\nu_{i,j+\frac{1}{2},k}\mathbf{x}^\nu_{i,j+\frac{1}{2},k}-\Phi^\mu_{i,j-\frac{1}{2},k}\mathbf{x}^\nu_{i,j-\frac{1}{2},k}\right)+ \\
&  \left.\left(\Phi^\zeta_{i,j,k+\frac{1}{2}}\mathbf{x}^\zeta_{i,j,k+\frac{1}{2}}-\Phi^\mu_{i,j,k-\frac{1}{2}}\mathbf{x}^\zeta_{i,j,k-\frac{1}{2}}\right) \right].\label{eqn:bflux2field_4}
\end{split}
\end{equation}
The above formulation is simple and works for any hexahedral cell defined by its corners. Note that $\nabla\cdot\mathbf{B}=0$ is required in the above equation calculating the Cartesian components of the magnetic fields located at cell centers.

\subsection{Time Stepping for Temporal Evolution}

\begin{figure}[b!] 
	\noindent\includegraphics[width=35pc]{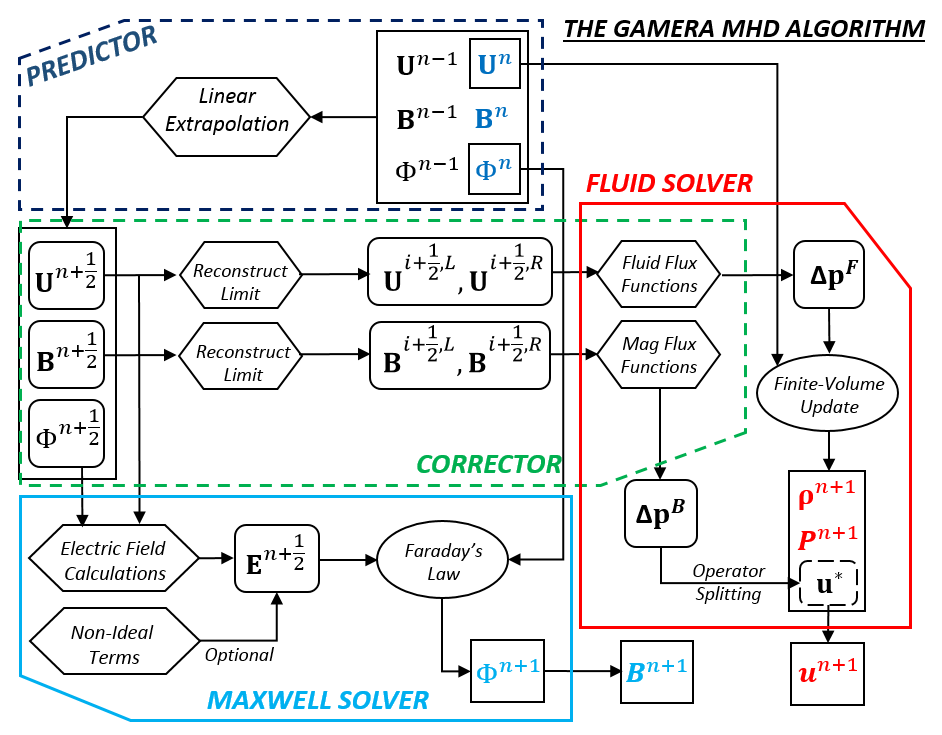}%
	\centering
	\caption{The main GAMERA MHD algorithm, including the predictor step, the corrector step, the fluid solver and the Maxwell solver.}\label{fig:code_structure}
\end{figure}

This section provides a general description of the numerical algorithms used in the GAMERA code for evolving the MHD equations (\ref{eqn:MHD1})-(\ref{eqn:MHD4}), including the time-stepping methods, the Courant condition calculations, the operator splitting technique for handling the magnetic terms $\mathbf{M(U)}$ in (\ref{eqn:FV-vector_form}) and the framework for handling system of equations in non-orthogonal, curvilinear geometries. Figure \ref{fig:code_structure} illustrates the main algorithms of the MHD solver including the predictor/corrector steps, the fluid solver for evolving the plasma variables in Equations (\ref{eqn:MHD1})-(\ref{eqn:MHD3}), and the Maxwell solver updating the magnetic flux/fields in Equation (\ref{eqn:MHD4}). 

For time-stepping we use the Adams-Bashforth second order scheme as the default algorithm. The predictor step is a linear extrapolation to the half time level ($t_{n+\frac{1}{2}}$) based on the state variables of the current ($t_n$) and previous ($t_{n-1}$) time steps. For the fluid part of the MHD equations, the corrector step is a full flux calculation for the integral form of Equation (\ref{eqn:MHD1})-(\ref{eqn:MHD3}) based on the predicted state using the Partial Interface Method (PIM) developed by \citet{Lyon2004}, which is denoted as the ``Fluid Solver'' and introduced in Section \ref{sec:pim}. For evolving the integral form of the Faraday's law, the corrector step is the calculation of electric fields along cell edges using a constrained transport method based on the predicted velocity and magnetic flux, which is denoted as the ``Maxwell Solver'' and described in Section \ref{sec:CT}. Extensions to multi-stage time stepping methods such as Runge-Kutta are possible based on the existing algorithms of the GAMERA code with appropriate modifications.

\subsubsection{The Predictor Step}

Assuming $Q^{n-1}$ is a state variable at time step $n-1$, $Q^{n}$ is the corresponding state variable at time step $n$. The predictor step for $Q^{n+\frac{1}{2}}$ at $n+\frac{1}{2}$ is computed using the following linear extrapolation in time:
\begin{equation}
Q^{n+\frac{1}{2}} = Q^{n} + \frac{\Delta t^n}{2\Delta t^{n-1}}\left(Q^{n} - Q^{n-1}\right),
\end{equation}
where $\Delta t^{n-1}$ and $\Delta t^n$ are the time steps at $t=n-1$ and $t=n$, respectively. In the GAMERA code, the predictor steps are applied to the primitive state variables $\mathbf{U}^P$ (for fluid flux/stress calculations), the cell centered Cartesian magnetic field components $\mathbf{B}^{x,y,z}$ (for magnetic stress calculations) and the face magnetic fluxes $\Phi^{\mu,\nu,\zeta}$ (for electric field calculations), respectively. Once the predictor step is done, the following corrector step is applied to the discrete form of Equation (\ref{eqn:FV_integral_form}) for evolving the fluid part of the MHD equations:
\begin{equation}
\begin{split}
\mathbf{U}^{V,n+1}  & =  \mathbf{U}^{V,n} + \Delta t^n\cdot\mathbf{F}\left(\mathbf{U}^{P,n+\frac{1}{2}}\right)+\Delta t^n\cdot\mathbf{M}\left(\mathbf{U}^{P,n+\frac{1}{2}},\mathbf{B}^{n+\frac{1}{2}}\right) \\
& = \mathbf{U}^{V,n} + \Delta\mathbf{U}_F + \Delta\mathbf{U}_B,
\end{split}\label{eqn:corrector}
\end{equation}
where $\mathbf{F}\left(\mathbf{U}^{P,n+\frac{1}{2}}\right)$ is a fluid flux calculation using the primitive variables $\mathbf{U}^{P,n+\frac{1}{2}}$ at half time step, $\mathbf{M}\left(\mathbf{U}^{P,n+\frac{1}{2}},\mathbf{B}^{n+\frac{1}{2}}\right)$  is the magnetic stress calculation using both $\mathbf{U}^{P,n+\frac{1}{2}}$ and $\mathbf{B}^{n+\frac{1}{2}}$. $\Delta\mathbf{U}_F$ is the volume-integrated changes in mass, momentum and energy from the fluid flux/forces, and $\Delta\mathbf{U}_B$ is the corresponding changes originating from the magnetic forces. The detailed algorithms for computing the $\Delta\mathbf{U}_F$ and $\Delta\mathbf{U}_B$ terms are discussed in the Section \ref{sec:flux_func}. For the integral form of the Faraday's law (\ref{eqn:faraday}), the corrector step follows
\begin{equation}
\Phi^{n+1}  = \Phi^{n} +\Delta t^n\oint\mathbf{E}^{n+\frac{1}{2}}\left( \Phi^{n+\frac{1}{2}},\mathbf{U}^{P,n+\frac{1}{2}} \right)\cdot d\mathbf{l},
\end{equation}
where $\mathbf{E}^{n+\frac{1}{2}}$ is the edge electric field calculated using the predicted variables $\Phi^{n+\frac{1}{2}}$ and $\mathbf{U}^{P,n+\frac{1}{2}}$.  The details of the electric field and magnetic flux evolution computations are described in Section \ref{sec:CT}.

This second-order Adams-Bashforth time stepping scheme is simple and robust with the presence of non-linear limiters with the Total variation diminishing (TVD) property \citep{Lyon2004}. The main advantage of using the Adams-Bashforth scheme is the small amount of memory and small number of computations needed for the predictor step. Thus the second-order Adams Bashforth method is chosen as the default method in the GAMERA code.  Similarly, a third-order Adams-Bashforth scheme for the predictor is written as
\begin{equation}
Q^{n+\frac{1}{2}} = \frac{15}{8}Q^{n} - \frac{5}{4}Q^{n-1}+\frac{3}{8}Q^{n-2},
\end{equation}
which has slightly better stability properties since the leading term in the temporal truncation error is a $4^{th}$-order derivative term. However, the third-order Adams-Bashforth scheme requires more memory comparied to the second-order method, since two previous time steps of the state vector $Q^{n-1}$ and $Q^{n-2}$ are involved in the calculation of the predictor, assuming $\Delta t_{n-2}=\Delta t_{n-1}=\Delta t_{n}$. 

Since the default Adams-Bashforth scheme is explicit in time, the time step $\Delta t$ is determined by the Courant-Friedrichs-Levy (CFL) condition. The explicit time step is calculated as:
\begin{equation}
\Delta t = N_{CFL}\cdot MIN\left(\frac{\left\langle\Delta x\right\rangle}{v}\right),
\end{equation}
where $N_{CFL}$ is the Courant Number between 0 and 1, $\left\langle\Delta x\right\rangle$ is an effective minimum ``length'' of a hexahedron cell as shown in Figure \ref{fig:FV-grid-def}. In a computational cell indexed as $(i,j,k)$, the corresponding minimum cell length $\left\langle\Delta x\right\rangle$ is approximated as the cell volume $V_{i,j,k}$  divided by the maximum face area:
\begin{equation}
\left\langle\Delta x\right\rangle_{i,j,k} = \frac{V_{i,j,k}}{MAX\left(A^\mu_{i\pm\frac{1}{2},j,k},A^\nu_{i,j\pm\frac{1}{2},k}, A^\zeta_{i,j,k\pm\frac{1}{2}}\right)},
\end{equation}
and $v$ is the maximum wave speed within the computational cell. For the ideal MHD equations, the maximum wave speed is calculated as 
\begin{equation}
v = |u| + \sqrt{V_A^2+C_S^2},
\end{equation}
where $|u|$ is the magnitude of the plasma bulk velocity, $\sqrt{V_A^2+C_S^2}$ is the magnetosonic speed calculated using the Alfv\'{e}n speed $V_A$ and the plasma sound speed $C_S$ within cell ($i,j,k$). The Courant number $N_{CFL}$ used in the default GAMERA solver is a function of numerical diffusion introduced in the default non-linear limiter \citep{Hain1987}, which is described in Appendix C.



\subsubsection{The Corrector Step}\label{sec:pim}

In the corrector step, we use the ``\textit{Partial Interface Method}'' (PIM) described by \citet{Lyon2004} in the LFM global magnetospheric model for handling the system of fluid equations in multi-dimensional, non-orthogonal geometries. The PIM provides a general framework that has the advantage of being readily adapted to arbitrary high-order spatial reconstruction methods as well as underlying numerical flux functions. The default reconstruction algorithms and flux functions used in the GAMERA code are described in the next section.

The PIM algorithm loops over each curvilinear direction and is split into the following steps:

\textit{Step 1}. Start from the predictor $\mathbf{U}^{n+\frac{1}{2}}_i$ and $\mathbf{B}^{n+\frac{1}{2}}_i$ in the $\mu$-direction;

\textit{Step 2}. Calculate provisional values $\mathbf{U}^{n+\frac{1}{2}}_{i+\frac{1}{2}}$ and $\mathbf{B}^{n+\frac{1}{2}}_{i+\frac{1}{2}}$ at interfaces;

\textit{Step 3}. Split the provisional values into left- and right-interface states $(\mathbf{U}_{i+\frac{1}{2}}^{L}, \mathbf{U}_{i+\frac{1}{2}}^{R})$ and $(\mathbf{B}_{i+\frac{1}{2}}^{L}, \mathbf{B}_{i+\frac{1}{2}}^{R})$;

\textit{Step 4}. Evaluate numerical flux $\mathbf{F}(\cdot)$ and $\mathbf{M}(\cdot)$ at cell interfaces using left- and right- states;

\textit{Step 5}. Integrate numerical flux $\mathbf{F}$ and $\mathbf{M}$ through $\mu$-interfaces and add to $\Delta\mathbf{U}_F$ and $\Delta\mathbf{U}_B$;

\textit{Step 6}. Repeat \textit{Step 2-5} in the $\nu$- and $\zeta$-direction.

\begin{figure}[b!]
	\noindent\includegraphics[width=35pc]{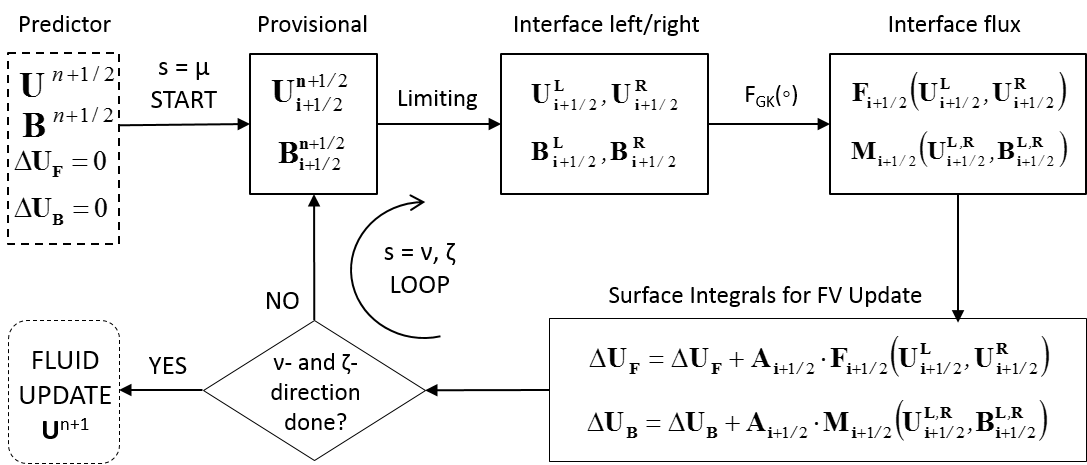}%
	\centering
	\caption{The Partial Interface Method for the corrector step. }\label{fig:pim}
\end{figure}

The algorithm of computing $\Delta\mathbf{U}_F$ and $\Delta\mathbf{U}_B$ based on the predictor state is summarized in Figure \ref{fig:pim}, where steps 2 and 3 are basically in the same vein as a typical one-dimensional TVD scheme for interface states. Note that in the PIM, the reconstruction algorithm is essentially one-dimensional; that is, the provisional values at the cell interfaces are reconstructed along each curvilinear grid coordinate $\mu,\nu,\eta$, and no multi-dimensional corrections are applied to modify the provisional value $\mathbf{U}^{n+\frac{1}{2}}_{i+\frac{1}{2}}$. As a consequence, the corresponding left-state ($\mathbf{U}^L_{i+\frac{1}{2}}$) and right-state ($\mathbf{U}^R_{i+\frac{1}{2}}$) at the cell interface $i+\frac{1}{2}$ are computed in a one-dimensional stencil along the curvilinear direction. In step 4, the default flux functions used in the PIM for both the fluid and magnetic terms are centered flux functions with the following form:
\begin{equation}
\mathbf{F}_{i+\frac{1}{2}}\left(\mathbf{U}^L_{i+\frac{1}{2}},\mathbf{U}^R_{i+\frac{1}{2}}\right)=\mathbf{F}_{GK}\left(\mathbf{U}^L_{i+\frac{1}{2}}\right) + \mathbf{F}_{GK}\left(\mathbf{U}^R_{i+\frac{1}{2}}\right)\label{eqn:PIM}
\end{equation}
\begin{equation}
\mathbf{M}_{i+\frac{1}{2}}\left(\mathbf{U}^L_{i+\frac{1}{2}},\mathbf{U}^R_{i+\frac{1}{2}};\mathbf{B}^L_{i+\frac{1}{2}},\mathbf{B}^R_{i+\frac{1}{2}}\right)=\mathbf{F}_{GK}^{B}\left(\mathbf{U}^L_{i+\frac{1}{2}},\mathbf{B}^L_{i+\frac{1}{2}}\right) + \mathbf{F}_{GK}^{B}\left(\mathbf{U}^R_{i+\frac{1}{2}},\mathbf{B}^R_{i+\frac{1}{2}}\right),\label{eqn:PIM2}
\end{equation}
where $\mathbf{F}_{i+\frac{1}{2}}\left(\cdot\right)$ denotes the fluid flux terms at the cell interface, $\mathbf{M}_{i+\frac{1}{2}}\left(\cdot\right)$ denotes the magnetic stress terms at the cell interface. $\mathbf{F}_{GK}\left(\cdot\right)$ is a Gas-Kinetic flux function (described in Section \ref{sec:flux_func}) for computing the fluid flux using $\mathbf{U}^L_{i+\frac{1}{2}}$ and $\mathbf{U}^R_{i+\frac{1}{2}}$, and $\mathbf{F}_{GK}^B\left(\cdot\right)$ is the corresponding magneto-gas kinetic flux function for the Lorentz terms. The sum of two Gas-Kinetic integrals $\mathbf{F}_{GK}\left(\mathbf{U}^L_{i+\frac{1}{2}}\right)$ and $\mathbf{F}_{GK}\left(\mathbf{U}^R_{i+\frac{1}{2}}\right)$ gives the numerical flux through the interface $i+\frac{1}{2}$. If we choose $\mathbf{U}^L_{i+\frac{1}{2}} = \mathbf{U}_{i}$ and $\mathbf{U}^R_{i+\frac{1}{2}} = \mathbf{U}_{i+1}$, the scheme becomes a robust but overly diffusive first-order Gas-Kinetic flux scheme developed by \citet{Croisille1995}. Note that if $\mathbf{U}^L = \mathbf{U}^R = \mathbf{U}$, the flux function (\ref{eqn:PIM}) returns $\mathbf{F}_{i+\frac{1}{2}}\left(\mathbf{U},\mathbf{U}\right)=\mathbf{F}\left(\mathbf{U}^C\right)$ as defined in equation (\ref{eqn:FV-vector_form}), which is the exact form of the fluid part of the ideal MHD equations. It is only when $\mathbf{U}^L\neq\mathbf{U}^R$ that any numerical dissipation is introduced, although the amount of numerical dissipation introduced by the Gas-Kinetic scheme $\mathbf{F}_{GK}\left(\cdot\right)$ is implicit. The numerical diffusion of that scheme is not as obvious as a first-order Rusanov scheme and has no simple description as does Rusanov. In the case of PIM, many different numerical flux functions can also be used to replace the default gas-kinetic scheme. For example a global heliosphere implementation of the LFM code used the HLL function function for solving the MHD equations \citep{NcNutt1999}. Note that the default GAMERA MHD solver uses a $7^{th}$-order reconstruction method to calculate $\mathbf{U}^L_{i+\frac{1}{2}}$ and $\mathbf{U}^R_{i+\frac{1}{2}}$. With such high-order reconstruction schemes, standard MHD test simulations suggest that the numerical solutions are not sensitive to the choice of the low-order flux function used in the PIM. 

In Step 5, the surface integral evaluations for the mass and energy flux are straightforward since the corresponding directions of mass flux and energy flux are normal to cell interfaces. The face integrals for momentum flux calculations in $\Delta\mathbf{U}_F$ require coordinate transform from the local face-normal coordinate system (defined in Section \ref{sec:face_normal_coord}) to the Base Cartesian System in order to evolve the Cartesian components of the plasma momentum vector. This coordinate transformation enables the fluid solver to track the Cartesian components of the plasma momentum without the requirement of grid orthogonality.

When $\Delta\mathbf{U}_F$ and $\Delta\mathbf{U}_B$ are calculated using the PIM algorithm, an operator splitting technique is used to evolve the fluid part of the MHD equations (\ref{eqn:MHD1})-(\ref{eqn:MHD3}). This operator splitting technique enables solving for the plasma pressure $P$ without implementing the $\mathbf{u}\cdot\nabla\cdot\left( \overline{\overline{\mathbf{I}}}\frac{B^2}{2} - \mathbf{BB} \right)$ term explicitly in the plasma energy equation. The algorithm shown in Figure \ref{fig:code_structure} illustrates the process of updating the plasma state vector using the operator splitting technique described above. The first step of the operator splitting is solving the ideal gas dynamics equations using $\Delta\mathbf{U}_F$ only:
\begin{equation}
\mathbf{U}^{V,n+1} = \mathbf{U}^{V,n} + \Delta\mathbf{U}_F,
\end{equation}
and get $\rho^{n+1}$ and $P^{n+1}$ with an intermediate velocity update $\mathbf{u^*}$, then apply the Lorentz force terms $\Delta\mathbf{U}_B$ to the intermediate velocity $\mathbf{u^*}$ for the final update of velocity $\mathbf{u}^{n+1}$:
\begin{equation}
\mathbf{u}^{n+1} = \mathbf{u^*} + \Delta t\frac{\Delta\mathbf{U}_B}{\rho^{n+1}}.
\end{equation}

The operator splitting technique is implemented by separating the fluid force terms $\nabla\cdot(\rho\mathbf{u}\mathbf{u}+\overline{\mathbf{I}}P)$ and the magnetic force terms $\nabla\cdot\left( \overline{\mathbf{I}}\frac{B^2}{2} - \mathbf{BB} \right)$ in the momentum equations, thus the plasma pressure $P$ can be solved directly from the ideal gas dynamics part of the MHD equations (\ref{eqn:MHD1})-(\ref{eqn:MHD3}) before applying the Lorentz force $\nabla\cdot\left( \overline{\mathbf{I}}\frac{B^2}{2} - \mathbf{BB} \right)$ in the momentum equations, and the $\mathbf{u}\cdot\nabla\cdot\left( \overline{\mathbf{I}}\frac{B^2}{2} - \mathbf{BB} \right)$ term is no longer needed in solving the thermal pressure $P$ from the plasma energy equation at each time step. The operator splitting technique simplifies the implementation of the semi-relativistic (Boris) correction \citep{Boris1970} and the multi-fluid extension \citep{Brambles2011} for planetary magnetosphere simulations significantly. 

\subsection{Calculation of Interface States}

In this section, we describe the default algorithms for computing the left-state variables ($\mathbf{U}^L$) and right-state variables ($\mathbf{U}^R$) at cell interfaces for flux evaluations in the PIM framework shown in Figure \ref{fig:pim}. These interface values are computed through two consecutive steps: 1) a high-order reconstruction step and 2) a limiting step. In the reconstruction step, the high-order reconstruction is performed on the volume-integrated state variables $\mathbf{U}^V$ (mass, momentum, plasma energy) in order to incorporate the variations in grid geometry. After the reconstruction step, the reconstructed $\mathbf{U}^V$ at cell interfaces are then converted to the conservative densities $\mathbf{U}^C$ with the geometry variations removed. In the limiting step, the reconstructed $\mathbf{U}^C$ at cell interface is first converted to the primitive state $\mathbf{U}^P$ (including density, velocity, pressure and magnetic fields). These interface primitive variables at cell interfaces are then split into left- and right-state using the non-linear limiter algorithm. In the GAMERA code, the reconstruction and the limiting modules are implemented separately in order to make the choice of numerical schemes flexible. In the following sections we introduce the high-order reconstruction schemes and the Partial Donor Cell Method (PDM) limiter implemented as the default in the MHD solver.

\subsubsection{High-Order Reconstruction\label{sec:recon}}

\begin{figure}[htb!]
	\noindent\includegraphics[width=30pc]{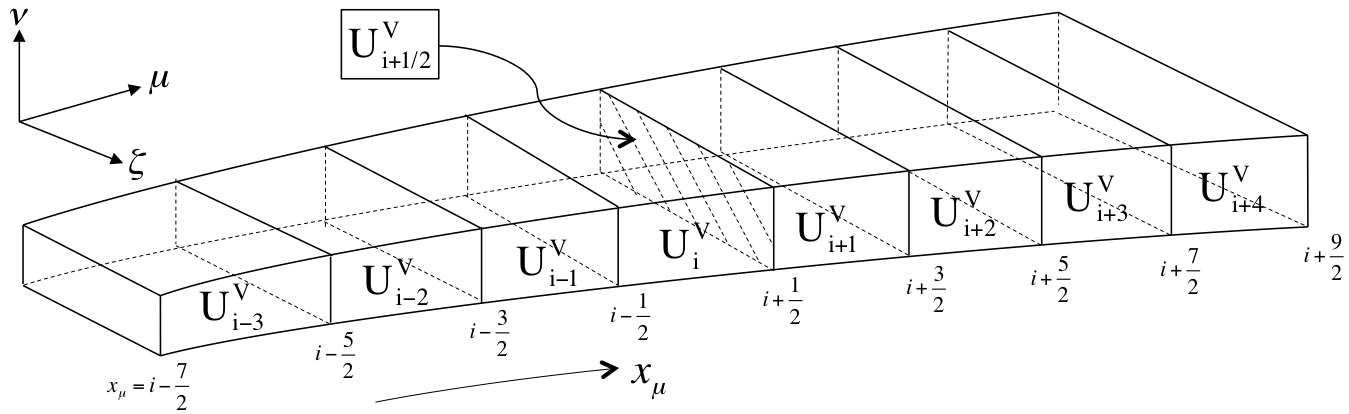}%
	\centering
	\caption{A one-dimensional, high-order reconstruction stencil for computing $\mathbf{U}^{V}_{i+\frac{1}{2},j,k}$ in the $\mu$-direction. }\label{fig:recon}
\end{figure}

The interface values used in the GAMERA code for flux calculations are high-order approximations of the primitive variables $\mathbf{U}^P$ and magnetic fields $\mathbf{B}^{xyz}$ at cell interfaces. To incorporate spatial variations originating from curvilinear geometries, the high-order reconstruction scheme operates on the volume-integrated fluid variables $\mathbf{U}^V$, $V\cdot\mathbf{B}^{xyz}$ defined at cell centers and $\Phi^{\mu\nu\zeta}$ at cell faces. In the PIM framework, the reconstruction process for computing interface values is one-dimensional along each curvilinear direction, without multi-dimensional corrections which simplifies the calculation in non-orthogonal geometries. For example, in order to estimate high-order approximations for interface states $\mathbf{U}^V_{i+\frac{1}{2}}$ along the $\mu$-direction, first define a one-dimensional integral function $G(x_\mu)$ along the $\mu$-direction as
\begin{equation}
G(x_\mu)=\int_{-\infty}^{x_\mu}\mathbf{U}^V(\xi)d\xi,
\end{equation}
where $\mathbf{U}^V$ is the vector form of the volume-integrated conservative variables, $x_\mu$ is the curvilinear spatial coordinate along the $\mu$-direction in the computational space as shown in Figure \ref{fig:recon}. Therefore the conserved quantity $\mathbf{U}^V_{i+\frac{1}{2}}$ at a cell interface $i+\frac{1}{2}$ along the $\mu$-direction is calculated as the first derivative of $G(x_\mu)$ evaluated at $x_\mu=\frac{1}{2}$:
\begin{equation}
\mathbf{U}^V_{i+\frac{1}{2}}=\frac{\partial}{\partial x_\mu}G\left(x_\mu\right)\biggr\vert_{x_\mu=i+\frac{1}{2}}.\label{derivative}
\end{equation}

The order of the reconstruction method to get interface values $\mathbf{U}^V_{i+\frac{1}{2}}$ depends on how accurately the first derivative term in Equation (\ref{derivative}) is approximated. An example of an eight-cell reconstruction stencil along the $\mu$-direction is shown in Figure \ref{fig:recon}. If applying a second order, centered approximation for the numerical derivative $\frac{\partial}{\partial x_\mu}G\left(x_\mu\right)$ at the interface $i+\frac{1}{2}$ as an example, the RHS of Equation (\ref{derivative}) is calculated as:
\begin{equation}
\mathbf{U}^V_{i+\frac{1}{2}}=\frac{\partial}{\partial x_\mu}G\left(x_\mu\right)\biggr\vert_{i+\frac{1}{2}} = \frac{G\left(i+\frac{3}{2}\right) - G\left(i-\frac{1}{2}\right)}{\left(i+\frac{3}{2}\right)-\left(i-\frac{1}{2}\right)}+\mathcal{O}\left(\Delta x_\mu^2\right) \approx \frac{\mathbf{U}^V_{i+1}+\mathbf{U}^V_{i}}{2}.
\end{equation}
Note that the integral function $G(x_\mu)$ is defined in a \textit{computational space} between $x_\mu = i-\frac{7}{2}$ and $x_\mu = i+\frac{9}{2}$ since the infomation of cell volume is already included in $G(x_\mu)$. After $\mathbf{U}^V_{i+\frac{1}{2}}$ is computed, the conserved state variables $\mathbf{U}^C_{i+\frac{1}{2}}$ at interface $i+1/2$ are calculated as 
\begin{equation}
\mathbf{U}^C_{i+\frac{1}{2}} =\frac{\mathbf{U}^V_{i+\frac{1}{2}} }{V_{i+\frac{1}{2}}}=\frac{1}{2V_{i+1/2}}\left( \mathbf{U}^V_{i+1}+\mathbf{U}^V_{i} \right),\label{eqn:2nd_order_recon}
\end{equation}
where $V_{i+\frac{1}{2}}$ is an estimation of ``cell volume'' at the corresponding cell interface $i+\frac{1}{2}$ assuming the cell volume $V_i$ varies smoothly in the $\mu$-direction. A $2^{nd}$-order centered approximation of interface volume for the $2^{nd}$-order interface reconstruction in (\ref{eqn:2nd_order_recon}) is calculated as
\begin{equation}
V_{i+\frac{1}{2}} = \frac{1}{2}\left(V_{i+1}+V_i\right)\label{eqn:vol_interp2}.
\end{equation}
Once $\mathbf{U}^C_{i+\frac{1}{2}}$ is calculated, the primitive state variable $\mathbf{U}^P_{i+\frac{1}{2}}$ is then computed for the limiting step described in the next section. The interface magnetic field $\mathbf{B}_{i+\frac{1}{2}}$ is calculated using the same algorithm by replacing $\mathbf{U}^V$ with $V_{i,j,k}\cdot\mathbf{B}_{i,j,k}$. In the original LFM code, the reconstruction module uses the eight-cell stencil shown in Figure \ref{fig:recon} with the following $8^{th}$-order centered reconstruction for computing high-order estimations of interface values:

\begin{equation}
\mathbf{U}^C_{i+\frac{1}{2}} = \frac{1}{V_{i+1/2}^{8th}}\left( -\frac{3}{840}\mathbf{U}^V_{i-3} + \frac{29}{840}\mathbf{U}^V_{i-2}  - \frac{139}{840}\mathbf{U}^V_{i-1} + \frac{533}{840}\mathbf{U}^V_{i} + \frac{533}{840}\mathbf{U}^V_{i+1} - \frac{139}{840}\mathbf{U}^V_{i+2} + \frac{29}{840}\mathbf{U}^V_{i+3} -\frac{3}{840}\mathbf{U}^V_{i+4}.\right), \label{eqn:recon_8}
\end{equation}

where $V_{i+\frac{1}{2}}^{8th}$ is an $8^{th}$-order approximation of the interface volume calculated as:
\begin{equation}
V_{i+\frac{1}{2}}^{8th} = -\frac{3}{840}V_{i-3} + \frac{29}{840}V_{i-2} - \frac{139}{840}V_{i-1} + \frac{533}{840}V_{i} + \frac{533}{840}V_{i+1} - \frac{139}{840}V_{i+2} + \frac{29}{840}V_{i+3} -\frac{3}{840}V_{i+4}. \label{eqn:vol_interp8}
\end{equation}
This $8^{th}$-order centered reconstruction scheme is chosen as the default method for computing interface state variables in the LFM code based on its high-resolving power of contact discontinuities according to \citet{Lyon2004}.
 
In fact, the reconstruction method for estimating a high-order interface value is not necessarily centered. High-order upwind reconstruction schemes are also implemented in the GAMERA code. Using the second-order centered reconstruction as an example, the interface variable is reconstructed as
\begin{equation}
\mathbf{U}^V_{i+\frac{1}{2}} = \frac{1}{2} \left(\mathbf{U}^V_{i}+\mathbf{U}^V_{i+1}\right).
\end{equation}

Combine the above second order approximation for $\mathbf{U}^V_{i+\frac{1}{2}}$ with a second-order centered approximation for the first derivative of $\mathbf{U}^V$ evaluated at $i+\frac{1}{2}$, the centered reconstruction becomes the first-order upwind method \citep{LeVeque2002}:
\begin{equation}
\mathbf{U}^V_{i+\frac{1}{2}} = \mathbf{U}^V\biggr\vert_{i+\frac{1}{2}}^{2nd} - \frac{1}{2}\frac{\partial}{\partial x}\mathbf{U}^V\biggr\vert_{i+\frac{1}{2}}^{2nd} = \frac{1}{2} \left(\mathbf{U}^V_{i}+\mathbf{U}^V_{i+1}\right) - \frac{1}{2} \left(\mathbf{U}^V_{i+1}-\mathbf{U}^V_{i}\right)=\mathbf{U}^V_{i}\label{eqn:upwind_derive}
\end{equation}
Equation (\ref{eqn:upwind_derive}) suggests that upwind reconstruction with order of $n-1$ can be derived from a $n^{th}$-order ($n$ even) centered stencil by canceling the outer most cell in the downwind direction using a numerical first derivative approximated to $n^{th}$-order accuracy at the cell interface. For example, combining the $8^{th}$ order centered interpolation for $\mathbf{U}^V_{i+\frac{1}{2}}$ with an $8^{th}$-order centered approximation of $\frac{1}{2}\frac{\partial\mathbf{U}^V}{\partial x_\mu}$, the following $7^{th}$-order upwind reconstruction method is obtained:
\begin{equation}
\mathbf{U}^V_{i+\frac{1}{2}} = -\frac{1}{40}\mathbf{U}^V_{i-3} + \frac{5}{84}\mathbf{U}^V_{i-2} - \frac{101}{420}\mathbf{U}^V_{i-1} + \frac{319}{420}\mathbf{U}^V_{i} + \frac{107}{210}\mathbf{U}^V_{i+1} - \frac{19}{210}\mathbf{U}^V_{i+2} + \frac{1}{105}\mathbf{U}^V_{i+3}.
\end{equation}

\begin{table}[t!]
\centering
\begin{tabular}{|l|l|l|l|l|l|l|l|l|l|l|l|l|}
\hline
         & $\mathbf{U}_{i-5}$     & $\mathbf{U}_{i-4}$    & $\mathbf{U}_{i-3}$    & $\mathbf{U}_{i-2}$   & $\mathbf{U}_{i-1}$      & $\mathbf{U}_{i}$       & $\mathbf{U}_{i+1}$     & $\mathbf{U}_{i+2}$       & $\mathbf{U}_{i+3}$    & $\mathbf{U}_{i+4}$     &  $\mathbf{U}_{i+5}$     & $\mathbf{U}_{i+6}$    \\ \hline
$1^{st}$   &                &                &                &              &                  &   1         &                 &                  &               &                &                  &               \\ \hline
$2^{nd}$  &                &                &                &              &                  & $\frac{1}{2}$         & $\frac{1}{2}$           &                  &               &                &                  &                \\ \hline
$3^{th}$   &                &                &                &              & $-\frac{1}{6}$         & $\frac{5}{6}$         & $\frac{1}{3}$           &                  &               &                &                  &                \\ \hline
$4^{th}$   &                &                &                &              & $-\frac{1}{12}$       & $\frac{7}{12}$       & $\frac{7}{12}$         & $-\frac{1}{12}$       &               &                &                  &                \\ \hline
$5^{th}$   &                &                &                & $\frac{1}{30}$       & $-\frac{13}{60}$       & $\frac{47}{60}$       & $\frac{9}{20}$         & $-\frac{1}{20}$       &               &               &                   &                \\ \hline
$6^{th}$   &                &                &                & $\frac{1}{60}$       & $-\frac{2}{15}$       & $\frac{37}{60}$       & $\frac{37}{60}$         & $-\frac{2}{15}$       & $\frac{1}{60}$        &               &                   &                \\ \hline
$7^{th}$   &                &                & $-\frac{1}{40}$     & $\frac{5}{84}$       &$ -\frac{101}{420}$      & $\frac{319}{420}$     & $\frac{107}{210}$       & $-\frac{19}{210}$     & $\frac{1}{105}$      &               &                   &                \\ \hline
$8^{th}$   &                &                & $-\frac{3}{840}$   & $\frac{29}{840}$     & $-\frac{139}{840}$     & $\frac{533}{840}$     & $\frac{533}{840}$       & $-\frac{139}{840}$     & $\frac{29}{840}$      & $-\frac{3}{840}$  &                   &                \\ \hline
$9^{th}$   &                & $\frac{1}{630}$       & $-\frac{41}{2520}$ & $\frac{199}{2520}$   & $-\frac{641}{2520}$   & $\frac{1879}{2520}$   & $\frac{275}{504}$       & $-\frac{61}{504}$     & $\frac{11}{504}$      & $-\frac{1}{504}$  &                   &                \\ \hline
$10^{th}$&                & $\frac{1}{1260}$     & $-\frac{23}{2520}$ & $\frac{127}{2520}$    & $-\frac{473}{2520}$   & $\frac{747}{1157}$   & $\frac{747}{1157}$     & $-\frac{473}{2520}$   & $\frac{127}{2520}$     & $-\frac{23}{2520}$& $\frac{1}{1260}$        &                \\ \hline
$11^{th}$&-$\frac{1}{2772}$    &$\frac{61}{13860}$     & $-\frac{58}{2287}$ & $\frac{371}{3960}$   & $-\frac{93}{353}$     & $\frac{260}{353}$     & $\frac{420}{737}$         & $-\frac{133}{921}$   & $\frac{339}{9923}$     & $-\frac{17}{3080}$&  $\frac{1}{2310}$       &              \\ \hline
$12^{th}$&$-\frac{1}{5544}$  &$\frac{67}{27720}$    & $-\frac{107}{6930}$ & $\frac{443}{6930}$    & $-\frac{529}{2594}$   & $\frac{356}{545}$     & $\frac{356}{545}$       & $-\frac{529}{2594}$   & $\frac{443}{6930}$    & $-\frac{107}{6930}$&$\frac{67}{27720}$       & $-\frac{1}{5544}$ \\ \hline


\end{tabular}
\caption{Centered and Upwind reconstruction coefficients for $\mathbf{U}_{i+\frac{1}{2}}$, up to $12^{th}$-order. The upwind coefficients are for the left-state interface values with the stencil shifted towards the left. }
\label{table:recon-coeff}
\end{table}

In the above $7^{th}$-order reconstruction, the right-most cell $i+4$ shown in Figure \ref{fig:recon} is removed from the reconstruction stencil, which makes the reconstruction shifted towards left (the upwind direction). Reconstruction coefficients for left interface states from $1^{st}$-order to $12^{th}$-order are listed in Table \ref{table:recon-coeff}. The right interface states are calculated using the same reconstruction coefficients by switching the upwind direction of a reconstruction stencil. For the centered reconstruction methods (e.g., $8^{th}$-order), the left- and right-state are computed using the same eight-cell stencil and coefficients due to symmetry; for the upwind reconstruction methods, (e.g., $7^{th}$-order), the left- and right-state are computed using two different seven-cell stencils shifted towards the left and right side of the interface, respectively. In the GAMERA code, the solver uses the $7^{th}$ order upwind reconstruction scheme as the default choice, while the original LFM MHD kernel uses the $8^{th}$ order centered reconstruction. The $8^{th}$-order reconstruction scheme has a leading truncation error of a dispersive $9^{th}$-order derivative, thus a sudden change in gradient (involving short wavelengths) may excite spurious unphysical oscillations due to the $9^{th}$-derivative dispersion term in the truncation error. However in the $7^{th}$-order reconstruction scheme, the leading truncation-error term is an $8^{th}$-order spatial derivative, which is dissipative instead of dispersive. This is an important improvement in the original LFM algorithm. Numerical experience over the past decade or more has shown that for numerical solutions of convection-dominated fluid flows, the leading truncation error in a convection algorithm should be dissipative (an even spatial derivative term) rather than dispersive (an odd spatial derivative term), which should also be of higher order than modeled physical diffusive terms (if any) \citep{Leonard1991}. Thus the odd-order ($order\geq 3$) reconstruction schemes are more natural choices satisfying the criteria of 1) dissipative truncation error terms and 2) higher-order than physical diffusion terms. Thus the GAMERA code uses a $7^{th}$-order reconstruction method as the default choice for upwind spatial reconstruction, and the $8^{th}$-order reconstruction method is chosen as the default centered method for reference. The reason for choosing such high-order reconstruction methods as the default in the MHD solver is explained in Appendix D, which is based on a simple measure of optimizing between low numerical diffusion and high computing efficiency as well as the need to resolve both physical and grid structure.

\subsubsection{Partial Donor Cell Method}

When discontinuities occur within a reconstruction stencil, a problem arises in using the high-order reconstruction schemes since spurious undershoots or overshoots may occur near discontinues. To avoid this problem, nonlinear switchers are used to ``correct'' the high-order reconstructed values on both sides of the interface when necessary. In the limiting module the \textit{Partial Donor Cell Method} (PDM) limiter developed by \citet{Hain1987} is implemented as the default choice, which does not depend on the numerical order of interface reconstruction. The basic idea of the PDM limiter is that the algorithm monitors sharp discontinuities; if a sharp discontinuity is identified, a ``limited'' value is used in order to the keep the solution from overshoots/undershoots; otherwise a high-order approximation is always chosen for the interface state to provide a more accurate estimation. The details about the PDM limiter can be found in \citet{Hain1987}. Here we provide a simplified description of how the PDM limiter works adapted from \citet{Huba2003}.

\begin{figure}[t!]
	\noindent\includegraphics[width=25pc]{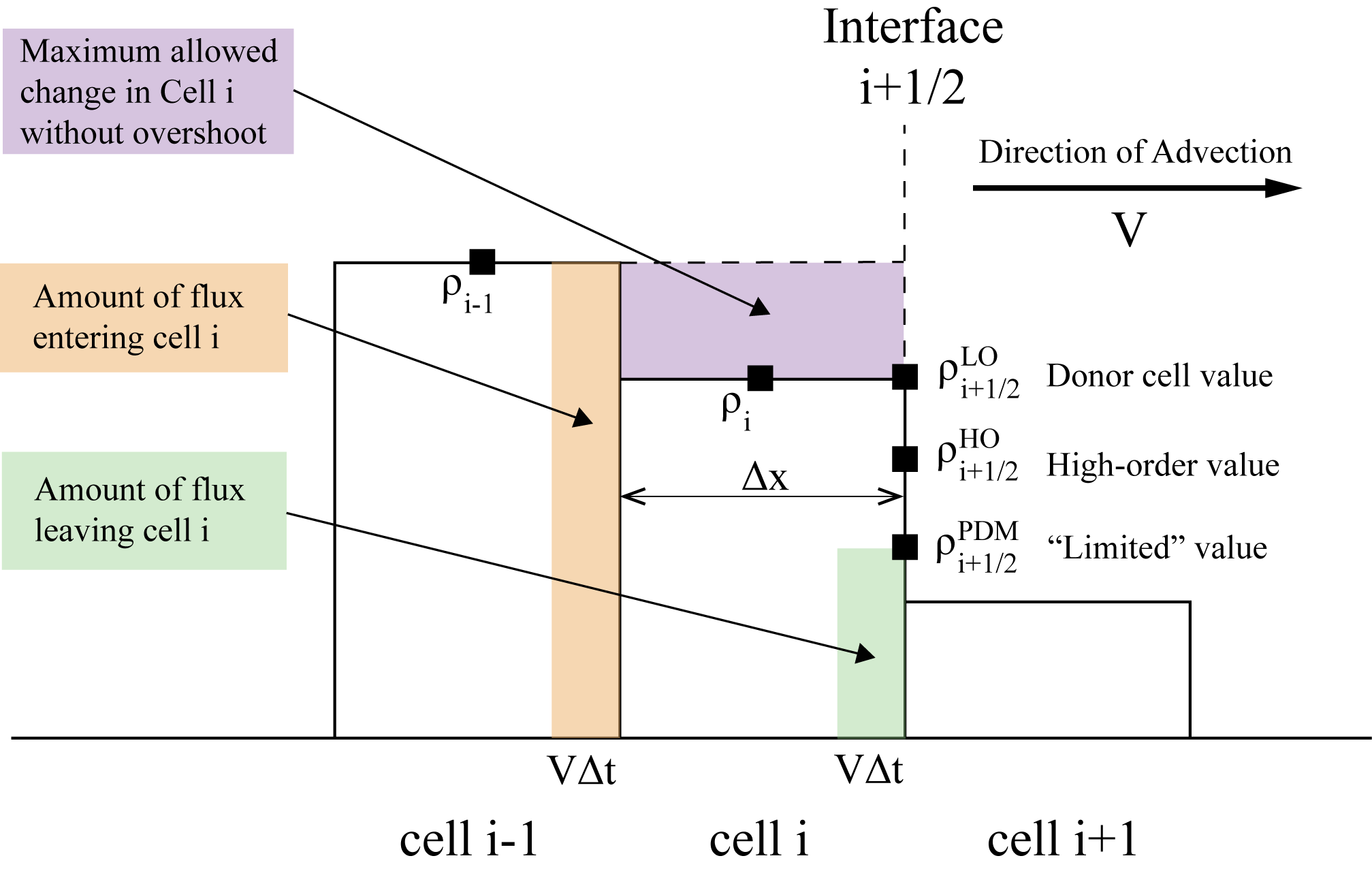}%
	\centering
	\caption{The calculation of the PDM value on the left-side of the interface $i+\frac{1}{2}$, using the advection of density in a one-dimensional stencil towards the right. The figure is adapted from Huba [2003].}\label{fig:PDM}
\end{figure}

Consider a density structure being advected at a constant velocity $V$ towards the right direction as shown in Figure \ref{fig:PDM}, the PDM limiter determines whether the high-order estimation of the left-state value $\rho_{L_{i+\frac{1}{2}}}^{HO}$ at interface $i+1/2$ needs to be ``limited'' for monotonicity preserving purpose. In Figure \ref{fig:PDM}, $\rho_{i-1}$ is the density in cell $i-1$; $\rho_{i}$ is the density in cell $i$, $\rho_{i+1/2}^{HO}$ is the high-order reconstructed density at cell interface $i+1/2$, and $\rho_{i+1/2}^{PDM}$ is the limited value (or the PDM value) of the interface density on the left side which will now be determined.

Assume the density structure is advanced for one time step $\Delta t$, it advects towards the right by a finite distance of $V\Delta t$. After one advection step, the amount of mass entering cell i from cell $i-1$ is $\rho_{i-1}V\Delta t$, which is illustrated by the orange shaded area in Figure \ref{fig:PDM}; the amount of mass leaving cell $i$ and entering cell $i+1$ is $\rho_{i+1/2}^{PDM}V\Delta t$ assuming that the left state is the PDM value that guarantees no  overshooting/undershooting, which is illustrated by the green shaded area in Figure \ref{fig:PDM}. Therefore the total density change in cell $i$ after one time step is calculated as $(\rho_{i-1}-\rho_{i+1/2}^{PDM} )V\Delta t$. Since the maximum density increase allowed in cell $i$ without spurious overshoot is  $(\rho_{i-1}-\rho_i )\Delta x$, which is denoted by the purple area, the left state of the PDM value at cell interface $i+1/2$ is derived by balancing the two quantities:
\begin{equation}
(\rho_{i-1}-\rho_{i+1/2}^{PDM} )V\Delta t = (\rho_{i-1}-\rho_i )\Delta x,
\end{equation}
which gives the PDM value:
\begin{equation}
\rho^{PDM}_{L_{i+\frac{1}{2}}} = \frac{1}{\epsilon}\rho_i - \left(1-\frac{1}{\epsilon}\right)\rho_{i-1},
\end{equation}
where $\epsilon=V\Delta t/\Delta x$ which is in the range between 0 and 1. By balancing the density entering and leaving cell $i$, it is clear that the $\rho_{i+1/2}^{PDM}$ is the \textit{minimum} value of the left state density allowed to leave the cell $i$ without spurious overshoot. If the left state density is lower than this value, spurious density overshoot occurs within cell $i$. By choosing the $\rho_{i+1/2}^{PDM}$ as the left state, the density on the left-side of the interface is ``limited''. Note that if $\epsilon=1$, then $\rho_{i+1/2}^{PDM}=\rho_i$, which is the first-order Donor Cell method with excessive amount of numerical diffusion for non-linear problems. By choosing $\epsilon<1$, the above scheme takes a portion of the Donor Cell solution as the limited value and reduces the amount of numerical diffusion significantly.

To achieve a high-order approximation for the left state, it is not necessary to use the PDM value all the time. As shown in Figure \ref{fig:PDM}, there are three density values to choose as the left state at interface $i+1/2$ : $\rho_i$ (first-order), $\rho_{i+1/2}^{HO}$ ($7^{th}$-order reconstruction as the default) and $\rho_{i+1/2}^{PDM}$ (the limited value). The rationale for choosing the final left state value is as follows. In general, one would want to use the high-order reconstructed value because it provides the best accuracy for estimating the left state. Since $\rho_{i+1/2}^{PDM}$ is the minimum amount of density allowed for a left state leaving cell $i$, as long as $\rho_{i+1/2}^{PDM}<\rho_{i+1/2}^{HO}<\rho_i$, the high-order value $\rho_{i+1/2}^{HO}$ is the correct choice for the left state. However, when $\rho_{i+1/2}^{HO}<\rho_{i+1/2}^{PDM}<\rho_i$, the PDM value $\rho_{i+1/2}^{PDM}$ must be chosen as the left value in order to avoid overshoots. In other words, the left state is always chosen as the median value of the three interface values: $\rho_i$, $\rho_{i+1/2}^{HO}$, and $\rho_{i+1/2}^{PDM}$. The PDM limiter is based on balancing the amount of flux moving through a controlled volume rather than on constraining the slope of the underlying reconstruction polynomial for computing interface values, which is similar to the ``Universal Limiter'' developed by \citet{Leonard1991}. Thus the interface value $\rho_{i+1/2}^{HO}$ can be replaced by arbitrary high-order reconstructed values.  The right-state density is derived in the same way through the advection of a density structure towards the left. The performance of the PDM limiter combined with different orders of reconstruction methods is demonstrated in Appendix D based on quantitative comparisons of 1-D linear advection test results.

\begin{figure}[t!]
	\noindent\includegraphics[width=38pc]{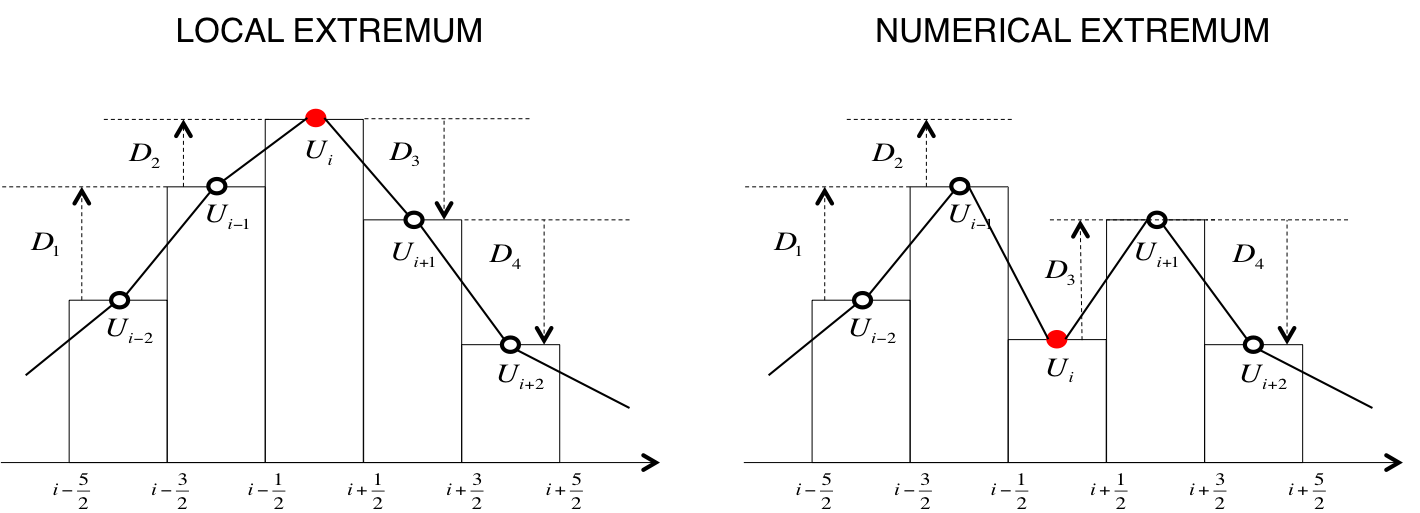}%
	\centering
	\caption{A local maximum versus numerical peaks within a stencil of five computational cells.}\label{fig:non-clip}
\end{figure}

For the linear advection equation, it is straightforward to show that the PDM limiter is TVD \citep{Hain1987}, thus it tends to ``clip'' local extrema as typical TVD limiters do. An optional non-clipping algorithm is implemented in the limiting module to preserve local extrema for simulations that are sensitive to the numerical errors introduced due to clipping of local extrema, e.g., the Harris current sheet equilibrium simulation. A simple non-clipping algorithm developed by \citet{Leonard1991} is implemented in the limiting process in order to distinguish between artificial numerical peaks and true physical extrema. Figure \ref{fig:non-clip} shows the difference between a physical extremum and a numerical extremum within one reconstruction stencil centered at cell $i$. If a local extremum is associated with short-wavelength oscillations, the limited value is chosen for the interface state; however, if the curvature of a peak follows the shape of a local extremum, the unlimited high-order value is chosen to avoid clipping. Using the stencil for calculating the left-state $\mathbf{U}_{i+\frac{1}{2}}$ at the interface $i+\frac{1}{2}$ shown in Figure \ref{fig:non-clip} as an example. The calculation of the local extrema indicator ($LEI$) is illustrated in Figure \ref{fig:non-clip}, by first computing the differences between each pair of consecutive cell center values:
\begin{equation}
D_1 = \mathbf{U}_{i-1} - \mathbf{U}_{i-2}, D_2 = \mathbf{U}_{i} - \mathbf{U}_{i-1}, D_3 = \mathbf{U}_{i+1} - \mathbf{U}_{i}, D_4 = \mathbf{U}_{i+2} - \mathbf{U}_{i+1}.
\end{equation}
Assuming the existence of one local maximum centered in cell $i$ as shown in the left panel of Figure \ref{fig:non-clip} (a minimum requires reversal of some of the subsequent inequalities), both $D_1$ and $D_2$ are positive and $D_3$ and $D_4$ are negative. Using a simple constraint for such a local maximum with decreasing gradient towards the peak, $|D_1|>|D_2|$ and $|D_3|<|D_4|$, the $LEI$ for the left state $\mathbf{U}_{i+\frac{1}{2}}$ in cell  $i$ is calculated as a Bool type of variable: 
\begin{equation}
LEI = (D_1>0)\&(D_2>0)\&(D_3<0)\&(D_4<0)\&(|D_1|>|D_2|)\&(|D_3|<|D_4|).
\end{equation}
When the non-clipping option is switched on, after the limiting step, the final interface state is selected based on the value of the $LEI$, if $LEI=TRUE$, the unlimited, high-order approximation (e.g., the $\rho_{i+1/2}^{HO}$) is used for the interface state regardless of the $\rho_{i+1/2}^{PDM}$ constraint; otherwise if $LEI=FALSE$, the limited value is chosen to avoid possible spurious overshooting/undershooting. 

Combining the reconstruction and the limiting step, the algorithm to compute the left and right-state variables along one-dimensional stencils (e.g., in the $\mu$-direction) is summarized as follows:

\textit{Step} 1. Compute the volume-integrated conserved quantities in each controlled volume using $\mathbf{U}^V_{i,j,k}=V_{i,j,k}\cdot\mathbf{U}_{i,j,k}^C$;

\textit{Step} 2. Compute a high-order (default $7^{th}$-order) reconstructed volume-integrated quantity $\mathbf{U}^V_{i+\frac{1}{2},j,k}$;

\textit{Step} 3. Estimate a high-order (default $8^{th}$-order) volume $V_{i+\frac{1}{2},j,k}$ at the interface location where $\mathbf{U}^V_{i+\frac{1}{2},j,k}$ is defined;

\textit{Step} 4. Compute the reconstructed densities $\mathbf{U}^C_{i+\frac{1}{2},j,k} = \mathbf{U}^V_{i+\frac{1}{2},j,k}/V_{i+\frac{1}{2},j,k}$, then convert to primitive variables $\mathbf{U}^P_{i+\frac{1}{2},j,k}$;

\textit{Step} 5. Split the interface primitive variable $\mathbf{U}^P_{i+\frac{1}{2},j,k}$ into the left- and right-state variables using the PDM limiter.

\textit{Step 6 (optional)}. Apply the non-clipping sweep to decide whether the PDM limiter is needed or not.

\subsection{Gas-Kinetic Flux Functions}\label{sec:flux_func}

The default flux functions for the fluid part of the MHD equations are based on one-dimensional gas-kinetic schemes for the Euler equations in the \textit{face-normal coordinate system} (detailed in Section \ref{sec:face_normal_coord}). In a local, orthogonal coordinate system $(x_n,x_1,x_2)$ with $x_n$ the direction normal to a cell interface and $x_1,x_2$ the two corresponding orthogonal directions tangential to the cell interface, the one-dimensional mass, momentum and plasma energy equations (\ref{eqn:MHD1})-(\ref{eqn:MHD3}) in this $(x_n,x_1,x_2)$ coordinate system are written as:
\begin{equation}
\begin{split}
\frac{\partial}{\partial t}\left[\begin{array}{c} \rho \\ \rho u_{x_n}\\ \rho u_{x_1} \\ \rho u_{x_2} \\ E_P \end{array}\right] = & -\frac{\partial}{\partial x_n}\left[\begin{array}{c} \rho u_{x_n} \\ \rho u_{x_n}^2+P \\ \rho u_{x_n}u_{x_1} \\ \rho u_{x_n}u_{x_2} \\ u_{x_n}(E_P+P) \end{array}\right] - \frac{\partial}{\partial x_n}\left[\begin{array}{c} 0 \\ B^2/2-B_{x_n}^2 \\ B_{x_n} B_{x_1} \\ B_{x_n} B_{x_2} \\ 0 \end{array}\right] + \\
& \left[\begin{array}{c} 0 \\ 0 \\ 0 \\ 0 \\ u_{x_n}\frac{\partial}{\partial x_n}(B^2/2-B^2_{x_n})+u_{x_1}\frac{\partial}{\partial x_1}B_{x_n}B_{x_1}+u_{x_2}\frac{\partial}{\partial x_2}B_{x_n}B_{x_2}\end{array}\right],\label{eqn:1d_mhd}
\end{split}
\end{equation}
where $E_P$ is the plasma energy defined in (\ref{eqn:E_P}), and $B^2/2$ is the magnetic energy calculated as $(B_{x_n}^2+B_{x_1}^2+B_{x_2}^2)/2$. The first part on the RHS of Equation (\ref{eqn:1d_mhd}) corresponds to the ideal gas dynamics with velocity components in the $(x_n,x_1,x_2)$ coordinate system, which gives the $\Delta\mathbf{U}_F$ terms in Equation (\ref{eqn:corrector}). The second part on the RHS of Equation (\ref{eqn:1d_mhd}) is the Lorentz force $\nabla\cdot\left(\overline{\mathbf{I}}\frac{B^2}{2} - \mathbf{BB} \right)$ in the momentum equation (\ref{eqn:MHD2}), which gives the $\Delta\mathbf{U}_B$ terms in Equation (\ref{eqn:corrector}). The third part on the RHS of Equation (\ref{eqn:1d_mhd}) is the work done by the Lorentz force $\mathbf{u}\cdot\nabla\cdot\left( \overline{\mathbf{I}}\frac{B^2}{2} - \mathbf{BB} \right)$ in the plasma energy equation (\ref{eqn:MHD3}), which is not computed explicitly for solving the plasma pressure $P$ due to the operator splitting technique as discussed in Section \ref{sec:pim}. The gas-hydro flux (fluid stress) and the magneto-hydro flux (magnetic stress) for the above one-dimensional MHD equations are calculated separately, which simplifies the implementation of the Semi-relativistic (Boris) correction for magnetospheric simulations. The default solver uses a Maxwellian-based gas-kinetic flux function to compute the fluid fluxes terms $\Delta\mathbf{U}_F$ and uses another Maxwellian-based magneto-gas kinetic flux function to compute the magnetic stresses terms $\Delta\mathbf{U}_B$. 

\subsubsection{Fluid Fluxes}

The default flux functions used in the solver for the fluid flux calculations are gas-kinetic schemes based on Maxwellian distributions, which considers the fluid on each side of a cell interface as a microscopic distributions of non-interacting particles \citep{Croisille1995}. These particles are allowed to free-stream across the interface for one time step, then the macroscopic conserved variables including mass, momentum, and energy are derived from the zeroth, first and second moment integral of the corresponding distribution functions on both sides of the interface. These updated macroscopic variables are then used to construct new distribution functions for the next time step.

Using the left primitive state variables $\mathbf{U}^{P,L} = \left( \rho^L, \mathbf{u}^L, P^L\right)^T$ calculated in the reconstruction module, the distribution function $f^L(v)$ for the plasma population on the left side of the cell interface is a Maxwell-Boltzmann distribution function:
\begin{equation}
f^L(v) = \displaystyle\sqrt[]{\displaystyle\frac{\rho^L}{2\pi P^L}}e^{-\frac{\rho^L}{2P^L}\left(v - u_{x_n}^L\right)^2}.
\end{equation}
Similarly, the corresponding distribution function $f^R(v)$ for the plasma population on the right side of the cell interface is
\begin{equation}
f^R(v) = \displaystyle\sqrt[]{\displaystyle\frac{\rho^R}{2\pi P^R}}e^{-\frac{\rho^R}{2P^R}\left(v - u_{x_n}^R\right)^2},
\end{equation}
where $\rho^L$ and $\rho^R$ are the left and right state of plasma density, $P^L$ and $P^R$ are the left and right state of plasma pressure, $u_{x_n}^L$ and $u_{x_n}^R$ are the left and right plasma bulk velocity in the $x_n$-direction normal to cell interfaces, respectively. The calculations of the $u_{x_n}^L$ and $u_{x_n}^R$ from the interface velocities $\mathbf{u}^L$ and $\mathbf{u}^R$ is described in the next section. For one-dimensional MHD flows described by Equation (\ref{eqn:1d_mhd}) with particles free-streaming in the $x_n$-direction, the moment integrals in this direction determine the form of flux functions. Thus in the local ($x_n,x_1,x_2$) coordinate system the mass flux, the fluid part of the momentum flux and the energy flux across the cell interface $i+\frac{1}{2}$ are calculated by integrating the corresponding distribution functions for the left-going particles on the right-side of the interface and the right-going particles on the left-side of the interface:
\begin{equation}
\mathbf{F}_\textsc{FLUID}^{n}=\left(\begin{array}{c} F_{\rho} \\ F_{\rho\mathbf{u}_{xn}} \\ F_{\rho\mathbf{u}_{x1}} \\ F_{\rho\mathbf{u}_{x2}} \\ F_{E_P} \end{array}\right) = \int_0^{+\infty}f^L(v)\left[\begin{array}{c} \rho^L v \\ \rho v^2+P^L \\ \rho vu_{x_1}^L \\ \rho vu_{x_2}^L \\ v(E_P^L+P^L) \end{array}\right]dv + \int_{-\infty}^{0}f^R(v)\left[\begin{array}{c} \rho^R v \\ \rho^R v^2+P^R \\ \rho vu_{x_1}^R \\ \rho vu_{x_2}^R \\ v(E_P^R+P^R) \end{array}\right] d v, \label{eqn:fluid_stress}
\end{equation}
where $\mathbf{F}_\textsc{FLUID}^{n}$ is the vector form of the fluid fluxes at interface $n$ ($n=\mu,\nu,\zeta$); $F_\rho$ is the mass flux, $\mathbf{F}_{\rho\mathbf{u}}$ is the momentum flux vector and $F_{E_P}$ is the plasma energy flux calculated in the ($x_n,x_1,x_2$) coordinates. $u_{x1,2}^L$ and $u_{x1,2}^R$ are the corresponding left- and right-interface \textit{tangential} velocity components transformed into the local ($x_n,x_1,x_2$) coordinate system, respectively. After the fluid fluxes are calculated, the $\Delta\mathbf{U}_F$ term for the correction step (\ref{eqn:corrector}) is computed as
\begin{equation}
\Delta\mathbf{U}_F = \Delta t\cdot\oint\mathbf{F}^n_\textsc{FLUID}\cdot d\mathbf{S} = \sum_{s=\mu,\nu,\zeta}\mathbf{F}_\textsc{FLUID}^s\cdot A_s\mathbf{n}_s, \label{eqn:delta_PF}
\end{equation}
where $A_s$ is the face area and $\mathbf{n}_s$ is the corresponding face-normal unit vector. If the left and right states are identical, i.e., $\mathbf{U}^L = \mathbf{U}^R \equiv \mathbf{U}$, it is straightforward to show that $f^L(v)\equiv f^R(v)$ and the RHS of one-dimensional ideal gas dynamics equations are recovered through the flux functions (\ref{eqn:fluid_stress}). The detailed derivations for evaluating of the moment integrals in Equation (\ref{eqn:fluid_stress}) can be found in \citep{Xu1999}. Here we only give the mathematical expressions of the Maxwellian distribution based gas-kinetic flux scheme for computing fluid fluxes:
\begin{equation}
\begin{split} 
\mathbf{F}_{\textsc{fluid}}=&\left\langle v^1 \right\rangle^L_{+}\left(\begin{array}{c} \rho_L \\ \rho_L u_{x_n}^L\\ \rho^L u_{x_1}^L \\ \rho^L u_{x_2}^L \\ E_P^L \end{array}\right)+\left(\begin{array}{c} 0 \\ P^L\left\langle v^0\right\rangle^L_{+} \\ 0 \\ 0 \\ P^L\left\langle v^1\right\rangle^L_{+}/2+P^Lu_{x_n}^L\left\langle v^0\right\rangle^L_{+} \end{array}\right)+ \\ & \left\langle v^1 \right\rangle^R_{-}\left(\begin{array}{c} \rho_R \\ \rho_R u_{x_n}^R\\ \rho^R u_{x_1}^R \\ \rho^R u_{x_2}^R \\ E_P^R \end{array}\right)+\left(\begin{array}{c} 0 \\ P^L\left\langle v^0\right\rangle^R_{-} \\ 0 \\ 0 \\ P^R\left\langle v^1\right\rangle^R_{-}/2+P^Ru_{x_n}^R\left\langle v^0\right\rangle^R_{-} \end{array}\right).\label{eqn:fluid_flux1}
\end{split}
\end{equation}
The zeroth and the first velocity moments of the left and right distribution functions used in Equations (\ref{eqn:fluid_flux1}) are calculated as:
\begin{equation}
\left\langle v^0\right\rangle^L_{+} = \int_0^{+\infty}v^0f^L(v)dv = \frac{1}{2}\mathrm{erfc}\left(-u^L_{x_n}\sqrt{\lambda^L}\right),\:\: \left\langle v^0\right\rangle^R_{-} = \int_{-\infty}^{0}v^0f^R(v)dv = \frac{1}{2}\mathrm{erfc}\left(-u^R_{x_n}\sqrt{\lambda^R}\right),
\end{equation}
\begin{equation}
\left\langle v^1\right\rangle^L_{+} = \int_0^{+\infty}v^1f^L(v)dv = u^L_{x_n}\left\langle v^0\right\rangle^L_{+} + \frac{1}{2}\displaystyle\frac{e^{-\lambda^L {u^L_{x_n}}^2}}{\sqrt{\pi\lambda^L}}, \:\: \left\langle v^1\right\rangle^R_{-} = \int_{-\infty}^{0}v^1f^R(v)dv = u^R_{x_n}\left\langle v^0\right\rangle^R_{-} + \frac{1}{2}\displaystyle\frac{e^{-\lambda^R {u^R_{x_n}}^2}}{\sqrt{\pi\lambda^R}},
\end{equation}
where $\lambda^L=\rho^L/2P^L$ and $\lambda^R=\rho^R/2P^R$. The erfc($\cdot$) function is the \textit{Complementary Error Function}. It is straightforward to show that if $f^L(v)= f^R(v)\equiv f(v)$, the following relationships hold:
\begin{equation}
\left\langle v^0\right\rangle_{+} + \left\langle v^0\right\rangle_{-}=1, 
\end{equation}
\begin{equation}
\left\langle v^1\right\rangle_{+} + \left\langle v^1\right\rangle_{-}=u_{x_n}. 
\end{equation}

The mass flux $\mathbf{F}_\rho$ and energy flux $\mathbf{F}_{E_P}$ in equations (\ref{eqn:fluid_flux1}) are in the $x_n$-direction normal to cell interfaces:
\begin{equation}
\mathbf{F}_\rho = \left( \left\langle v^1 \right\rangle^L_{+}\rho_L + \left\langle v^1 \right\rangle^R_{-}\rho_R \right)\mathbf{n}_{x_n},\label{eqn:mass_flux_func}
\end{equation}
\begin{equation}
\mathbf{F}_{E_P} = \left( \left\langle v^1 \right\rangle^L_{+}E_P^L + P^L\left\langle v^1\right\rangle^L_{+}/2+P^Lu_{x_n}^L\left\langle v^0\right\rangle^L_{+} + P^R\left\langle v^1\right\rangle^R_{-}/2+P^Ru_{x_n}^R\left\langle v^0\right\rangle^R_{-}\left\langle v^1 \right\rangle^R_{-}E_P^R \right)\mathbf{n}_{x_n}.\label{eqn:energy_flux_func}
\end{equation}
Thus the mass and energy flux are used to compute the surface integrals in Equation (\ref{eqn:delta_PF}) directly:
\begin{equation}
\Delta\mathbf{U}_{F}\big|_\rho = \Delta t\sum_{s=\mu,\nu,\zeta}\mathbf{F}_\rho^s\cdot A_s\mathbf{n}_s \label{eqn:delta_U_rho}
\end{equation}
\begin{equation}
\Delta\mathbf{U}_{F}\big|_{E_P} = \Delta t\sum_{s=\mu,\nu,\zeta}\mathbf{F}_{E_P}^s\cdot A_s\mathbf{n}_s. \label{eqn:delta_U_Ep}
\end{equation}
However, the momentum flux vector $\mathbf{F}_{\rho\mathbf{u}}$ calculated using equation (\ref{eqn:fluid_flux1}) is defined in the local ($x_n,x_1,x_2$) coordinate system:
\begin{equation}
\begin{split}
\mathbf{F}_{\rho\mathbf{u}} =  & \left( \left\langle v^1 \right\rangle^L_{+}\rho_L u_{x_n}^L + P^L\left\langle v^0\right\rangle^L_{+} + \left\langle v^1 \right\rangle^R_{-}\rho_R u_{x_n}^R + P^L\left\langle v^0\right\rangle^R_{-} \right)\mathbf{n}_{x_n}+ \\
& \left( \left\langle v^1 \right\rangle^L_{+}\rho^L u_{x_1}^L + \left\langle v^1 \right\rangle^R_{-}\rho^R u_{x_1}^R \right)\mathbf{n}_{x_1}+ \\
& \left( \left\langle v^1 \right\rangle^L_{+}\rho^L u_{x_2}^L + \left\langle v^1 \right\rangle^R_{-}\rho^R u_{x_2}^R \right)\mathbf{n}_{x_2}, \\
\end{split}\label{eqn:momentum_flux_func},
\end{equation}
which cannot be used to compute the surface integrals in Equation (\ref{eqn:delta_PF}) directly in order to update the base Cartesian components of the momentum ($\rho u_x, \rho u_y, \rho u_z$). Thus after evaluating the fluid fluxes in the face-normal coordinate system ($x_n,x_1,x_2$) using the one-dimensional flux function form (\ref{eqn:fluid_flux1}), the momentum flux vector $\mathbf{F}_{\rho\mathbf{u}}$ at cell interface is then rotated back to the base Cartesian coordinate system for updating the Cartesian component of the momentum ($\rho u_x, \rho u_y,\rho u_z$). Then the changes to the plasma momentum $\Delta\mathbf{U}_{F}\big|_{\rho\mathbf{u}}$ in the base Cartesian system is computed as:
\begin{equation}
\Delta\mathbf{U}_{F,\rho\mathbf{u}} = \Delta t\sum_{s=\mu,\nu,\zeta}{\overline{\mathbf{T}}}^{-1}\cdot\mathbf{F}_{\rho\mathbf{u}}^s A_s, \label{eqn:dalta_U_mom}
\end{equation}
where $\overline{\mathbf{T}}^{-1}$ is a matrix for transforming the momentum flux vector from the ($x_n,x_1,x_2$) system to the base Cartesian system ($x,y,z$). The detailed calculation of the transform matrix $\overline{\mathbf{T}}^{-1}$ is described in Section \ref{sec:face_normal_coord}. This rotation method for evaluating momentum fluxes in the base Cartesian coordinate system $(x,y,z)$ does not require the orthogonality of the physical grid and works with any microscopic distribution functions in the gas-kinetic flux scheme. 

The gas-hydro distribution function for the fluid flux calculations does not depend on the local Alfv\'{e}n speed, therefore the numerical diffusion speed near discontinuities, although non-explicit, is only related to the bulk flow and plasma thermal speed. In very low-beta plasma simulations (e.g., $\beta < 10^{-4}$) such that magnetosonic waves dominate the dynamics, the use of thermal speed in the gas-hydro distribution functions above is not adequate for describing the behavior of plasma associated with Alfv\'{e}n waves. Thus additional numerical diffusion  is needed to damp spurious oscillations at the Alfv\'{e}n speed when the left and right states are not equal. For gas-kinetic fluid flux functions, the numerical diffusion speed for the fluid flux is adjusted based on the average Alfv\'{e}n speed at the interface, which is simply implemented via incorporating an additional diffusion term that follows a Rusanov type of numerical flux calculation centered at cell interfaces:
\begin{equation}
\Delta\mathbf{U}_F = \Delta\mathbf{U}_F + \frac{1}{2}V_{A_{i+\frac{1}{2}}}\left(\begin{array}{c} \rho_L - \rho_R \\ \rho_L u_x^L - \rho_R u_x^R \\ \rho_L u_y^L - \rho_R u_y^R \\ \rho_L u_z^L -\rho_R u_z^R \\ E_P^L - E_P^R \end{array}\right)\cdot\Delta t A_s,\label{eqn:hogs}
\end{equation}
where $V_{A_{i+\frac{1}{2}}}$ is chosen to be the average Alfv\'{e}n speed $V_A$ at the cell interface $i+\frac{1}{2}$:
\begin{equation}
V_{A_{i+\frac{1}{2}}} = \frac{1}{2}\left( V_{A^L}+V_{A^R}\right).
\end{equation}
Note that the change of the momentum flux vector $\Delta\mathbf{U}_{F}\big |_{\rho\mathbf{u}}$ in Equation (\ref{eqn:hogs}) has already been transformed back to the base ($x,y,z$) system. Thus the diffusion terms for the momentum fluxes are computed using the Cartesian components of the interface bulk velocities ($\mathbf{u}^L$ and $\mathbf{u}^R$) directly. The additional diffusion terms in Equation (\ref{eqn:hogs}) only apply to the interfaces where the left-state and right-state variables are not equal. In smooth flow regions, the left- and right-state variables are equal or the differences are negligibly small, no numerical diffusion is introduced to the interface fluxes through the above flux function.

\subsubsection{Magnetic Stresses\label{sec:mag_stress}}

The calculations of magnetic stress terms go through a similar process as computing the fluid fluxes, using similar Maxwellian-based, magneto-gas kinetic distribution functions for the magnetic fields on the left- and right-side of the interfaces \citep{Xu1999}:
\begin{equation}
f^L_B(v) = \displaystyle\sqrt{\displaystyle\frac{\rho^L}{2\pi P^L_{\textsc{tot}}}}\displaystyle e^{-\frac{\rho^L}{2P^L_{\textsc{tot}}}\left(v - u_{x_n}^L\right)^2},\label{eqn:mag_distri1}
\end{equation}
\begin{equation}
f^R_B(v) = \displaystyle\sqrt{\displaystyle\frac{\rho^R}{2\pi P^R_{\textsc{tot}}}}\displaystyle e^{-\frac{\rho^R}{2P^R_{\textsc{tot}}}\left(v - u_{x_n}^R\right)^2}.\label{eqn:mag_distri2}
\end{equation}
where $P_\textsc{tot}^L$ and $P_\textsc{tot}^R$ are the corresponding the left and right total pressure defined as $P_\textsc{tot}^L=\frac{1}{2}\left({B_x^L}^2+{B_y^L}^2+{B_z^L}^2\right)+P^L$ and $P_\textsc{tot}^R=\frac{1}{2}\left({B_x^R}^2+{B_y^R}^2+{B_z^R}^2\right)+P^R$, with ($B_x^L$, $B_y^L$,$B_z^L$) and ($B_x^L$, $B_y^L$,$B_z^L$) the interface magnetic fields reconstructed from the cell-centered magnetic fields $\mathbf{B}^{xyz}$. This choice for the magnetic distribution function is arbitrary since the actual magnetic field is not a physically distributed quantity in the microscopic plasma velocity space $v$. The idea of introducing a velocity distribution function for the magnetic field is to spread the plasma information in the magnetic field such that the bulk plasma and the magnetic fields are coupled through the Alfv\'{e}n speed. 

In the magnetic flux functions, the magnetic stress tensor applied at the interface is calculated through similar moment integrals as in Equation (\ref{eqn:fluid_stress}) using the magneto-gas kinetic distribution functions  $f^L_B(v)$ and $f^R_B(v)$: 
\begin{equation}
\begin{split}
\overline{\mathbf{S}}_\textsc{mag} &= \int_0^{+\infty}\left[\frac{1}{2}\left(B^L\right)^2\overline{\mathbf{I}} - \mathbf{B}^L\mathbf{B}^L\right]f^L_B(v)dv + \int_{-\infty}^{0}\left[\frac{1}{2}\left(B^R\right)^2\overline{\mathbf{I}} - \mathbf{B}^R\mathbf{B}^R\right]f^R_B(v)dv \\
& = \left\langle v_B^0\right\rangle^L_{+}\left[\frac{1}{2}\left(B^L\right)^2\overline{\mathbf{I}} - \mathbf{B}^L\mathbf{B}^L\right] + \left\langle v_B^0\right\rangle^R_{-}\left[\frac{1}{2}\left(B^R\right)^2\overline{\mathbf{I}} - \mathbf{B}^R\mathbf{B}^R\right],
\end{split}\label{eqn:stress_calc}
\end{equation}
where $\mathbf{B}^L$ and $\mathbf{B}^R$ are the Cartesian components of interface magnetic fields on the left- and right-side of the interface, respectively. $(B^L)^2/2=\frac{1}{2}\left[(B_x^L)^2+(B_y^L)^2+(B_z^L)^2\right]$ is the magnetic energy on the left side of the interface, and $(B^R)^2/2=\frac{1}{2}\left[(B_x^R)^2+(B_y^R)^2+(B_z^R)^2\right]$ is the magnetic energy on the right side of the interface. $\left\langle v_B^0\right\rangle^L_{+}$ and $\left\langle v_B^0\right\rangle^R_{-}$ are zeroth moments of the left and right magneto-gas kinetic distribution functions:
\begin{equation}
\left\langle v^0_B\right\rangle^L_{+} = \int_0^{+\infty}v^0f^L_B(v)dv = \frac{1}{2}\mathrm{erfc}\left(-u^L_{x_n}\sqrt{\lambda_B^L}\right),\:\: \left\langle v_B^0\right\rangle^R_{-} = \int_{-\infty}^{0}v^0f^R_B(v)dv = \frac{1}{2}\mathrm{erfc}\left(-u^R_{x_n}\sqrt{\lambda_B^R}\right),
\end{equation}
with $\lambda_B^L=\rho^L/2P^L_\textsc{tot}$ and $\lambda_B^R=\rho^R/2P^R_\textsc{tot}$. The evaluation of the moment integrals in Equation (\ref{eqn:stress_calc}) for the magnetic stress $\overline{\mathbf{S}}_\textsc{mag}$ is straightforward since the magnetic stress tensor $\left(\frac{1}{2} B^2\bar{\mathbf{I}}-\mathbf{B}\mathbf{B}\right)$ is not a function of the bulk velocity. Thus the left and right magnetic stresses applied on a cell interface are calculated only using the zeroth moment of the magneto-gas kinetic distribution functions $f_B^L(v)$ and $f_B^L(v)$. 

Unlike the momentum flux calculations, Equation (\ref{eqn:stress_calc}) for the magnetic stress $\overline{\mathbf{S}}_\textsc{mag}$ is evaluated directly in the base ($x,y,z$) coordinate system, because only zeroth-moment of the magneto-gas kinetic distribution functions are involved. Thus no face-normal coordinate transform/inverse transform are needed as used in the calculations of fluid stress terms. According to Gauss's law, the volume-integrated Lorentz force $\Delta\mathbf{U}_B$ within a controlled cell $V$ is the surface integral of the magnetic stress tensor:
\begin{equation}
\begin{split}
\Delta\mathbf{U}_B &=\int_V \nabla\cdot\left(\frac{1}{2} B^2\bar{\mathbf{I}}-\mathbf{B}\mathbf{B}\right)dV = \oint\overline{\mathbf{S}}_\textsc{mag}\cdot d\mathbf{S}=\sum_{s=\mu,\nu,\zeta}\left(\overline{\mathbf{S}}_\textsc{mag}\cdot\mathbf{n}_sA_s\right)\\
&=\sum_{s=\mu,\nu,\zeta}\left[ \left\langle v_B^0\right\rangle^L_{+}\left[\frac{1}{2}\left(B^L\right)^2\overline{\mathbf{I}} - \mathbf{B}^L\mathbf{B}^L\right]\cdot\mathbf{n}_sA_s + \left\langle v_B^0\right\rangle^R_{-}\left[\frac{1}{2}\left(B^R\right)^2\overline{\mathbf{I}} - \mathbf{B}^R\mathbf{B}^R\right]\cdot\mathbf{n}_sA_s\right]\\
&=\sum_{s=\mu,\nu,\zeta}\left( \left\langle v_B^0\right\rangle^L_{+}\left[\frac{1}{2}\left(B^L\right)^2\mathbf{n}_sA_s - \mathbf{B}^L\Phi^s\right] + \left\langle v_B^0\right\rangle^R_{-}\left[\frac{1}{2}\left(B^R\right)^2\mathbf{n}_sA_s - \mathbf{B}^R\Phi^s\right]\right).
\end{split}\label{eqn:stress_calc_face}
\end{equation}

In the above flux function, $\mathbf{B}^L\cdot\mathbf{n}_s = \mathbf{B}^R\cdot\mathbf{n}_s = \Phi^s$ is the interface magnetic flux tracked in the Maxwell solver, which does not require the reconstruction and splitting procedure. The fact that $\mathbf{B}^L\cdot\mathbf{n}_s = \mathbf{B}^R\cdot\mathbf{n}_s$ is consistent with the requirement of the continuity condition of $B_{x_n}$ in the one-dimensional flux function as discussed in \citep{Xu1999}. 

The above calculations for the magnetic stress terms can be adapted to incorporate background magnetic field terms for very low-$\beta$ plasma simulations such as the inner magnetosphere of the Earth. The details of implementing background fields can be problem specific but in general, a background field $\mathbf{B}_0$ can be included in the magnetic stress calculations using the following algorithm:


\begin{align}
\Delta\mathbf{U}_B & =\int_V \nabla\cdot\left[\frac{1}{2} \left(\mathbf{B}+\mathbf{B}_0\right)^2\bar{\mathbf{I}}-\left(\mathbf{B}+\mathbf{B}_0\right)\left(\mathbf{B}+\mathbf{B}_0\right)\right]dV = \oint\overline{\overline{\mathbf{S}'}}_\textsc{mag}\cdot d\mathbf{S} \\ 
&=\sum_{s=\mu,\nu,\zeta}\left(\overline{\overline{\mathbf{S}'}}_\textsc{mag}\cdot\mathbf{n}_sA_s\right)\\
&=\sum_{s=\mu,\nu,\zeta}\left( \left\langle v_B^0\right\rangle^L_{+}\left[\frac{1}{2}\left({B^L}^2+2\mathbf{B}^L\cdot\mathbf{B}_0+B_0^2\right)\mathbf{n}_sA_s - \left(\mathbf{B}^L+\mathbf{B}_0\right)\left(\Phi^s + \Phi_{0s}\right)\right]\right) \nonumber \\ & +\sum_{s=\mu,\nu,\zeta}\left( \left\langle v_B^0\right\rangle^R_{-}\left[\frac{1}{2}\left({B^R}^2+2\mathbf{B}^R\cdot\mathbf{B}_0+B_0^2\right)\mathbf{n}_sA_s - \left(\mathbf{B}^R+\mathbf{B}_0\right)\left(\Phi^s + \Phi_{0s}\right)\right]\right),
\label{eqn:bzero_stress}
\end{align}

where $\frac{1}{2}B_0^2 = \left(B_{0x}^2+B_{0y}^2+B_{0z}^2\right)/2$ and $\Phi_{0s} =\mathbf{B}_0\cdot\mathbf{n}_s$ are the magnetic energy and the magnetic flux of the background field at interface $s$, respectively. The background magnetic field components $\mathbf{B}_0$ on cell interfaces used in Equation (\ref{eqn:bzero_stress}) are computed using a $12^{th}$-order 2-D Gaussian quadrature rather than point evaluations to improve the implementation of force-free background magnetic fields, i.e., $\nabla\times\mathbf{B}_0=0$. For example, at $\mu$-faces:
\begin{equation}
\mathbf{B}_{0_{i+\frac{1}{2},j,k}}^\mu=\frac{1}{A_\mu}\int^{j+\frac{1}{2}}_{j-\frac{1}{2}}\int^{k+\frac{1}{2}}_{k-\frac{1}{2}}\mathbf{B}_0\left(\mu=i+\frac{1}{2},\nu,\zeta\right)d\nu d\zeta,
\end{equation}
and the magnetic energy of the background at corresponding cell interfaces are computed as
\begin{equation}
\left(B_{0_{i+\frac{1}{2},j,k}}^\mu\right)^2=\frac{1}{A_\mu}\int^{j+\frac{1}{2}}_{j-\frac{1}{2}}\int^{k+\frac{1}{2}}_{k-\frac{1}{2}}B_0^2\left(\mu=i+\frac{1}{2},\nu,\zeta\right)d\nu d\zeta,
\end{equation}
and the cross terms of the background magnetic fields in Equation \ref{eqn:bzero_stress} are calculated as
\begin{equation}
\mathbf{B}_{0_{i+\frac{1}{2},j,k}}^\mu B_{0_{i+\frac{1}{2},j,k}}^\mu=\frac{1}{A_\mu}\int^{j+\frac{1}{2}}_{j-\frac{1}{2}}\int^{k+\frac{1}{2}}_{k-\frac{1}{2}}B_0\left(\mu=i+\frac{1}{2},\nu,\zeta\right)\mathbf{B}_0\left(\mu=i+\frac{1}{2},\nu,\zeta\right) d\nu d\zeta.
\end{equation}

\subsubsection{Face-Normal Coordinate Systems}\label{sec:face_normal_coord}

The MHD solver in the GAMERA code only tracks Cartesian components of the plasma velocity ($u_x, u_y, u_z$) regardless of the grid geometry. However, as discussed in the previous section, the calculations of fluid flux are done in the local \textit{face-normal coordinate system} ($x_n, x_1, x_2$), which is usually not the same as the base Cartesian system. Thus after the interface reconstruction of the Cartesian velocity components ($u_x, u_y,u_z$) and magnetic field ($B_x,B_y,B_z$) in the $\mu$-, $\nu$- and $\zeta$-directions, corresponding coordinate transforms at cell interfaces are needed to rotate the left and right velocity and magnetic field from the base Cartesian coordinate system ($x,y,z$) into the face-normal coordinate system ($x_n,x_1,x_2$), in order to evaluate the numerical fluxes using the Gas-Kinetic flux functions described in Section \ref{sec:flux_func}. For updating the plasma momentum after computing the momentum flux vector in local face-normal coordinate systems at each interface, inverse transforms are needed to rotate the face-integrated momentum flux from the ($x_n,x_1,x_2$) system to the base Cartesian system ($x,y,z$). Then the Cartesian components of the plasma momentum ($\rho u_x,\rho u_y,\rho u_z$) are updated as described in the previous section. This transform - inverse transform algorithm enables the MHD solver to evolve the plasma velocity components in the base Cartesian coordinate system, which is independent of the curvilinear coordinate system used to define the computational grid. Therefore orthogonality of the curvilinear grid is not required in the MHD solver, and the numerical grid can be adapted to simulate problem specific flow patterns such as planetary magnetospheres.

\begin{figure}[htb!]
	\noindent\includegraphics[width=23pc]{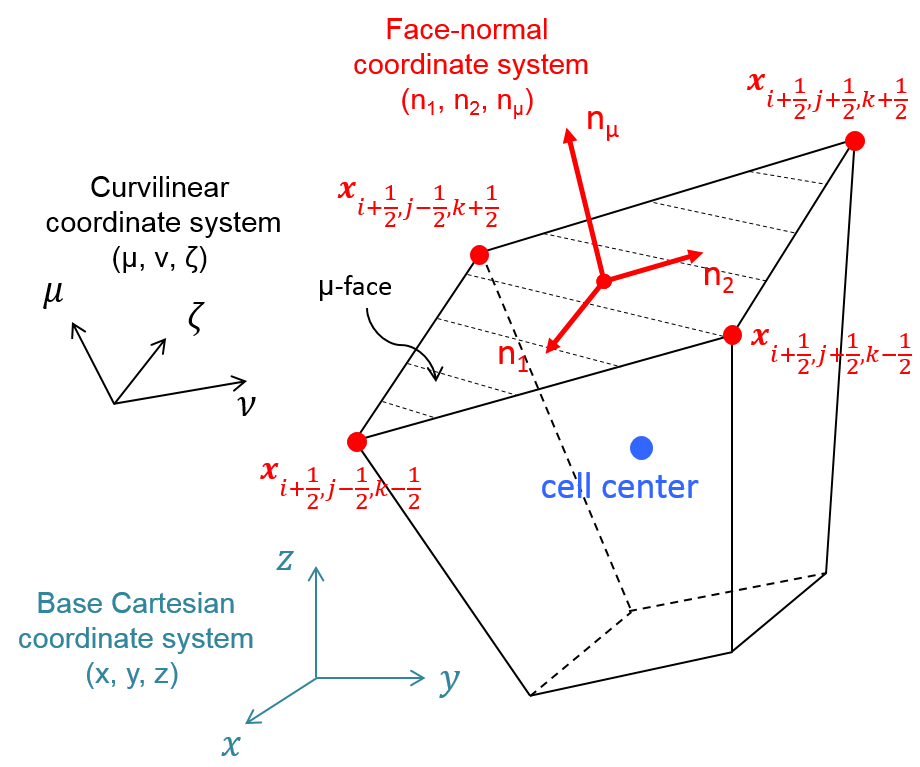}%
	\centering
	\caption{The local $\mu$-face normal coordinate system ($\mathbf{n}_\mu$,$\mathbf{n}_1$,$\mathbf{n}_2$) in a finite volume cell indexed as ($i,j,k$).} \label{fig:face_norm_coord}
\end{figure}

Using the orthogonal, $\mu$-face normal coordinate system ($\mathbf{n}_\mu$,$\mathbf{n}_1$,$\mathbf{n}_2$) at cell interfaces indexed by $(i+\frac{1}{2},j,k)$ illustrated in Figure \ref{fig:face_norm_coord} as an example, where $\mathbf{n}_\mu$ is the unit vector normal to the $\mu$-face, $\mathbf{n}_1$ and $\mathbf{n}_2$ are the two unit vectors perpendicular to $\mathbf{n}_\mu$. The $\mathbf{n}_1$,$\mathbf{n}_2$ and $\mathbf{n}_\mu$ vectors from an orthogonal system, which are calculated using the four corner grid points forming the $\mu$-face shown in Figure \ref{fig:face_norm_coord}:
\begin{equation}
\mathbf{n}_\mu = \displaystyle\frac{\left(\mathbf{x}_{i+\frac{1}{2},j+\frac{1}{2},k+\frac{1}{2}}-\mathbf{x}_{i+\frac{1}{2},j+\frac{1}{2},k-\frac{1}{2}}\right)\times\left(\mathbf{x}_{i+\frac{1}{2},j+\frac{1}{2},k+\frac{1}{2}}-\mathbf{x}_{i+\frac{1}{2},j-\frac{1}{2},k+\frac{1}{2}}\right)}{\biggr|\mathbf{x}_{i+\frac{1}{2},j+\frac{1}{2},k+\frac{1}{2}}-\mathbf{x}_{i+\frac{1}{2},j+\frac{1}{2},k-\frac{1}{2}}\biggr|\times\biggr|\mathbf{x}_{i+\frac{1}{2},j+\frac{1}{2},k+\frac{1}{2}}-\mathbf{x}_{i+\frac{1}{2},j-\frac{1}{2},k+\frac{1}{2}}\biggr|}\label{eqn:n_mu}
\end{equation}
\begin{equation}
\mathbf{n}_2=\displaystyle\frac{\mathbf{n}_\mu\times\left(\mathbf{x}_{i+\frac{1}{2},j+\frac{1}{2},k+\frac{1}{2}}-\mathbf{x}_{i+\frac{1}{2},j-\frac{1}{2},k+\frac{1}{2}}\right)}{\biggr|\mathbf{x}_{i+\frac{1}{2},j+\frac{1}{2},k+\frac{1}{2}}-\mathbf{x}_{i+\frac{1}{2},j-\frac{1}{2},k+\frac{1}{2}}\biggr|}
\end{equation}
\begin{equation}
\mathbf{n}_1=\mathbf{n}_2\times\mathbf{n}_\mu.
\end{equation}
Since $||\mathbf{n}_1|| = ||\mathbf{n}_2|| = ||\mathbf{n}_\mu||\equiv 1$ and $\mathbf{n}_1\perp\mathbf{n}_2\perp\mathbf{n}_\mu$, the transform of the interface velocity vectors from the base Cartesian $(x,y,z)$ coordinate system to the face-normal $(\mathbf{n}_1,\mathbf{n}_2,\mathbf{n}_\mu)$ coordinate system is done as follows:
\begin{equation}
\mathbf{u}_{n_1,n_2,n_\mu} = \overline{\mathbf{T}}\cdot\mathbf{u}_{x,y,z}\Rightarrow \left(\begin{array}{c} u_{n_1} \\ u_{n_2} \\ u_{n_\mu} \end{array}\right)=\left[\begin{array}{ccc} n_1^x & n_1^y & n_z^z \\ n_2^x & n_2^y & n_2^z \\ n_\mu^x & n_\mu^y & n_\mu^z \end{array}\right]\cdot\left(\begin{array}{c} u_{x} \\ u_{y} \\ u_{z} \end{array}\right),
\end{equation}
where $\mathbf{u}_{n_1,n_2,n_\mu}=\left(u_{n_1},u_{n_2},u_{n_\mu}\right)$ is the velocity vector in the face-normal coordinate system $(\mathbf{n}_1,\mathbf{n}_2,\mathbf{n}_\mu)$ at $\mu$-faces, and $\mathbf{u}_{x,y,z}=\left(u_{x},u_{y},u_{z}\right)$ is the corresponding velocity vector in the base Cartesian coordinate system $(x,y,z)$. Since the transform matrix $\overline{\mathbf{T}}$ in the above equation is an orthogonal matrix, the inverse transform to rotate the stress $\mathbf{F}_\textsc{FLUID}$ from the $(\mathbf{n}_1,\mathbf{n}_2,\mathbf{n}_\mu)$ system to the base $(x,y,z)$ system is simply done using the transpose of $\overline{\mathbf{T}}$:
\begin{equation}
\mathbf{F}^{x,y,z}_{\rho\mathbf{u}} = \overline{\mathbf{T}}^T\cdot\mathbf{F}^{n_1,n_2,n_\mu}_{\rho\mathbf{u}}\Rightarrow \left(\begin{array}{c} F_{x} \\ F_{y} \\ F_{z} \end{array}\right) =\left[\begin{array}{ccc} n_1^x & n_2^x & n_\mu^x \\ n_1^y & n_2^y & n_\mu^y \\ n_1^z& n_2^z & n_\mu^z \end{array}\right]\cdot\left(\begin{array}{c} F_{n_1} \\ F_{n_2} \\ F_{n_\mu} \end{array}\right),
\end{equation}
where $\mathbf{F}^{n_1,n_2,n_\mu}_{\rho\mathbf{u}}$ is the momentum flux vector in the $(\mathbf{n}_1,\mathbf{n}_2,\mathbf{n}_\mu)$ coordinate system described in Equation (\ref{eqn:momentum_flux_func}), and $\mathbf{F}^{x,y,z}_{\rho\mathbf{u}}$ is the corresponding vector of momentum flux in the base Cartesian coordinate system ($x,y,z$). With the coordinate transform, the surface integral of the fluid stress for $\Delta\mathbf{U}_F\big|_{\rho\mathbf{u}}$ is simply calculated using (\ref{eqn:dalta_U_mom}). However the surface integrals of mass flux and energy flux are in the $\mathbf{n}_\mu$ direction which are used to compute the surface integrals directly without coordinate transforms as shown in (\ref{eqn:delta_U_rho}) and (\ref{eqn:delta_U_Ep}). For the magnetic stresses, since only the zeroth moment of the distribution function is involved, the stress calculations on cell interfaces are done directly using the $x$-, $y$- and $z$-components of the interface magnetic fields and the  $\mathbf{n}_1,\mathbf{n}_2,\mathbf{n}_\mu$ vectors in the base Cartesian system as shown in Equation (\ref{eqn:stress_calc_face}), without the above transform/reverse transform procedures. 


\subsection{Constrained Transport for Non-Orthogonal Grids\label{sec:CT}}

\begin{figure}[t!]
	\noindent\includegraphics[width=32pc]{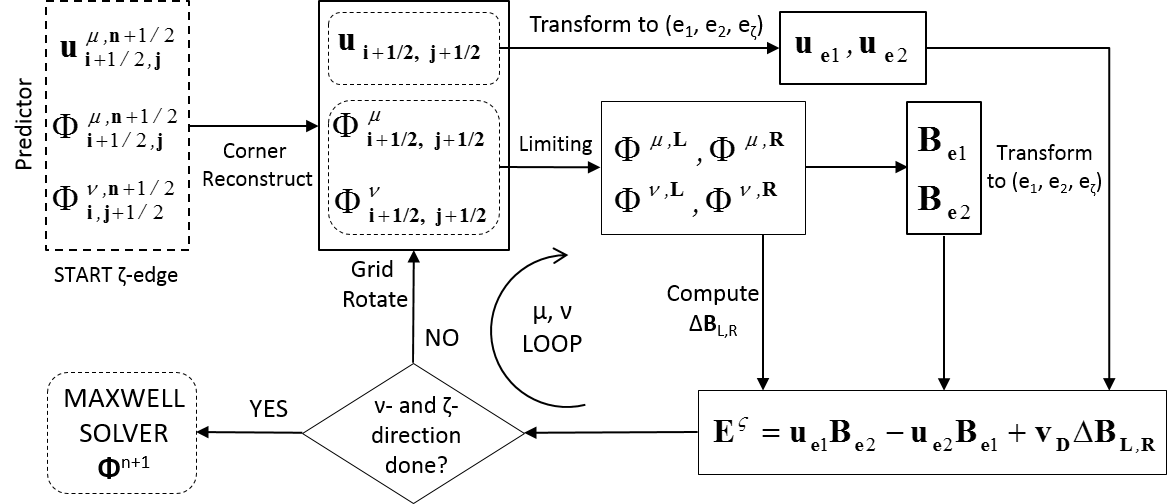}%
	\centering
	\caption{The algorithm for calculating edge-aligned electric field using the predictors. }\label{fig:maxwell_solver}
\end{figure}

In this section we describe the numerical schemes for calculating the electric fields defined at cell edges for evolving the magnetic flux $\Phi$ through Faraday's law. The calculation of the electric field is handled through the implementation of a constrained transport method based on the same high-order reconstruction schemes (described in Section \ref{sec:recon}) adapted to non-orthogonal, staggered grids. As a main part of the corrector step illustrated in Figure \ref{fig:code_structure}, the predicted velocity $\mathbf{u}^{i+\frac{1}{2}}$ and magnetic flux $\Phi^{i+\frac{1}{2}}$ are used in the Maxwell solver for computing the electric fields. Figure \ref{fig:maxwell_solver} shows the algorithms of computing the electric field defined at cell edges using the predicted bulk velocity $\mathbf{u}^{n+\frac{1}{2}}$ and magnetic flux $\Phi^{n+\frac{1}{2}}$, with details described in the following sections. 

\subsubsection{Calculation of Electric Fields}

\begin{figure}[b!]
	\noindent\includegraphics[width=38pc]{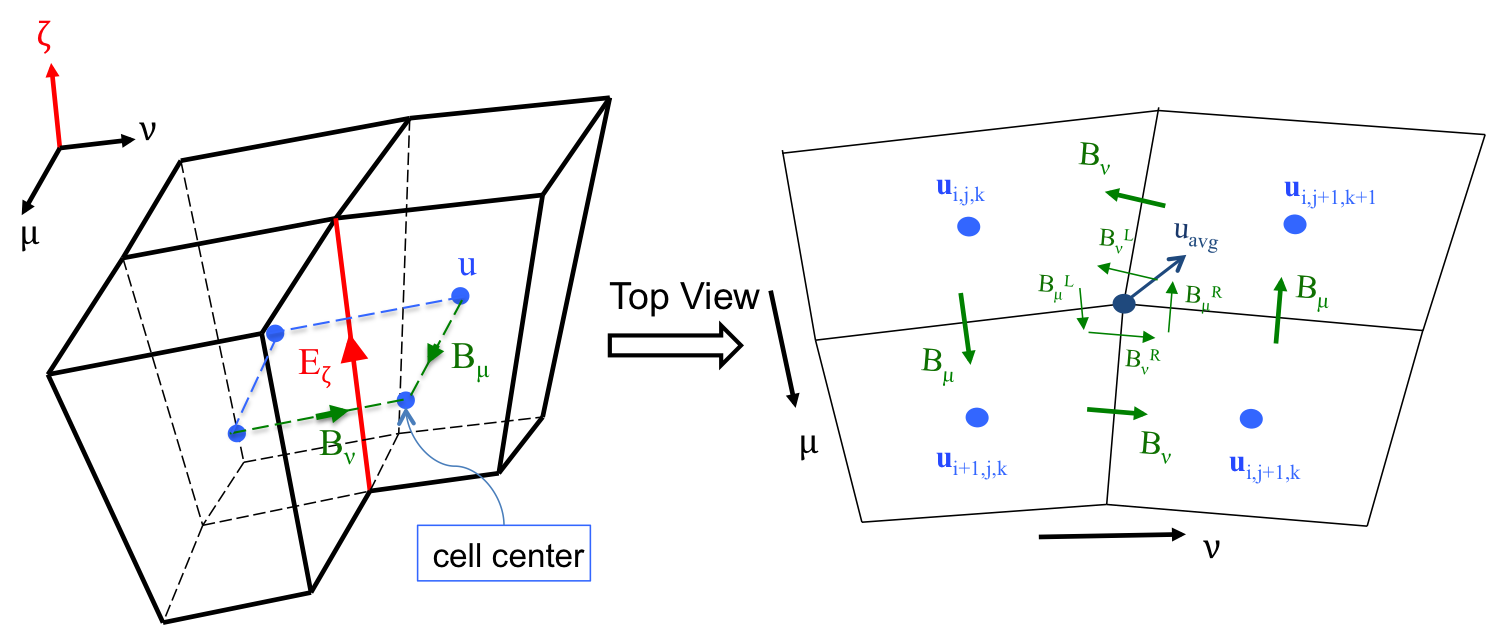}%
	\centering
	\caption{A schematic showing locations of various quantities needed for calculation of the electric field at cell edges.}\label{fig:ef_coord}
\end{figure}

Figure \ref{fig:ef_coord} shows a schematic of the grid geometry for computing the electric field component $E_\zeta$ at $\zeta$-edges. The $E_\mu$ and $E_\nu$  components along $\mu$- and $\nu$-edges are calculated using the same algorithm as $E_\zeta$, with corresponding rotations of the computational grid. Using the example shown in Figure \ref{fig:ef_coord}, the electric field $E_\zeta$ is located at the cell edge indexed as $\left(i+\frac{1}{2},j+\frac{1}{2},k\right)$, which is surrounded by four neighboring control volumes whose cell centers are indexed as $(i,j,k)$, $(i+1,j,k)$, $(i,j+1,k)$ and $(i+1,j+1,k)$, respectively. To get a high-order estimation of the electric field component $E_\zeta$ at the $\zeta$-edge, both face-centered magnetic flux and cell-centered plasma velocity are reconstructed at cell edges using the high-order reconstruction method described in Section \ref{sec:recon}. For the velocity $\mathbf{u}\left(i+\frac{1}{2},j+\frac{1}{2},k\right)$ at cell edges, this is done by first doing a high order reconstruction to $\mu$-interfaces and then reconstructing these $\mu$-interface values in the $\nu$-direction towards the edge (or corner as it appears in the top view of the 2D slice shown in the right panel of Figure \ref{fig:ef_coord}):
\begin{equation}
\rho\mathbf{u}\left(i,j,k\right)\xrightarrow{\mathrm{interp\; in\;} \mu}\rho\mathbf{u}\left(i+\frac{1}{2},j,k\right)\xrightarrow{\mathrm{interp\; in\;} \nu}\rho\mathbf{u}\left(i+\frac{1}{2},j+\frac{1}{2},k\right)\xrightarrow{\mathrm{divide\; by\;} \rho}\mathbf{u}\left(i+\frac{1}{2},j+\frac{1}{2},k\right).
\end{equation}

\begin{figure}[h!]
	\noindent\includegraphics[width=35pc]{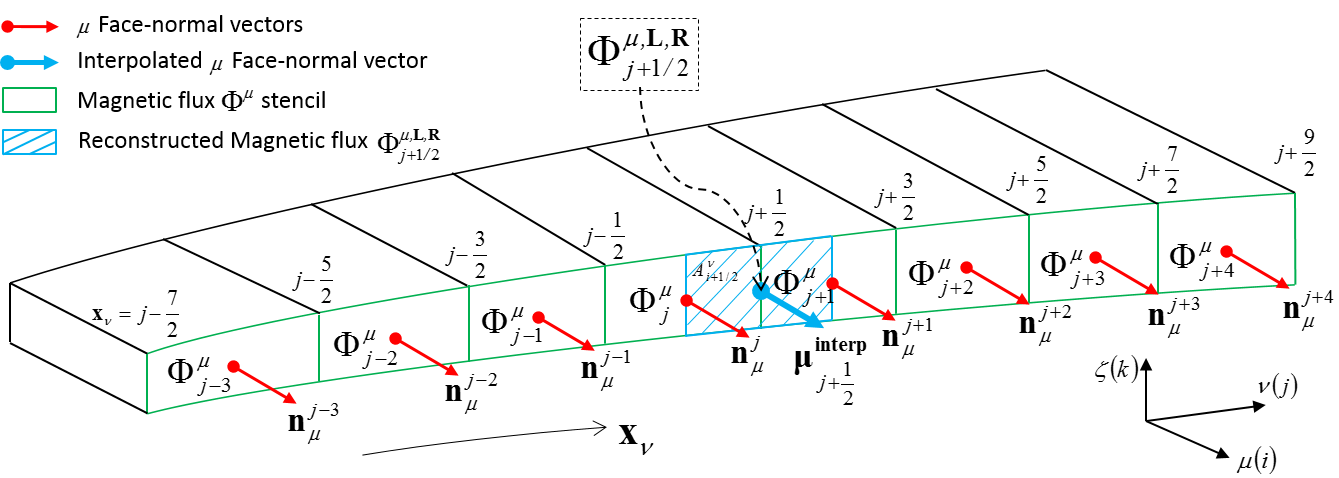}%
	\centering
	\caption{The one-dimensional reconstruction stencil for $\Phi^\mu$ and $\mathbf{n}_\mu$ in the $\nu$-direction for estimating $\mathbf{B}_{avg}^{\mu}$.}\label{fig:b_recon}
\end{figure}

The default choice for reconstructing edge velocity is an $8^{th}$-order centered scheme described in Equation (\ref{eqn:recon_8}). 
The estimation of magnetic field vectors at the cell edge is more complicated than computing the velocity vectors, which requires two reconstruction steps for both magnetic flux and face-normal vectors. The first step is to reconstruct the magnetic flux at cell edges. Since magnetic flux $\Phi^\mu$ and $\Phi^\nu$ are already defined on cell interfaces, it only requires one single reconstruction to the corner along the $\nu$- and $\mu$-direction, respectively. Using the $\Phi^\mu$ component of magnetic flux along the $\nu$-direction as an example shown in Figure \ref{fig:b_recon}, after applying the one-dimensional reconstruction algorithm to the magnetic flux $\Phi_{i+\frac{1}{2},j,k}^\mu$ in the $\nu$-direction, two ``interface'' states of magnetic flux, $\Phi_{i+\frac{1}{2},j+\frac{1}{2},k}^{(\mu,L)}$ and $\Phi_{i+\frac{1}{2},j+\frac{1}{2},k}^{(\mu,R)}$ are computed at the cell edge $\left(i+\frac{1}{2},j+\frac{1}{2},k\right)$. To estimate the magnetic field strength at cell edges, an eighth-order interpolation is applied to the face area $A_{i+\frac{1}{2},j,k}^\mu$ along the $\nu$-direction to find an estimation for the area $A_{i+\frac{1}{2},j+\frac{1}{2},k}^\mu$ at the cell edge for the estimated interface fluxes $\Phi_{i+\frac{1}{2},j+\frac{1}{2},k}^{(\mu,L)}$ and $\Phi_{i+\frac{1}{2},j+\frac{1}{2},k}^{(\mu,R)}$, as shown by the shaded area in Figure \ref{fig:b_recon}. Then the magnetic field strength in the $\mu$-direction at the cell edge $\left(i+\frac{1}{2},j+\frac{1}{2},k\right)$ is calculated as the average of the left and right state of the magnetic flux divided by the estimated interface area:
\begin{equation}
B_{avg}^\mu = \frac{1}{2}\displaystyle\frac{ \Phi_{i+\frac{1}{2},j+\frac{1}{2},k}^{(\mu,L)} + \Phi_{i+\frac{1}{2},j+\frac{1}{2},k}^{(\mu,R)} } {A_{i+\frac{1}{2},j+\frac{1}{2},k}^\mu}.\label{eqn:b_mu}
\end{equation}

The next step is to estimate the direction of $B_{avg}^\mu$ at the cell edge $\left(i+\frac{1}{2},j+\frac{1}{2},k\right)$, which is denoted as $\hat{\mu}^{\mathrm{interp}}_{i+\frac{1}{2},j+\frac{1}{2},k}$ in Figure \ref{fig:b_recon}. This is done by interpolating the face-normal unit vectors $\hat{\mathbf{n}}_\mu$ to cell edge using the same $8^{th}$-order reconstruction algorithm. The expression for the $\hat{\mathbf{n}}_\mu$ vectors are defined in equation (\ref{eqn:n_mu}), and the high-order reconstructed $\hat{\mathbf{n}}_\mu$ vector at the cell edge is denoted as $\hat{\mu}_{i+\frac{1}{2},j+\frac{1}{2},k}^{\mathrm{interp}}$. Similarly, the average magnetic field in the $\nu$-direction is calculated by applying the same reconstruction algorithm on the magnetic flux component $\Phi_{i,j+\frac{1}{2},k}^\nu$ and the corresponding $\nu$-face area $A_{i,j+\frac{1}{2},k}^\nu$ the $\mu$-direction:
\begin{equation}
B_{avg}^\nu = \frac{1}{2}\displaystyle\frac{ \Phi_{i+\frac{1}{2},j+\frac{1}{2},k}^{(\nu,L)} + \Phi_{i+\frac{1}{2},j+\frac{1}{2},k}^{(\nu,R)} } {A_{i+\frac{1}{2},j+\frac{1}{2},k}^\nu}.\label{eqn:b_nu}
\end{equation}
The direction of $B_{avg}^\nu$ is also interpolated from the face-normal $\hat{\mathbf{n}}_\nu$ vectors using the same high-order interpolation schemes. The high-order interpolated $\hat{\mathbf{n}}_\nu$ vector at the cell edge is denoted as $\hat{\nu}_{i+\frac{1}{2},j+\frac{1}{2},k}^{\mathrm{interp}}$.

Using the reconstructed edge velocity and average magnetic field vector, the calculation of the $\zeta$-component of the electric field is basically a Rusanov scheme adapted to a cell corner. With the edge velocity vector $\mathbf{u}_{i+\frac{1}{2},j+\frac{1}{2},k}$ and mean magnetic fields $\mathbf{B}_{avg}^{\mu,\nu}$ , the electric field component $E_\zeta$ at the $\zeta$-edge is calculated as:
\begin{equation}
E_\zeta = -\mathbf{u}_{i+\frac{1}{2},j+\frac{1}{2},k}\times\mathbf{B}_{avg}^{\mu,\nu} + \eta_A\mathbf{j}_\zeta,\label{eqn:efield_calc}
\end{equation}
where $\mathbf{j}_\zeta$ is the component of $\nabla\times\mathbf{B}$ along the $\zeta$-edge computed using the left- and right-states of the $\mu$- and $\zeta$-edge magnetic field as shown in the right panel of Figure \ref{fig:ef_coord}, and $\eta_A$ is a numerical resistivity dealing with the propagation of Alfv\'{e}n waves near discontinuities. If computing the edge electric field using Equation (\ref{eqn:efield_calc}) without the ``resistive'' term, simulations tend to develop spurious oscillations in MHD flows where Alfv\'{e}n waves are dominant, especially when discontinuities occur. We should note that this ``resistive'' term $\eta_A\mathbf{j}$ in Equation (\ref{eqn:efield_calc}) only operates when the limiter detects a discontinuous situation. When the flow is smooth, $\mathbf{B}_\mu^L=\mathbf{B}_\mu^R$ and $\mathbf{B}_\nu^L=\mathbf{B}_\nu^R$, the $\mathbf{j}_\zeta$ component in the diffusion term is essentially zero. Therefore in smooth regions the electric field $E_\zeta$ is just a high-order approximation of $-\mathbf{u}\times\mathbf{B}$ computed at cell edges. 

\begin{figure}[b!]
	\noindent\includegraphics[width=25pc]{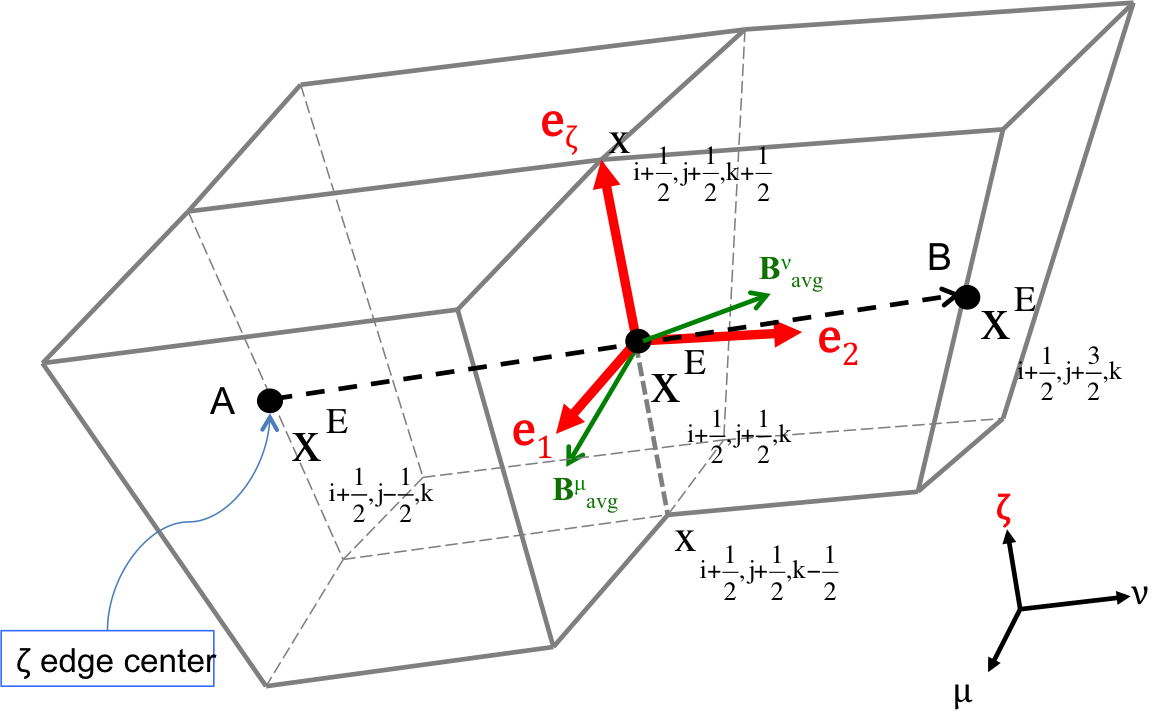}%
	\centering
	\caption{The edge-aligned coordinate system $\left(\mathbf{e}_1,\mathbf{e}_2,\mathbf{e}_\zeta\right)$ for computing the electric field.}\label{fig:ef_coord3}
\end{figure}

In Equation (\ref{eqn:efield_calc}), the interpolated velocity $\mathbf{u}_{i+\frac{1}{2},j+\frac{1}{2},k}$ and the average magnetic field $\mathbf{B}_{avg}$ at the $\zeta$-edge are not in the same coordinate system: the velocity $\mathbf{u}_{i+\frac{1}{2},j+\frac{1}{2},k}$ is in the base Cartesian coordinate system $(x,y,z)$, while the magnetic field $\mathbf{B}_{avg}$  is in a local non-orthogonal coordinate system $\left(\hat{\mu}_{i+\frac{1}{2},j+\frac{1}{2},k}^{\mathrm{interp}},\hat{\nu}_{i+\frac{1}{2},j+\frac{1}{2},k}^{\mathrm{interp}},\zeta\right)$ as illustrated by the green arrows in Figure \ref{fig:ef_coord3}. Therefore to calculate the cross product between the velocity and magnetic field at cell edges, we define a new orthogonal coordinate system $\left(\mathbf{e}_1,\mathbf{e}_2,\mathbf{e}_\zeta\right)$ with one axis $\mathbf{e}_\zeta$ aligned with the direction of the cell edge and transform $\mathbf{u}_{i+\frac{1}{2},j+\frac{1}{2},k}$ and $\mathbf{B}_{avg}^{\mu,\nu}$ into this new coordinate system. Using the $\zeta$-edge aligned orthogonal system shown in Figure \ref{fig:ef_coord3} as an example, the local coordinate system $\left(\mathbf{e}_1,\mathbf{e}_2,\mathbf{e}_\zeta\right)$ for computing the $\zeta$-component of the electric field using Equation (\ref{eqn:efield_calc}) is shown in red, where $E_\zeta$ is along the direction of $\mathbf{e}_\zeta$. For the $\zeta$-edge indexed by $\left(i+\frac{1}{2},j+\frac{1}{2},k\right)$ the $\left(\mathbf{e}_1,\mathbf{e}_2,\mathbf{e}_\zeta\right)$ coordinate system is computed as follows. First compute the $\zeta$-edge unit vector $\mathbf{e}_\zeta$:
\begin{equation}
\mathbf{e}_\zeta = \displaystyle\frac{\mathbf{x}_{i+\frac{1}{2},j+\frac{1}{2},k+\frac{1}{2}}- \mathbf{x}_{i+\frac{1}{2},j+\frac{1}{2},k-\frac{1}{2}}}{\bigg\vert \mathbf{x}_{i+\frac{1}{2},j+\frac{1}{2},k+\frac{1}{2}}- \mathbf{x}_{i+\frac{1}{2},j+\frac{1}{2},k-\frac{1}{2}} \bigg\vert}.
\end{equation}
Then calculate the direction $\mathbf{e}_1$ normal to the $\zeta$ vector:
\begin{equation}
\mathbf{e}_1 = \displaystyle\frac{ \left(\mathbf{x}^E_{i+\frac{1}{2},j+\frac{3}{2},k}-\mathbf{x}^E_{i+\frac{1}{2},j-\frac{1}{2},k} \right)\times\mathbf{e}_\zeta   }{\bigg\vert \mathbf{x}^E_{i+\frac{1}{2},j+\frac{3}{2},k}-\mathbf{x}^E_{i+\frac{1}{2},j-\frac{1}{2},k} \bigg\vert},
\end{equation}
where $\mathbf{x}^E_{i+\frac{1}{2},j+\frac{3}{2},k}$ and $\mathbf{x}^E_{i+\frac{1}{2},j-\frac{1}{2},k}$ are the edge-centered positions $\mathbf{A}$ and $\mathbf{B}$ at two neighboring $\zeta$-edges shown in Figure \ref{fig:ef_coord3}:
\begin{equation}
\mathbf{x}^E_{i+\frac{1}{2},j-\frac{1}{2},k} = \displaystyle\frac{1}{2}\left(\mathbf{x}_{i+\frac{1}{2},j-\frac{1}{2},k+\frac{1}{2}}+\mathbf{x}_{i+\frac{1}{2},j-\frac{1}{2},k-\frac{1}{2}} \right),
\end{equation}
\begin{equation}
\mathbf{x}^E_{i+\frac{1}{2},j+\frac{3}{2},k} = \displaystyle\frac{1}{2}\left(\mathbf{x}_{i+\frac{1}{2},j+\frac{3}{2},k+\frac{1}{2}}+\mathbf{x}_{i+\frac{1}{2},j+\frac{3}{2},k-\frac{1}{2}} \right).
\end{equation}
then the third direction $\mathbf{e}_2$, orthogonal to both $\mathbf{e}_1$ and $\mathbf{e}_\zeta$, is obtained using the following cross product:
\begin{equation}
\mathbf{e}_2 = \mathbf{e}_\zeta\times\mathbf{e}_1.
\end{equation}
Since the reconstructed edge velocity $\mathbf{u}_{i+\frac{1}{2},j+\frac{1}{2},k}$ is in the base Cartesian system, it is straightforward to map the reconstructed edge velocity vector $\mathbf{u}_{i+\frac{1}{2},j+\frac{1}{2},k}$ into the new edge-aligned coordinate system $\left(\mathbf{e}_1,\mathbf{e}_2,\mathbf{e}_\zeta\right)$:
\begin{equation}
u_{\mathbf{e}1} = \mathbf{u}_{i+\frac{1}{2},j+\frac{1}{2},k}\cdot\mathbf{e}_1,
\end{equation}
\begin{equation}
u_{\mathbf{e}2} = \mathbf{u}_{i+\frac{1}{2},j+\frac{1}{2},k}\cdot\mathbf{e}_2,
\end{equation}
where $u_{\mathbf{e}_1}$ and $u_{\mathbf{e}_2}$ are the velocity components in the directions of $\mathbf{e}_1$ and $\mathbf{e}_2$, respectively. The $u_{\mathbf{e}_\zeta}$ component is along the $\zeta$-edge which does not contribute to the calculation of the $E_\zeta$ component. The edge magnetic field components $\mathbf{B}_{avg}^\mu$ and $\mathbf{B}_{avg}^\nu$ calculated using Equation (\ref{eqn:b_mu}) and (\ref{eqn:b_nu}) are in a non-orthogonal local coordinate system $\left(\hat{\mu}_{i+\frac{1}{2},j+\frac{1}{2},k}^{\mathrm{interp}},\hat{\nu}_{i+\frac{1}{2},j+\frac{1}{2},k}^{\mathrm{interp}},\hat{\zeta}\right)$, thus the coordinate transform is more complicated than the velocity components, which requires solving the following $2\times2$ system:
\begin{equation}
B^{\mu}_{avg} = \xi^\mu_1B_{avg}^{\mathbf{e}1} + \xi^\mu_2B_{avg}^{\mathbf{e}2},
\end{equation}
\begin{equation}
B^{\nu}_{avg} = \xi^\nu_1B_{avg}^{\mathbf{e}1} + \xi^\nu_2B_{avg}^{\mathbf{e}2},
\end{equation}
where $\mathbf{\xi}^\mu$ and $\mathbf{\xi}^\nu$ are the projections of the directions of magnetic field $\mathbf{B}^{\mu,\nu}_{avg}$ in the $(\mathbf{e}_1,\mathbf{e}_2,\mathbf{e}_\xi )$ coordinate system:
\begin{equation}
\xi^\mu_1=\hat{\mu}_{i+\frac{1}{2},j+\frac{1}{2},k}^{\mathrm{interp}}\cdot\mathbf{e}_1 \quad \xi^\mu_2=\hat{\mu}_{i+\frac{1}{2},j+\frac{1}{2},k}^{\mathrm{interp}}\cdot\mathbf{e}_2,
\end{equation}
\begin{equation}
\xi^\nu_1=\hat{\nu}_{i+\frac{1}{2},j+\frac{1}{2},k}^{\mathrm{interp}}\cdot\mathbf{e}_1 \quad \xi^\nu_2=\hat{\nu}_{i+\frac{1}{2},j+\frac{1}{2},k}^{\mathrm{interp}}\cdot\mathbf{e}_2.
\end{equation}

After obtaining the average magnetic field components $B_{avg}^{\mathbf{e}1}$ and $B_{avg}^{\mathbf{e}2}$ in the new edge-aligned coordinate system $\left(\mathbf{e}_1,\mathbf{e}_2,\mathbf{e}_\zeta\right)$, the electric field $E_\zeta$ along the edge $\mathbf{e}_\zeta$ is calculated based on Equation (\ref{eqn:efield_calc}):
\begin{equation}
E_\zeta = -\left(u_{\mathbf{e}1}B_{avg}^{\mathbf{e}2} - u_{\mathbf{e}2}B_{avg}^{\mathbf{e}1}\right) + v_D\left(B_R^\mu - B_L^\mu +B_L^\nu - B_R^\nu\right)\label{eq:e_calc}.
\end{equation}
Then the electric potential along the $\zeta$-edge is calculated as 
\begin{equation}
\mathcal{E}^\zeta_{i+\frac{1}{2},j+\frac{1}{2},k} = E_\zeta\bigg\vert \mathbf{x}_{i+\frac{1}{2},j+\frac{1}{2},k+\frac{1}{2}} - \mathbf{x}_{i+\frac{1}{2},j+\frac{1}{2},k-\frac{1}{2}} \bigg\vert,
\end{equation}
which is the edge centered electric field $E_\zeta$ multiplied by the length of the corresponding $\zeta$ edge. $\mathcal{E}^\zeta_{i+\frac{1}{2},j+\frac{1}{2},k}$ is the actual component used in the integral form of the Faraday's law to evolve the face-centered magnetic fluxes $\Phi$. The diffusive speed $v_D$ in Eq. (\ref{eq:e_calc}) is defined as
\begin{equation}
v_D = \frac{1}{2}\left(V_A+|\mathbf{u}|\right)\frac{\bigg\vert\hat{\mu}^{\mathrm{interp}}_{i+\frac{1}{2},j+\frac{1}{2},k}\times\hat{\nu}^{\mathrm{interp}}_{i+\frac{1}{2},j+\frac{1}{2},k}\bigg\vert}{\bigg\vert\hat{\mu}^{\mathrm{interp}}_{i+\frac{1}{2},j+\frac{1}{2},k}\bigg\vert \bigg\vert\hat{\nu}^{\mathrm{interp}}_{i+\frac{1}{2},j+\frac{1}{2},k}\bigg\vert}.
\end{equation}
The magnitude of the diffusive speed is usually set to reflect the magnitude of the local convection speed and Alfv\'{e}n speed. The $\big\vert\mathbf{\mu}^{\mathrm{interp}}_{i+\frac{1}{2},j+\frac{1}{2},k}\times\mathbf{\nu}^{\mathrm{interp}}_{i+\frac{1}{2},j+\frac{1}{2},k}\big\vert/\big\vert\mathbf{\mu}^{\mathrm{interp}}_{i+\frac{1}{2},j+\frac{1}{2},k}\mathbf{\nu}^{\mathrm{interp}}_{i+\frac{1}{2},j+\frac{1}{2},k}\big\vert$ term originates from the $\nabla\times$ operation in the coordinate transform of the non-orthogonal geometry when the diffusive current $\mathbf{j}_\zeta$ in Equation (\ref{eqn:efield_calc}) is computed.

\subsubsection{Evolution of Magnetic fluxes}

The corresponding $\nu$-edge and $\mu$-edge aligned orthogonal coordinate systems are calculated in a similar way as shown in the $\zeta$-edge. The electric potential $\mathcal{E}^\mu$ and $\mathcal{E}^\nu$ along the $\mu$-edge and $\nu$-edge are also computed in the same way using the edge-aligned coordinate system. These electric potential are calculated using the same subroutine as for $\mathcal{E}^\zeta$ through proper rotation of the computational grid. Once all the electric fields are calculated the magnetic flux threading a face can be updated via Faraday's law:
\begin{equation*}
\Phi^{\zeta,n+1}_{i,j,k+\frac{1}{2}} = \Phi^{\zeta,n}_{i,j,k+\frac{1}{2}}+\Delta t\left(\mathcal{E}^\nu_{i,j-\frac{1}{2},k+\frac{1}{2}} + \mathcal{E}^\mu_{i+\frac{1}{2},j,k+\frac{1}{2}} - \mathcal{E}^\nu_{i,j+\frac{1}{2},k+\frac{1}{2}} - \mathcal{E}^\mu_{i-\frac{1}{2},j,k+\frac{1}{2}} \right),
\end{equation*}
\begin{equation*}
\Phi^{\mu,n+1}_{i+\frac{1}{2},j,k} = \Phi^{\mu,n}_{i+\frac{1}{2},j,k}+\Delta t\left(\mathcal{E}^\zeta_{i+\frac{1}{2},j,k-\frac{1}{2}} + \mathcal{E}^\nu_{i+\frac{1}{2},j+\frac{1}{2},k} - \mathcal{E}^\zeta_{i+\frac{1}{2},j,k+\frac{1}{2}} - \mathcal{E}^\nu_{i+\frac{1}{2},j-\frac{1}{2},k} \right),
\end{equation*}
\begin{equation}
\Phi^{\nu,n+1}_{i,j+\frac{1}{2},k} = \Phi^{\nu,n}_{i,j+\frac{1}{2},k}+\Delta t\left(\mathcal{E}^\mu_{i-\frac{1}{2},j+\frac{1}{2},k} + \mathcal{E}^\zeta_{i,j+\frac{1}{2},k+\frac{1}{2}} - \mathcal{E}^\mu_{i+\frac{1}{2},j+\frac{1}{2},k} - \mathcal{E}^\zeta_{i,j+\frac{1}{2},k-\frac{1}{2}} \right).\label{eq:faraday}
\end{equation}
Based on Equation (\ref{eq:faraday}), it is straightforward to show that the divergence of the magnetic field defined by the flux follows
\begin{equation}
\nabla\cdot\mathbf{B}\biggr\vert^{n+1}_{i,j,k} = \nabla\cdot\mathbf{B}\biggr\vert^{n}_{i,j,k},
\end{equation}
ensuring that the volume integrated $\nabla\cdot\mathbf{B}$ is unmodified during the magnetic field evolution step, regardless of changes in the local electric fields. Therefore, as long as the initial divergence of the magnetic field is zero, the Maxwell solver keeps the $\nabla\cdot\mathbf{B}$ term to round-off error automatically.

\subsection{Implementation Considerations\label{sec:perf}}

Undertaking the task of completely rebuilding our simulation framework, while costly, presents significant opportunities.  We have used the opportunity to modernize the software design, improve computational performance, and incorporate numerous algorithmic advances.  Redesigning our code to prepare for the multicore era of supercomputing while maintaining a flexible and extensible code-base has been a primary focus of this effort.  The production implementation of Gamera has been rewritten entirely from scratch in modern (2003/2008 standard) Fortran.  Utilities that make up the pre- and post-processing ecosystem are implemented in \textit{Python}, with "heavy" data stored in the HDF5 format and interpreted via XDMF.  The computational considerations necessary for multicore performance can often increase code complexity.  We alleviate these difficulties by incorporating into our development the use of unit testing using the \textit{pFUnit} framework \citep{rilee2014towards}.

The core computational routines and data structures of Gamera have been designed to expose the multiple layers of heterogeneous parallelism necessary for modern multicore architectures.  At the highest level, spatial domain and model decomposition is undertaken via traditional MPI parallelism.  However, in a multicore framework it is inappropriate to use this higher-level parallelism across the light-weight compute cores within a socket.  Instead we utilize shared memory parallelism, OpenMP, to create finer-grained parallelism within the individual computations done on the compute domain of an MPI rank.  Typically we utilize $1$ MPI rank per compute socket, and $1$ thread per \emph{virtual} core.  Finally, and often most importantly, is the necessity of vector or SIMD (single-instruction multiple-data) parallelism.  Modern computational architecture is optimized to move through memory in a predictable, ideally linear, manner and perform the same computations on each memory location.  Code written to take advantage of this can easily outperform by an order of magnitude naive implementations.  

Two major design choices allow Gamera to fully take advantage of vector parallelism.  The first is HPC-friendly data structures, namely the use of large, contiguous arrays.  The main data structure within Gamera is a 5D array of size $N_{s} \times N_{v} \times N_{k} \times N_{j} \times N_{i}$, where the dimensions represent the number of ion species, the number of plasma flow variables, and the spatial dimensions of the domain respectively.  The second choice is the manner in which our "heavy" compute routines are written, e.g. the interface flux calculation.  These routines are written to work on memory-aligned blocks of data, typically twice the vector length of the architecture.  This ensures that even very complex computational routines will be properly converted to vector instructions by the compiler while avoiding difficult to maintain manually-inlined code.

The main computational workload of Gamera, and typically most codes, is done in a series of loop nests.  As an example, a sweep of the computational domain may take the ordering, from outer- to inner-most, of $(k,j,i)$.  In Gamera, the inner-most loop would be separated into $i_B$ ranging over $[1,N_{i}/N_{B}]$, with $N_{i}$ the number of cells in the $i$-direction and $N_{B}$ the blocking size, and $\delta i$ which ranges over $[1,N_{B}]$.  This results in a loop ordering $(k,j,i_B,\delta i)$, with the $\delta i$ loop within a subroutine called from the $i_B$ loop.  In all of our computationally-intensive loop nests the inner-most loop is over $\delta i$ regardless of the coordinate direction of the sweep, i.e. reconstruction in the $j$ and $k$ directions still utilize $\delta i$ as the inner-most loop although the outer-most loops may be reordered to improve cache locality. As an example, calculating fluxes in the $j$ direction uses a loop nest of the form $(k,i_B,j,\delta i)$.

The result of these considerations is a code that significantly outperforms the original LFM and has been used to do production science simulations up to $10k$ CPU-cores on NCAR's Cheyenne supercomputer.  Performance optimization, however, is an iterative process and one largely undertaken upon completion of the core algorithm.  The performance metrics we will discuss here are not intended to be definitive or final, but merely a snapshot of the code's performance upon the completion of main algorithm development.  Even so, we find that the code scales well in both OMP and MPI and is capable of production scientific simulations at $O(10k)$ CPU-cores.

\begin{figure}[htb!]
	\noindent\includegraphics[width=30pc]{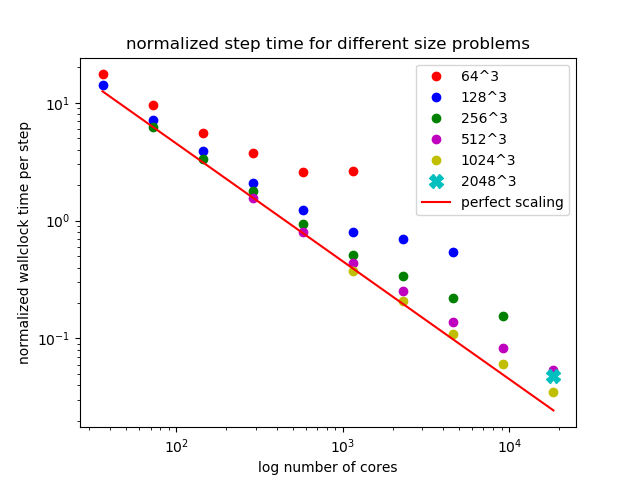}
	\centering
	\caption{Strong-scaling for a variety of problem sizes.} \label{fig:mpiscale}
\end{figure}

Our performance analysis has been undertaken on NCAR's Cheyenne supercomputer which utilizes Intel Xeon E5-2697V4 (Broadwell) processors, featuring dual-socket nodes with $18$ CPU-cores per socket.  Using $1$ MPI rank/socket and $2$ OMP threads per CPU core, we find OMP scaling at $85\%$ of optimal and a performance of approximately $500k$ zone-cycles per second per core.  The OpenMP scaling can likely be improved by merging thread-parallel regions to reduce fork/join overhead, however this can come at a cost in code complexity and maintainability.  Turning to cross-socket performance, Figure~\ref{fig:mpiscale} shows strong-scaling for a collection of problem sizes using triply-periodic MHD.  This includes both computation and communication times, with the former scaling better individually.  We note that we aren't currently using OMP threading in our communication routines, and expect that threaded packing/unpacking of buffers and utilizing thread-safe asynchronous MPI calls will further improve our scaling.  Even so, we see near-linear scaling for all problems sizes until we cross the threshold of approximately $16^3$ cells per compute core.  

\section{Test Results} \label{sec:test}

In this section, we show one set of test simulations of two-dimensional linear advection and four sets of simulation results from two-dimensional standard MHD test problems in both Cartesian and non-Cartesian geometries, including field-loop advection, circularly-polarized non-linear Alfv\'{e}n waves, the Orszag-Tang vortex and spherical blast waves in strong magnetic fields. The 2-D linear advection tests demonstrate the the choice of high-order reconstruction methods discussed in Section \ref{sec:recon}. Both the field-loop advection and non-linear Alfv\'{e}n wave simulations provide quantitative assessment for the numerical solutions, while the Orszag-Tang and blast wave simulations are less quantitative. The latter two sets of test simulations are used to demonstrate the effectiveness of the numerical schemes on handling highly non-linear MHD flows in non-orthogonal, distorted grid geometries.

\subsection{Circular Advection}

\begin{figure}[b!]
	\noindent\includegraphics[width=34pc]{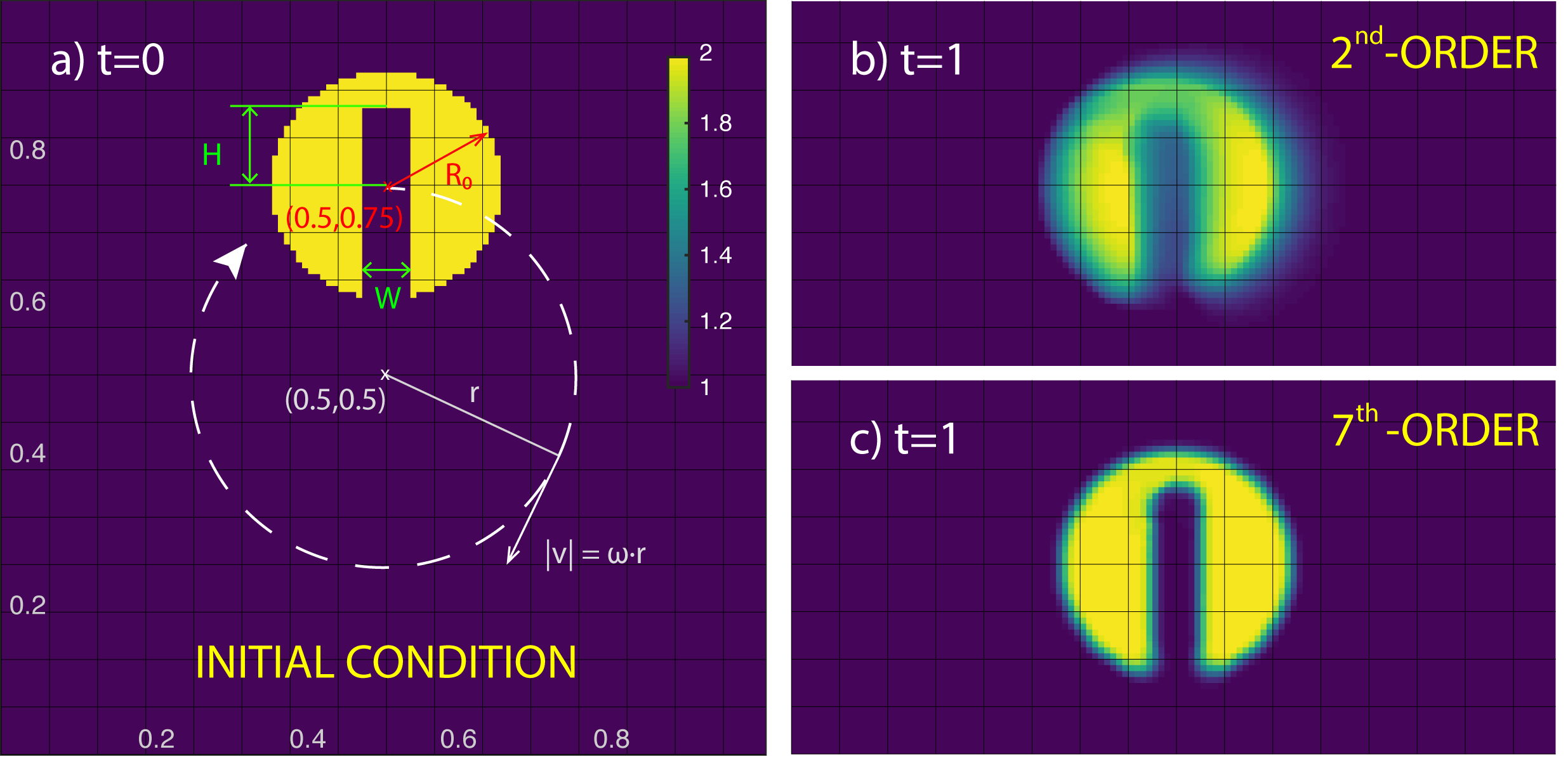}%
	\centering
	\caption{a) The initial distribution of a slotted cylinder in Cartesian geometry with $128\times 128$ cells at $t=0$; b) the simulated density distribution at $t=1$ using a $2^{nd}$-order reconstruction scheme; and c) the the simulated density distribution at $t=1$ using a $7^{th}$-order reconstruction scheme. Each ``grid cell'' shown in panels a)-c) using thin black lines contains $8\times 8$ actual computational cells in the $x$- and $y$-direction, respectively.}\label{fig:cadv1}
\end{figure}

We use the circular advection of a slotted cylinder problem used by \citet{Colella2011} as a multi-dimensional linear test to show the effectiveness of the advection scheme, which is originally developed by \citet{Zalesak1979}. This test problem is also used to demonstrate the necessity of high-order reconstruction method in the numerical algorithms. In this two-dimensional circular advection test, only the mass continuity equation is solved:
\begin{equation}
\frac{\partial\rho}{\partial t} = -\nabla\cdot\left( \rho \mathbf{v}_0\right), \label{eqn:masseqn}
\end{equation}
where $\mathbf{v}_0$ is a time-stationary circular velocity field defined as $\mathbf{v}_0=-2\pi\omega\hat{\mathbf{\theta}}$ with $\omega=1$ the angular velocity and $r = \sqrt{(x-0.5)^2+(y-0.5)^2}$ as shown in Figure \ref{fig:cadv1}a. The simulation domain is $0\le x\le 1$, $0 \le y\le 1$ with periodic boundary conditions in both the $x$- and $y$-direction. The initial density distribution at $t=0$ is defined as a slotted cylinder of radius $R_0 = 0.15$ centered at $(x,y) = (0.5,0.75)$, with slot width $W= 0.06$ and slot height $H = 0.1$:
\begin{equation}
\rho(x,y)\bigg|_{t=0} = \left\{ \begin{tabular}{cl}
  1, & $r \geq R_0$,  \\
  1, & $|x - 0.5|\le W\: and\: y \le 0.75+H$, \\
  2, & otherwise.
  \end{tabular} \right.
\end{equation}
At $t=1$, the center of the slotted cylinder returns to the initial location. The initial solution on a uniform Cartesian mesh with $128\times 128$ cells is plotted in Figure \ref{fig:cadv1}a. The solution of $\rho$ (at $t=1$) using the second-order centered reconstruction method introduced in Section \ref{sec:recon} is shown in Figure \ref{fig:cadv1}b, while the solution of $\rho$ at $t=1$ using the default seventh-order reconstruction scheme is shown in Figure \ref{fig:cadv1}c. At $t=1$, although the second-order reconstruction scheme preserves the basic shape of the slotted cylinder, the initially sharp edges of the cylinder is smeared significantly with noticeable fill-in of the slot. When using the seventh-order reconstruction scheme, the density distribution at $t=1$ is preserved more accurately compared to the second-order case, with sharper edges of the cylinder and an empty slot.

\begin{figure}[htb!]
	\noindent\includegraphics[width=30pc]{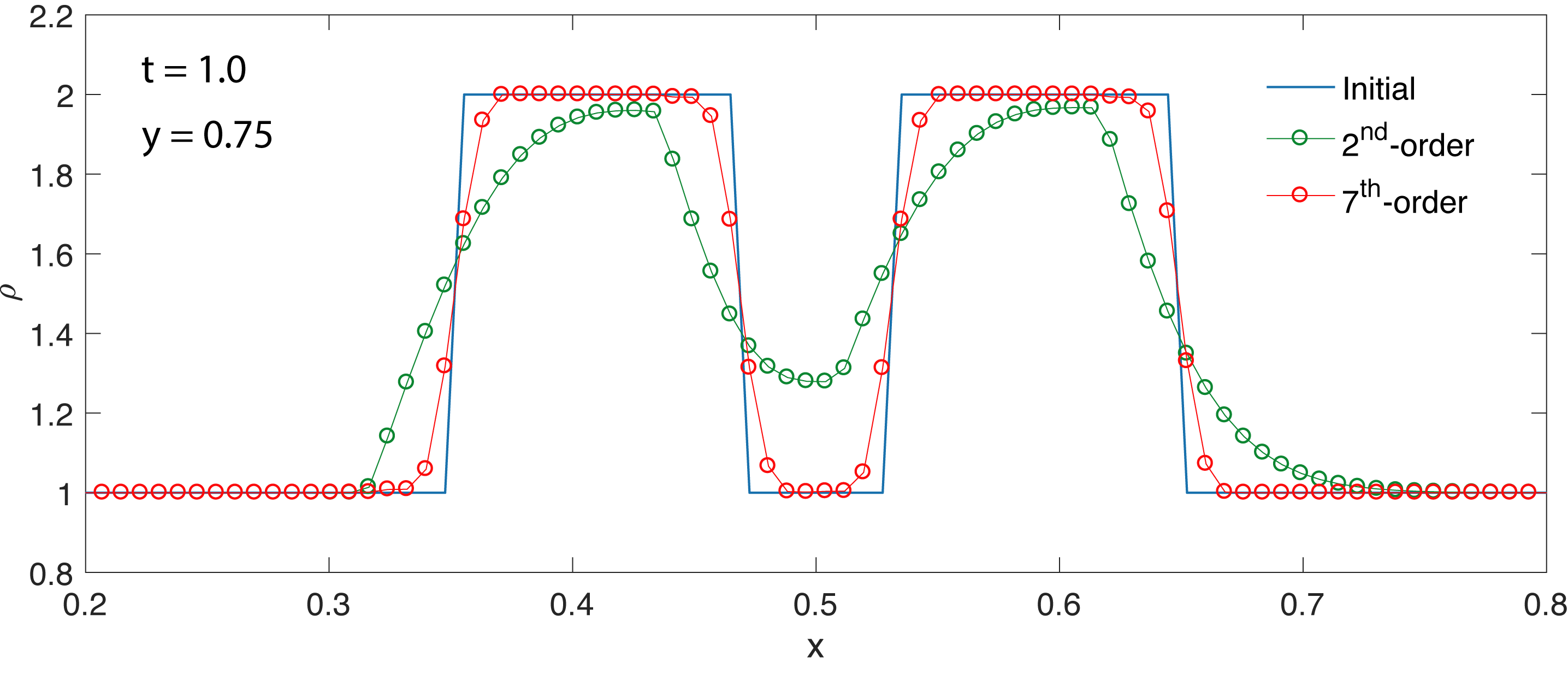}
	\centering
	\caption{The comparison of density profiles at $y=0.75$. The blue profile is the reference density distribution at $t=0$, the green and the red profiles are the corresponding density distribution at $t=1$ using the $2^{nd}$- and $7^{th}$-order schemes, respectively.} \label{fig:cadv2}
\end{figure}

Figure \ref{fig:cadv2} shows more quantitative comparisons of the circular advection results using line profiles of simulated $\rho$ at $y=0.75$. The blue profile shows the initial density distribution at $t=0$ as a reference, while the green and the red profiles show the corresponding density profiles at $t=1$ using the $2^{nd}$- and the $7^{th}$-order reconstruction scheme, respectively. The comparison shows that the $2^{nd}$-order reconstruction scheme introduces a significant amount of numerical diffusion. At $t=1$, the profile of the slotted density is smeared using the $2^{nd}$-order reconstruction, which has approximately 14 cells resolving the contact discontinuity at the edge of the cylinder ($x=0.5\pm R_0$) and a filled slot with density $\approx$1.3 ($|x-0.5|\le R_0$). On the other hand, the $7^{th}$-order reconstruction scheme preserves the initial profile well, with only 4 cells resolving the sharp edge of the cylinder and an empty slot (density $\approx$1.0). The comparison of the density profiles between the $2^{nd}$- and the $7^{th}$-order reconstruction scheme demonstrates the necessity of using very-high order reconstruction methods in order to reduce the numerical diffusion and resolve sharp contact discontinuity in multi-dimensional flow simulations. The additional computing cost in the $7^{th}$-order reconstruction scheme is approximately $24\%$ compared to the $2^{nd}$-order scheme. When using the $8^{th}$-order reconstruction scheme, the additional computational cost is approximately $12\%$. 
\begin{figure}[t!]
	\noindent\includegraphics[width=38pc]{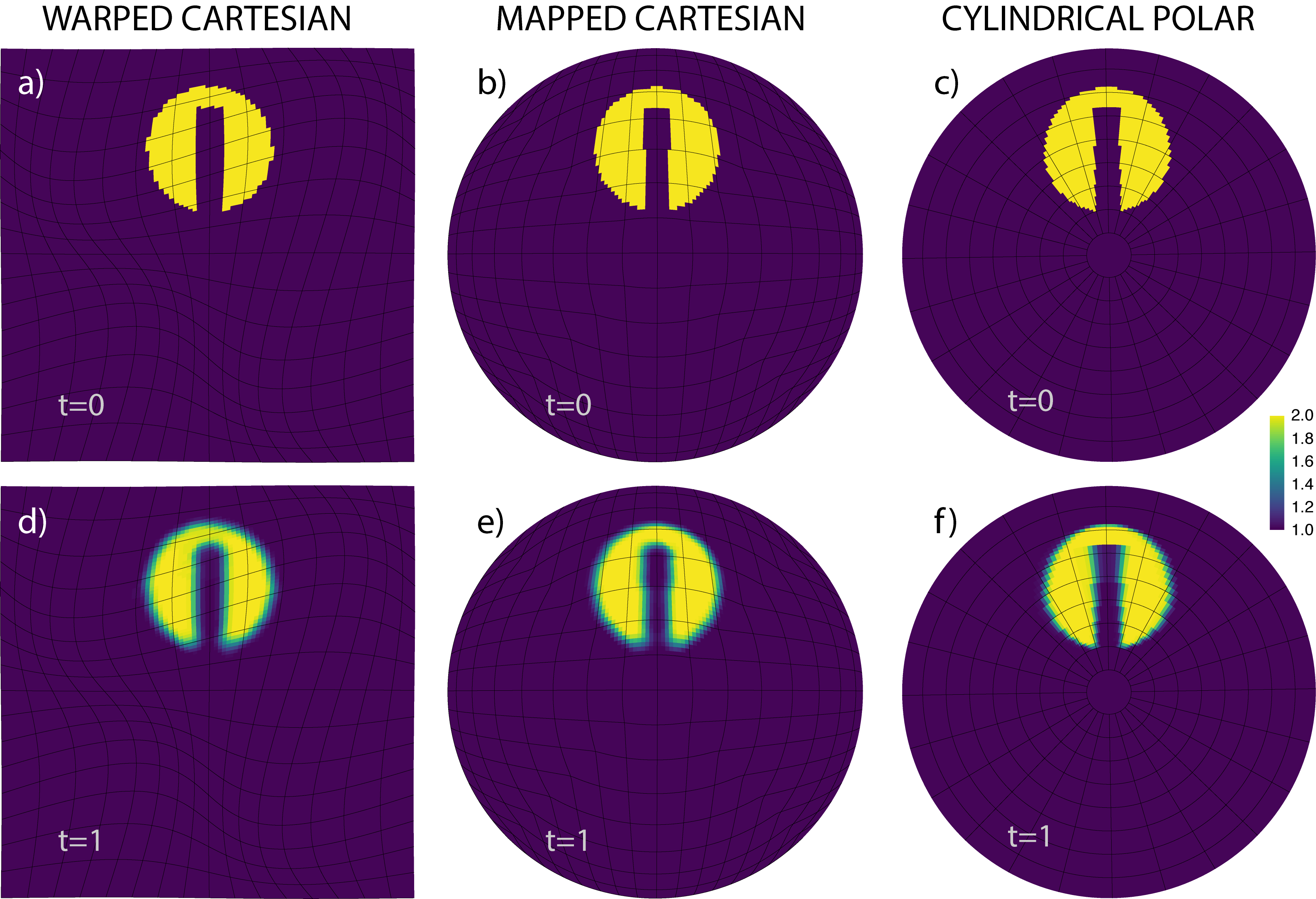}
	\centering
	\caption{The circular advection results using three different curvilinear grids.} \label{fig:cadv2_cut}
\end{figure}

The initial and final solution for the same two-dimensional circular advection simulation on three non-Cartesian grids are shown in Figure \ref{fig:cadv2_cut}. Figure \ref{fig:cadv2_cut}a and \ref{fig:cadv2_cut}d are the results using a distorted Cartesian grid with $128\times 128$ cells \citep{Colella2011}. Figure \ref{fig:cadv2_cut}b and \ref{fig:cadv2_cut}e use a mapped grid deforms a Cartesian domain with $128\times 128$ cells into a cylindrical domain \citet{Calhoun2008}; and Figure \ref{fig:cadv2_cut}c and \ref{fig:cadv2_cut}f are from a spherical polar grid  (this simulation uses $72\times 192$ cells in order to obtain similar spatial resolution near the slotted cylinder as the other two grids using $128\times 128$ cells). This set of simulation results shows that the advection scheme with the $7^{th}$-order reconstruction method preserves the shape of the slotted cylinder reasonably well even in non-uniform, non-orthogonal grid geometries.
\begin{figure}[t!]
	\noindent\includegraphics[width=38pc]{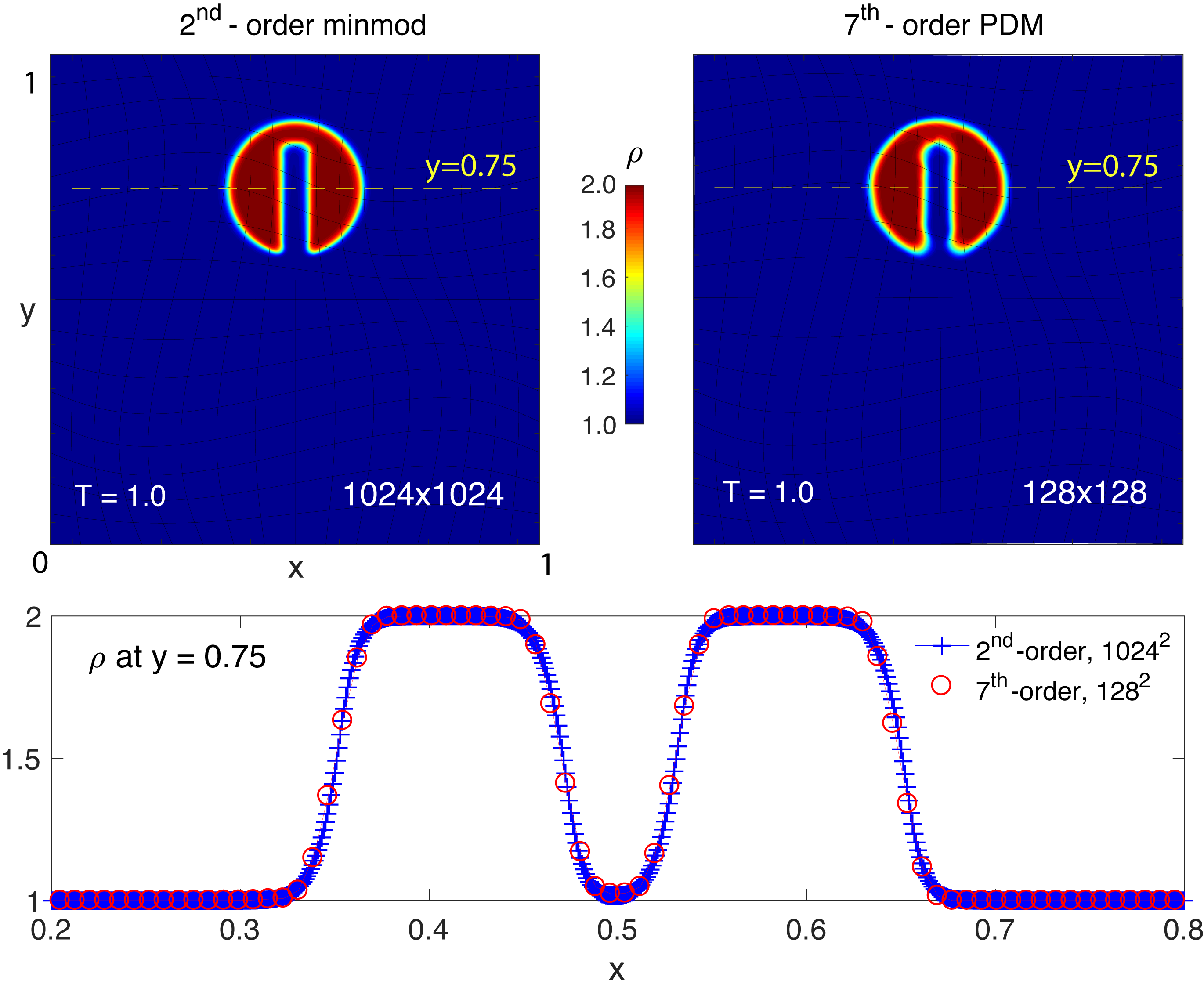}
	\centering	\caption{The circular advection results using three different curvilinear grids.} \label{fig:pdm_tvd}
\end{figure}
The use of high-order reconstruction schemes is essential for the GAMERA code to resolve substantial flow structures with relatively small amount of computational cells compared to commonly-used $2^{nd}$-order TVD schemes, especially when the grid geometry varies significantly in the computation space. Figure {\ref{fig:pdm_tvd}} shows the 2-D circular advection simulation using a $2^{nd}$-order TVD scheme and the PDM scheme with the $7^{th}$-order reconstruction method. 

The top left panel of Figure \ref{fig:pdm_tvd} shows the 2-D circular advection simulation results at $t=1.0$ using the 2nd-order TVD scheme (minmod limiter) with 1024x1024 cells. The computation grid is the same distorted Cartesian as in Figure \ref{fig:cadv2_cut}a, which is non-uniform and non-orthogonal. The top right panel of Figure \ref{fig:pdm_tvd} shows the corresponding distribution of the slotted cylinder using the GAMERA scheme with $7^{th}$-order reconstruction. The comparisons between the two simulations suggest that qualitatively, the spatial distribution of the slotted density in the $2^{nd}$-order TVD scheme resembles that in the $7^{th}$-order PDM scheme. The comparison is more quantitative in the line cut profiles of the slotted density as shown in the bottom panel of Figure \ref{fig:pdm_tvd}, which are taken right through the center of the slotted cylinder at $y=0.75$. The comparisons between the two cut profiles suggest that in order to get a similar density gradient in the simulation using the PDM scheme with $128\times128$ cells, $1024\times1024$ cells are needed while using the $2^{nd}$-order TVD scheme, which is $64\times 8$ times more computations per unit simulation time for the 2-D test problem. Given that the computational cost for the $7^{th}$-order PDM scheme is about $24\%$ more than a second order TVD scheme, the total computational cost for the $2^{nd}$-order TVD scheme is more than two orders of magnitude higher than the $7^{th}$-order PDM scheme. For 3-D simulations, the difference in computation time is even more. 

\subsection{Field-Loop Advection}

The field-loop advection test consists of the advection of a circular magnetic field loop by a constant initial velocity in a two-dimensional, periodic simulation domain. The initial conditions used in this study is adapted from \citet{Stone2008}. The simulation domain is a 2-D Cartesian box with $-L_x\le x\le L_x$, $-L_y\le y\le L_y$. The plasma flow is inclined at $30^\circ$ to the horizontal direction, ensuring that the numerical fluxes in the $x$- and $y$-directions are different, which makes the test truly multi-dimensional. In order to test the effectiveness of the MHD solver in non-orthogonal grid geometries, three additional sets of field-loop advection simulations are performed using the Cartesian grid with increasing level of distortion, in which the numerical fluxes in $x$- and $y$-directions are different regardless of the direction of the flow. The distorted Cartesian grids used in this section are generated using the following mapping functions:

\begin{equation}
x_{i,j} =  L_x\left(2\left[\frac{i}{N_i} + w_0\sin(\frac{i\pi}{N_i})\sin(\frac{j\pi}{N_j})\right]-1\right),\label{eqn:cart_distort1}
\end{equation}
\begin{equation}
y_{i,j} =  L_y\left(2\left[\frac{j}{N_j} + w_0\sin(\frac{i\pi}{N_i})\sin(\frac{j\pi}{N_j})\right]-1\right),\label{eqn:cart_distort2}
\end{equation}

where $L_x=1/\sin\frac{\pi}{3}$, $L_y=1/\cos\frac{\pi}{3}$ and $w_0$ is the parameter determines the amount of grid distortion. The integers $i=1,2,...,N_i$ and $j=1,2,...,N_j$ are the grid index in the $\mu$- and $\nu$-directions, respectively, with $N_i$ and $N_j$ the total number of cells in each direction. A similar test for this magnetic field loop advection using much more complicated warped gridding schemes can be found in \citet{Zilhao2014}. 

The initial plasma density $\rho$ and thermal pressure $P$ are both 1.0, with $\gamma=\frac{5}{3}$. The magnitude of the initial flow velocity is 1.0, with $v_x=\cos(\pi/6)$,  $v_y=\sin(\pi/6)$ and $v_z=0$. The magnetic field loop is initialized using the following vector potential $\mathbf{A}=(A_x,A_y,A_z)$:
\begin{equation}
 \begin{tabular}{ll}
  $A_x = A_y = 0$, &  \\
  $A_z = MAX\left(\left[A_0\left( R_0 - r \right)\right],0\right),$ & 
  \end{tabular}\label{eqn:loop2d}
\end{equation}
where $A_0$ is the magnitude of the field loop, $R_0$ is the center of the loop and $r = \sqrt{x^2+y^2}$. To ensure that the magnetic field loop is in quasi magnetostatic equilibrium ($\beta \gg 1$), we use $A_0 = 1.0\times 10^{-3}$ for the loop with $R_0 = 0.3$. Face-centered magnetic fluxes $\Phi_{i,j,k}$ are computed using the face-integrated form of $\nabla\times\mathbf{A}$ to guarantee $\nabla\cdot\mathbf{B}=0$ initially. Note that the vector potential (\ref{eqn:loop2d}) has a discontinuous first derivative. Thus the initial condition has a line current at the center of the loop and a surface return current. These non-smooth currents make the simulation more difficult and provide excellent tests for the effectiveness of the MHD solver handling non-orthogonal, curvilinear geometries.

\begin{figure}[t!]
	\noindent\includegraphics[width=40pc]{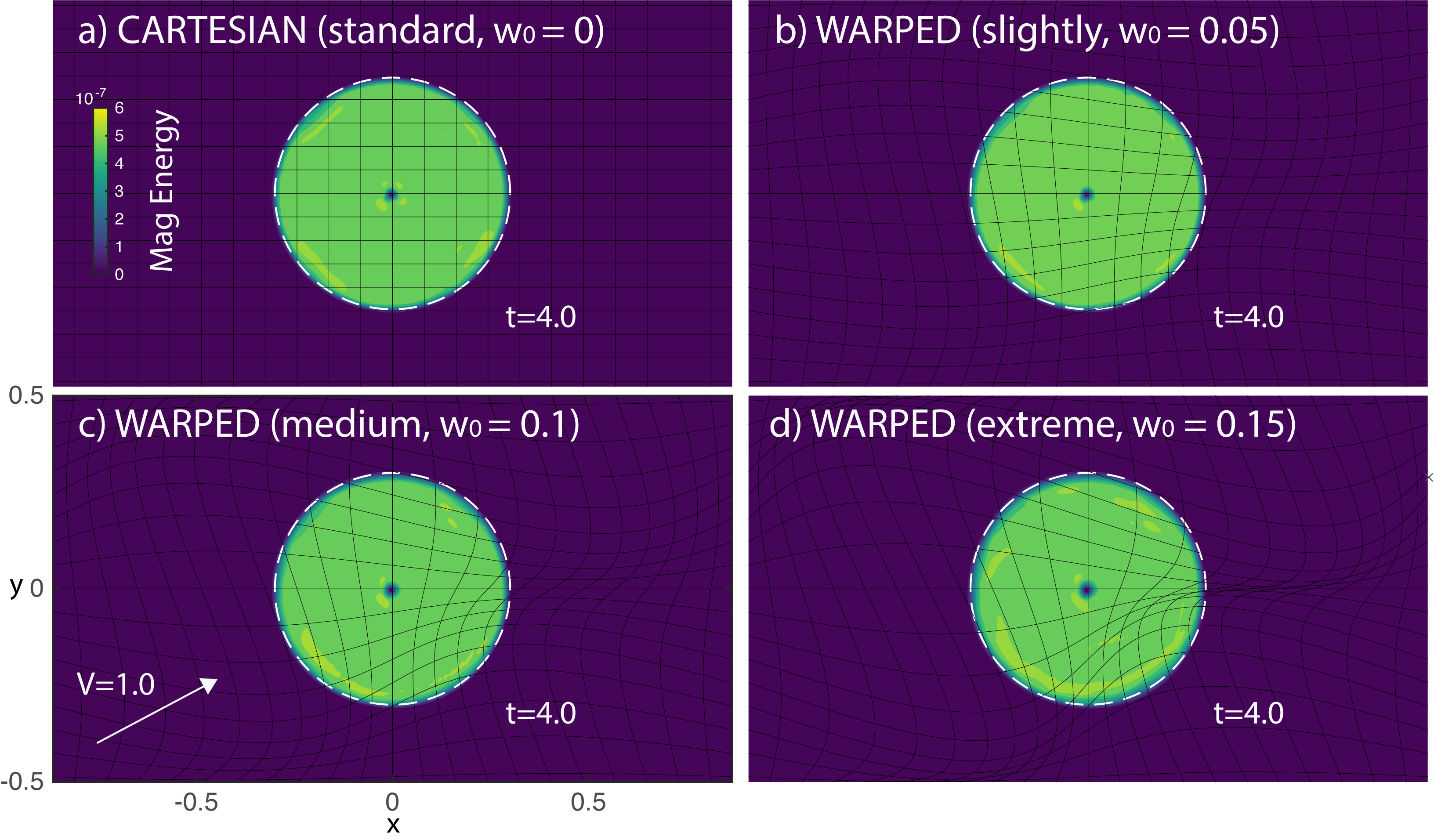}%
	\centering
	\caption{The spatial distribution of magnetic energy at $t=4$ from four curvilinear grids.} \label{fig:loop2d}
\end{figure}

Figure \ref{fig:loop2d} shows the spatial distributions of magnetic energy ($\frac{1}{2}B^2$) at $t=4.0$ from four sets of simulations using the same initial conditions but different computational grids. At $t=4.0$, the field loop has been advected around the grid twice. The grid resolution used in each simulation is $256\times 128$. The boundary of the magnetic field loop ($r=0.3$) at $t=0$ is shown using dashed white contours. Figure \ref{fig:loop2d}a shows the spatial distribution of magnetic energy at $t=4.0$ using the standard Cartesian grid as a reference, while Figure \ref{fig:loop2d}b-d show the corresponding simulation results in the distorted Cartesian grids with $w_0=0.05,0.1,0.15$, respectively. The comparisons in Figure \ref{fig:loop2d} show that the spatial distributions of the magnetic energy at $t=4$ from distorted Cartesian grids resemble that in the standard Cartesian grid, without noticeable distortions or asymmetries in the shape/boundary of the field loop, even using an extremely distorted grid ($w_0=0.15)$ as shown in Figure \ref{fig:loop2d}d.

\begin{figure}[t!]
	\noindent\includegraphics[width=40pc]{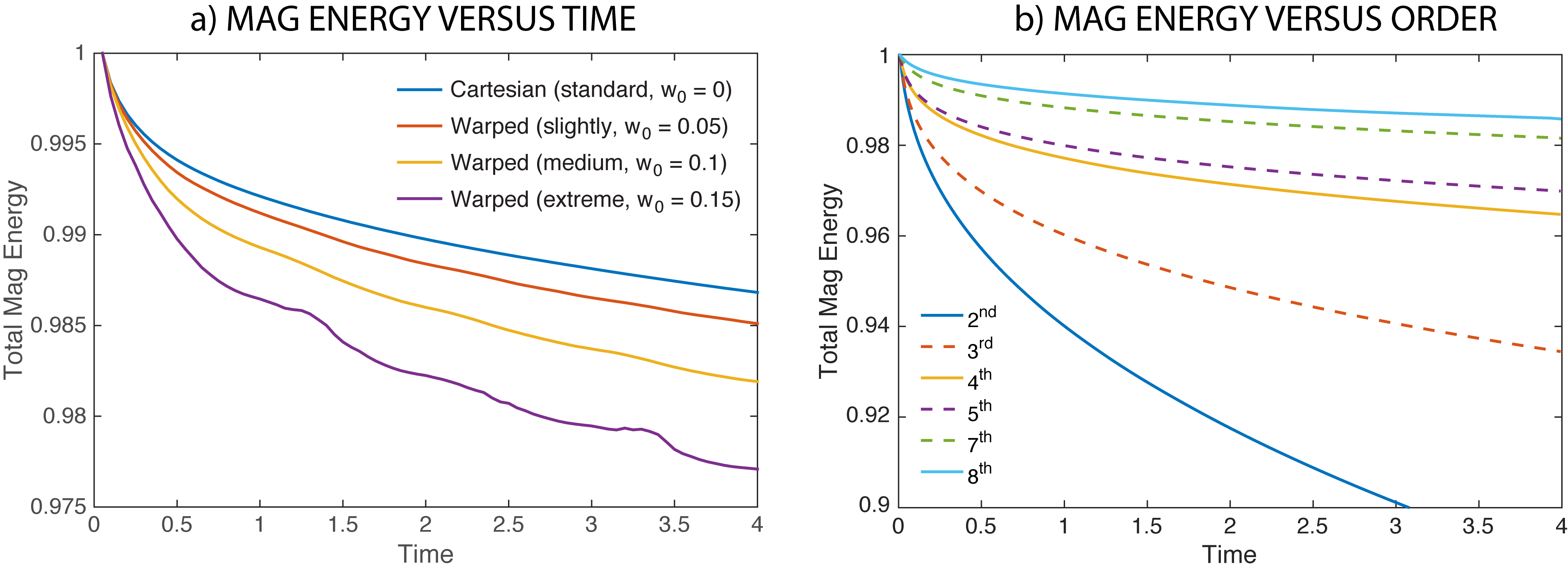}%
	\centering
	\caption{a) The decay of total magnetic energy as a function of time derived from the four field-loop advection simulations using different grids. b) The decay of total magnetic energy as a function of time from six simulations with different orders of reconstruction in Cartesian geometries. } \label{fig:loop2d_time}
\end{figure}

Figure \ref{fig:loop2d_time}a shows the decay of the volume-integrated magnetic energy as a function of simulation time derived from the four sets of simulations. In the standard Cartesian case, after crossing the simulation box twice ($t=4.0$), the total magnetic energy decreases to approximately $98.7\%$ of the original value. The simulations using distorted Cartesian grids exhibit similar behaviors in the decay of magnetic energy, but with decreasing final magnetic energy around $98.5\%$ $98.2\%$ and $97.7\%$ of the original value as the distortion parameter $w_0$ increases from $0.05$ to $0.15$. As the grid becomes more distorted (from case b) to case d)), the rate of magnetic energy decay varies as a function of location, because the field loop samples grid resolution non-uniformly when it advects along the diagonal. An animation of the field-loop advection simulations in four different grids is included in the Supplemental material.

\begin{figure}[t!]
	\noindent\includegraphics[width=44pc]{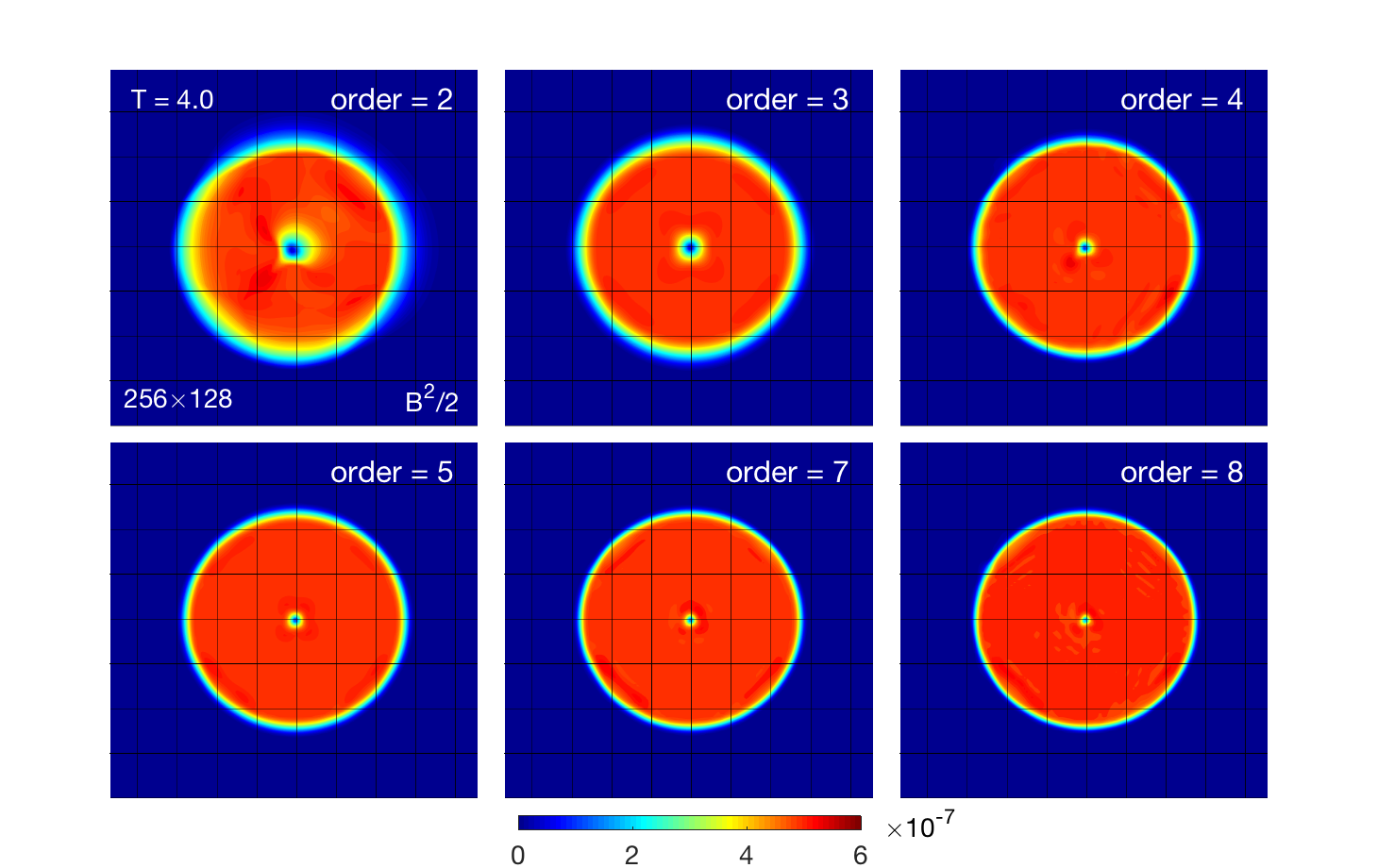}%
	\centering
	\caption{The corresponding spatial distributions of magnetic energy at $t=4$ using six different orders of reconstruction.} \label{fig:loop2d_degrade}
\end{figure}

Figure \ref{fig:loop2d_time}b shows the decay of total magnetic energy with time using six different orders of reconstruction. The grid geometries used in the six field loop simulations were all standard Cartesian ($w_0 = 0$) with $256\times128$ cells. By the time when the magnetic field loop crossed the simulation box twice ($t=4.0$), the total magnetic energy decreases with the order of reconstruction used in the solver. The total magnetic energy in the second-order reconstruction method is about $88.5\%$, while in the $7^{th}$-order reconstruction method is about $98.1\%$. Although the total magnetic energy in the end of the simulation using a second-order reconstruction only degraded approximately $10\%$, the spatial distribution is significantly distorted compared to the high-order reconstruction schemes ($7^{th}$- and $8^{th}$-order), which is shown in Figure \ref{fig:loop2d_degrade}. As the order of reconstruction decreases from 8 to 2, the thickness of the edge of the magnetic field loop increases significantly, which is a consequence of the increased numerical ``diffusion width'' with decreasing order of reconstruction, as analyzed in detail in Appendix D.

For the 2-D MHD test simulations, the additional computing cost using the $7^{th}$-order reconstruction scheme is approximately $24\%$ compared to the $2^{nd}$-order scheme. When using the $8^{th}$-order reconstruction scheme, the additional computational cost is only about $12\%$ more compared to the $2^{nd}$-order scheme. For 3-D field-loop advection simulations (not shown in this section), the computational costs for going from $2^{nd}$-order to $7^{th}$-order and $8^{th}$-order are approximately $23\%$ and $5\%$, respectively. 

\subsection{Non-linear Alfv\'{e}n Wave}

We run the non-linearly polarized circular Alfv\'{e}n wave simulation from \citet{Toth1997} to test the performance of the numerical MHD schemes in nonlinear regime. Since the Alfv\'{e}n wave simulation is smooth, it is also a reasonable test for convergence of the MHD solver. The simulation domain is chosen to be a 2-D Cartesian box with $0\le x\le 1$ and  $0\le y\le \frac{1}{\cos\alpha}$, where $\alpha = \frac{\pi}{3}$ is the angle at which the Alfv\'{e}n wave propagates with respect to the $x$-axis. Thus the waves do not propagate along the diagonal of the simulation domain, which guarantees the multi-dimensional nature of the test simulation by making the numerical flux different in the $x$- and $y$-direction. The initial conditions are $\rho = 1$, $P = 0.1$, $u_\perp = 0.1 \sin2\pi x_\|$, $B_\perp = 0.1 \sin2\pi x_\|$, and $u_z = B_z = 0.1 \cos2\pi x_\|$ with $\gamma = \frac{5}{3}$ and $x_\| = (x \cos\alpha + y \sin\alpha)$, where $u_\perp$ and $B_\perp$ are the components of velocity and magnetic field perpendicular to the wavevector. The $B_\|$ and $B_\perp$ components are calculated using cell centered $B_x$ and $B_y$ via $B_\perp = B_y \cos\alpha - B_x \sin\alpha$, and $B_\| = B_x \cos\alpha + B_y \sin\alpha$. The initial magnetic fluxes $\Phi$ are calculated using the following vector potential $\mathbf{A}$:

\begin{align}
A_x\left(x,y,z\right) & =  B_{\|0}\sin\alpha z \\
A_y\left(x,y,z\right) & = -B_{\|0}\cos\alpha z+\frac{0.1}{2\pi\cos\alpha}\sin\left[2\pi(x\cos\alpha+y\sin\alpha)\right] \\
A_z\left(x,y,z\right) & =  \frac{0.1}{2\pi}\cos\left[2\pi(x\cos\alpha+y\sin\alpha)\right],
\end{align}
where $B_{\|0}=1.0$ is the strength of the initial magnetic field parallel to the wave vector.
 
\begin{figure}[b!]
	\noindent\includegraphics[width=40pc]{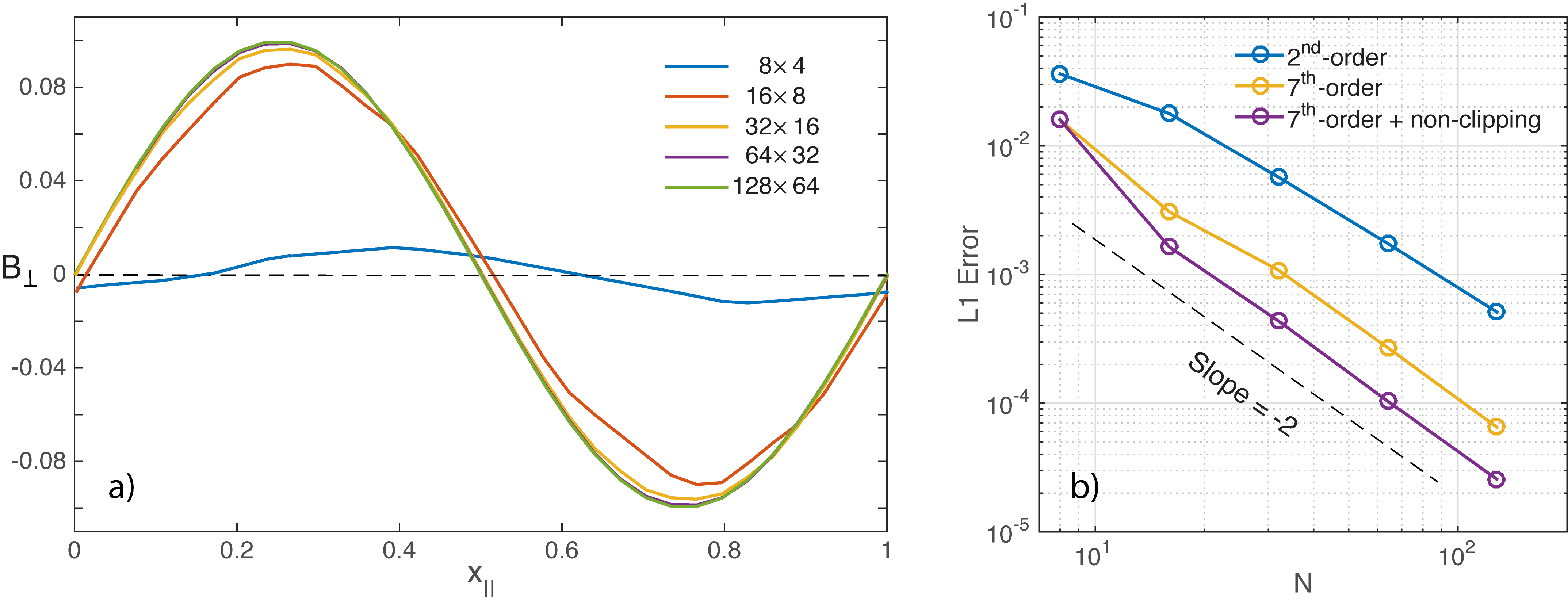}%
	\centering
	\caption{a) The distribution of $B_\perp$ versus $x_\|$ at t=5.0 in nonlinear Alfv\'{e}n wave simulations, computed from five runs with increasing resolution. b) the average $L1$ error as a function of grid resolution $N$.} \label{fig:aflven_converge}
\end{figure}

Figure \ref{fig:aflven_converge}a shows the simulated $B_\perp$ component along the direction of wave propagation $x_\|$ at $t=5.0$, using the default scheme with the $7^{th}$-order spatial reconstruction and the PDM limiter. Six grid resolutions with $N\times\frac{N}{2}$ cells are used to test the convergence of the solution at $t=5.0$, by increasing the grid parameter $N$ from 8 to 256. When the grid resolution is $64\times 32$ and above, the solution of $B_\perp$ along $x_\|$ starts to converge to the analytical solution $B_\perp=0.1\sin 2\pi x_\|$. Figure \ref{fig:aflven_converge}b shows the average $L1$ error norm as a function of grid size $N$ using two different reconstruction methods ($2^{nd}$- and $7^{th}$-order reconstruction), We find a $2^{nd}$-order convergence rate based on this smooth but nonlinear Alfv\'{e}n wave simulation, regardless of the size of the reconstruction stencil, suggesting that the formal order of convergence for the MHD solver is two. This is expected since no high-order quadrature methods are used in the evaluation of face fluxes in the finite-volume solver as discussed in Section \ref{sec:flux_func}. However, the formal order of convergence ($2^{nd}$-order) does not mean that the high-order reconstruction ($7^{th}$-order and above) is not necessary in the solver. In fact, it is important to note that when using the default $7^{th}$-order reconstruction scheme, the magnitude of average $L1$ error is approximately a factor of ten lower than that from a $2^{nd}$-order reconstruction scheme, suggesting the necessity of using high-order reconstructions for achieving highly-accurate solutions with moderate number of computational cells. This reduction in average $L1$ error is consistent with the comparisons in Figure \ref{fig:cadv2_cut}, which shows significant improvement in preserving the shape of initial density structures. Note that when the non-clipping option is switched on, the average $L1$ error is further reduced by approximate a factor of $3$ due to the preserving of local extrema.

\begin{figure}[h!]
	\noindent\includegraphics[width=42pc]{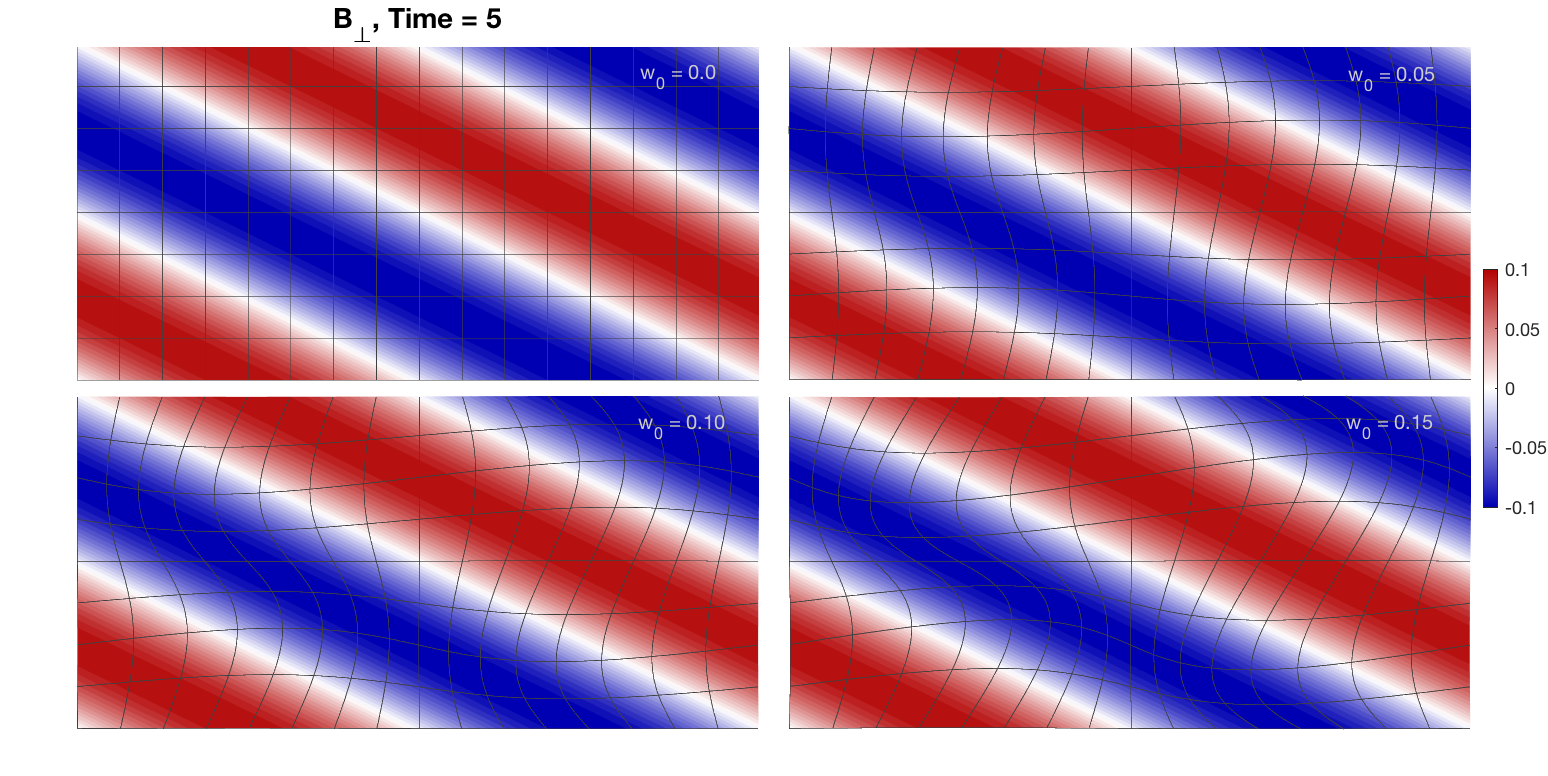}%
	\centering
	\caption{The spatial distribution of $B_\perp$ at $t=5.0$ derived from four nonlinear Alfv\'{e}n wave simulations Cartesian geometries with different levels of distortion indicated by the $w_0$ parameter in Eqn. \ref{eqn:cart_distort1}-\ref{eqn:cart_distort2}. The grid resolution used in each simulation is $256\times128$ cells.}\label{fig:alfven_2grid}
\end{figure}

\begin{figure}[t!]
	\noindent\includegraphics[width=35pc]{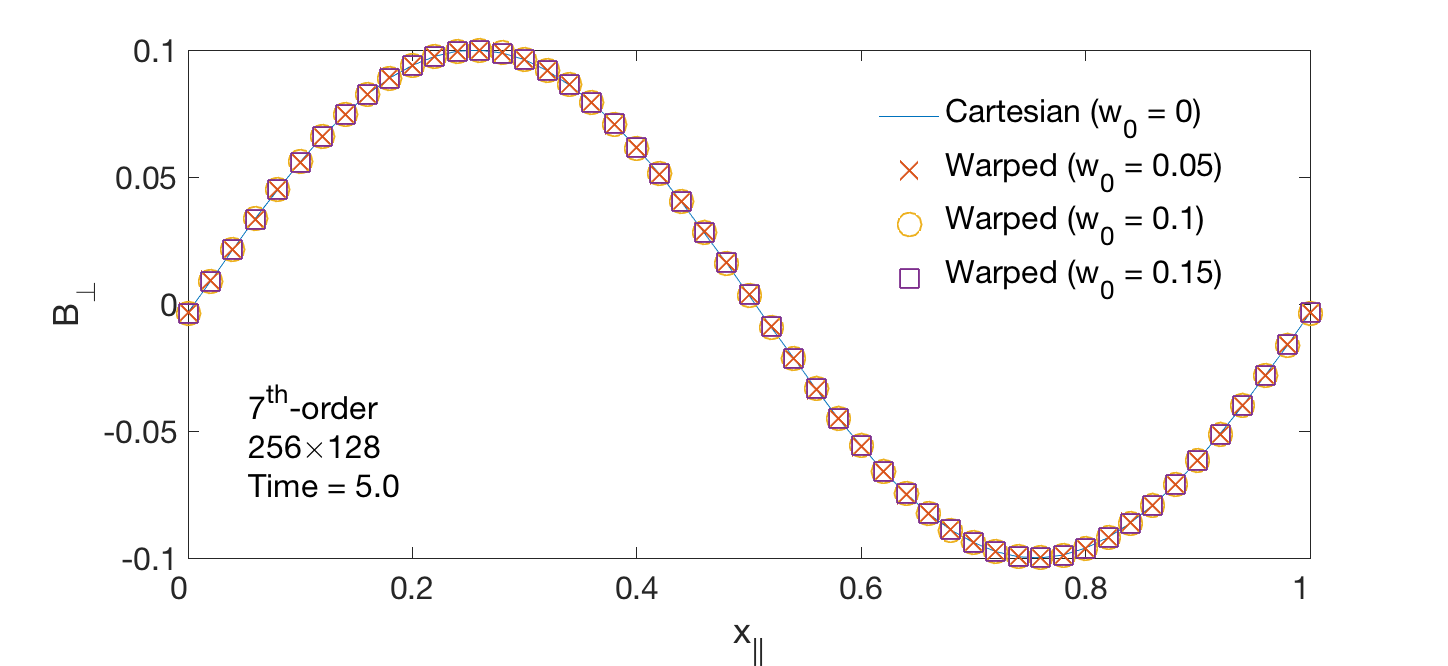}%
	\centering
	\caption{The spatial distribution of $B_\perp$ at $t=5.0$ along the wave vector derived from the four simulations with different levels of grid distortion.} \label{fig:bperp_cut}
\end{figure}

Figure \ref{fig:alfven_2grid} compares the spatial distributions of $B_\perp$ at $t=5.0$ from four sets of non-linear Alfv\'{e}n wave simulations using exactly the same initial and boundary conditions in Cartesian geometries with different levels of distortion. The grid size used in the test simulations are $256\times128$. Figure \ref{fig:alfven_2grid}a is from a standard, orthogonal Cartesian grid, and Figure \ref{fig:alfven_2grid}b-d are from distorted, non-orthogonal Cartesian grids defined in Equation (\ref{eqn:cart_distort1}) - (\ref{eqn:cart_distort2}) with $w_0=0.05, 0.1$ and $0.15$, respectively. Compared to the Cartesian case in Figure \ref{fig:alfven_2grid}a, the simulated wavefronts indicated by $B_\perp$ in the distorted, non-orthogonal grids are perpendicular to the direction of propagation, without distortions introduced by the non-orthogonal grids. Figure \ref{fig:bperp_cut} shows the comparison on the perpendicular component of the magnetic field ($B_\perp$) along the wave vector derived from the four test simulations with different grids. The cut profiles were taken at the center of each simulation domain for one parallel wavelength. These comparisons shown in Figure \ref{fig:alfven_2grid} and \ref{fig:bperp_cut} also demonstrate the capability of the numerical schemes handling smooth MHD solutions in the non-linear regime when using non-orthogonal grids.

\subsection{Orszag-Tang Vortex}

\begin{figure}[t!]
	\noindent\includegraphics[width=34pc]{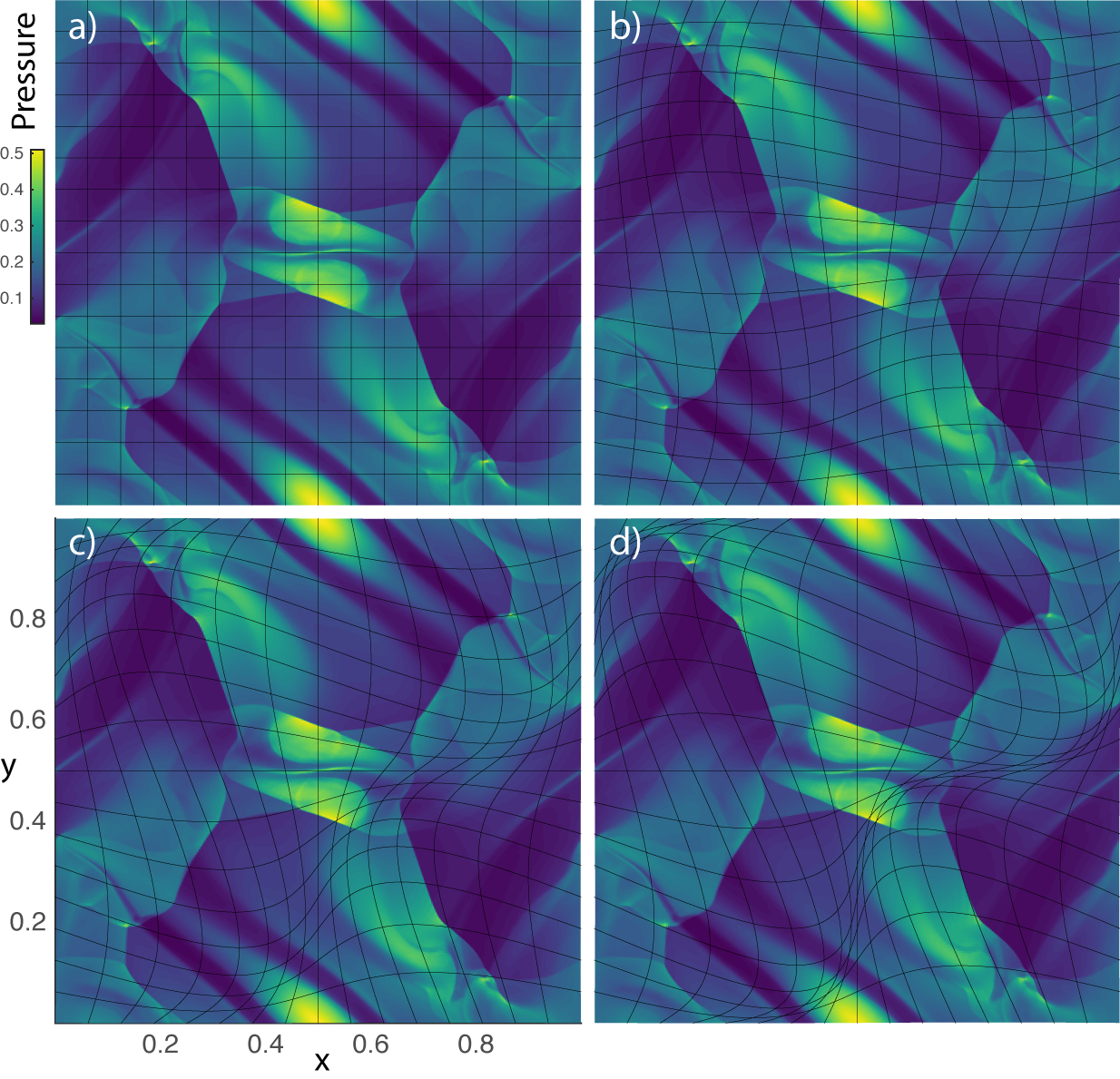}%
	\centering
	\caption{The spatial distribution of plasma pressure $P$ at $t=0.48$ in four Orszag-Tang simulations using different grids. Each simulation has $512\times 512$ cells. } \label{fig:ot2d}
\end{figure}

\begin{figure}[htb!]
	\noindent\includegraphics[width=30pc]{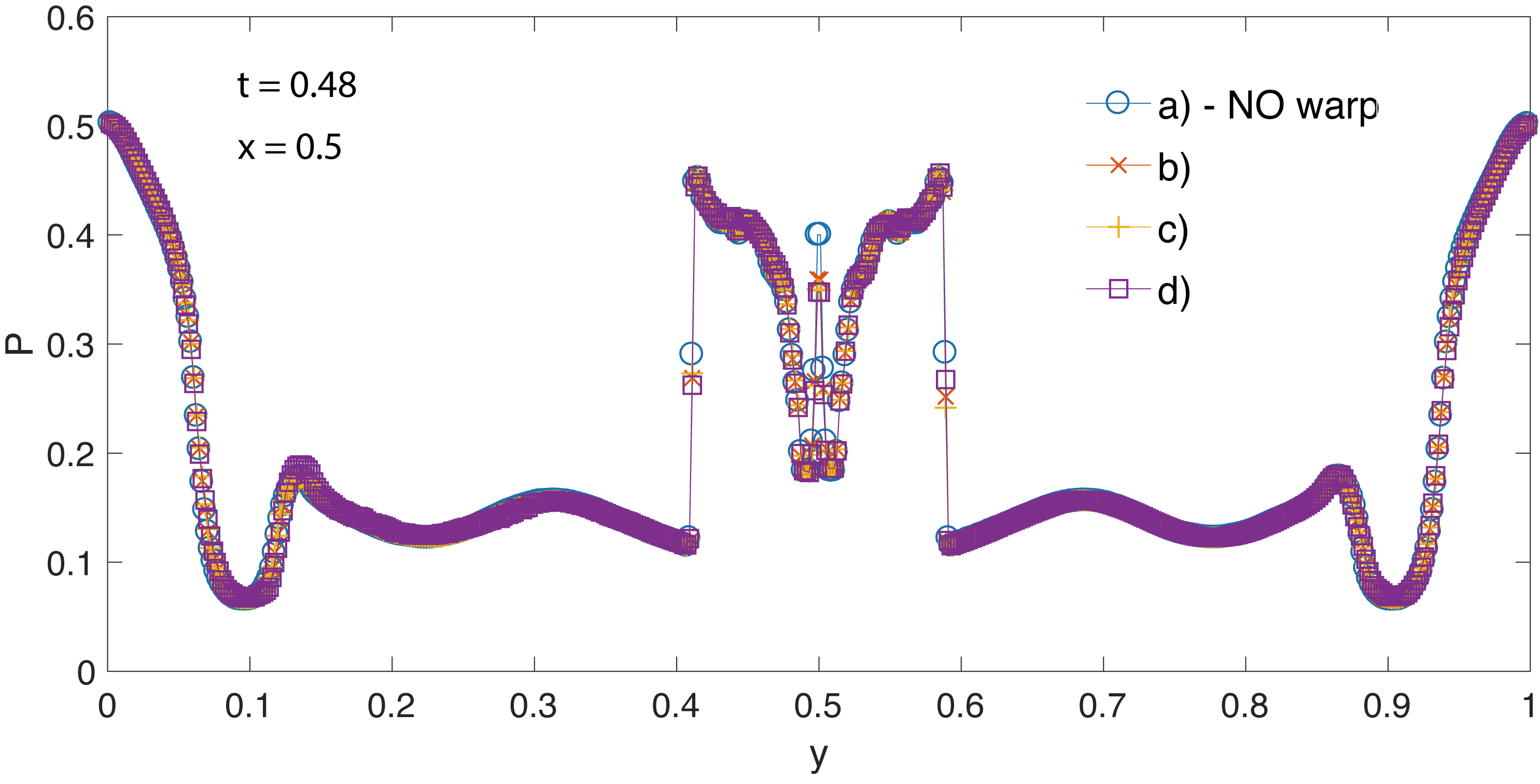}%
	\centering
	\caption{The line profiles of plasma pressure at $x=0.5$ from four Orszag-Tang simulations using different grids, taken from same simulation snapshots with $t=0.48$.} \label{fig:ot2d_cut}
\end{figure}

We use the Orszag-Tang vortex simulation to test how robust the code is for handling the formation and interaction of non-linear MHD shocks in non-orthogonal, curvilinear geometry. The Orszag-Tang vortex is a well-known two-dimensional problem for testing the transition to supersonic MHD turbulence, which is a very common test of numerical MHD schemes used in many previous studies as a basis for consistent comparison of codes. In this test, we focus on the effectiveness of the numerical schemes in handling non-orthogonal grids and compare the results in non-orthogonal geometries with that from a standard uniform Cartesian grid. For simulating the Orszag-Tang vortex, we use a 2-D square simulation domain with $0\le x\le 1$ and  $0\le y\le 1$. The initial density $\rho = \frac{25}{36}\pi$ and the initial pressure is $P=\frac{5}{12}\pi$ distributed uniformly in the simulation domain, with $\gamma = \frac{5}{3}$. The initial velocities are set as $v_x = - \sin(2\pi y)$ and $v_y = \sin(2\pi x)$. The magnetic fluxes $\Phi$ are initialized using the following vector potential function $\mathbf{A}$:
\begin{equation}
 \begin{tabular}{ll}
  $A_x = A_y = 0$, &  \\
  $A_z = B_0 (\frac{1}{4\pi}\cos(4\pi x) + \frac{1}{2\pi}\cos(2\pi y)),$ & 
  \end{tabular}\label{eqn:ot2d}
\end{equation}
where $B_0 = 1.0$. The above vector potential function gives $B_x = -B_0\sin(2\pi y)$ and $B_y = B_0\sin(4\pi x)$ initially. The boundary conditions are periodic everywhere. All the test simulations shown in Figure \ref{fig:ot2d} have a computational grid with $512\times 512$ cells. The reconstruction method is the default $7^{th}$-order scheme without the non-clipping option switched on. 

Figure \ref{fig:ot2d}a shows the simulated spatial distributions of plasma pressure at simulation time $t=0.48$ in a uniform Cartesian grid, when the MHD shocks start to interact near the center of the simulation domain. Figure \ref{fig:ot2d}b-d shows the corresponding spatial distributions of plasma pressure at the same simulation time using three sets of non-orthogonal, distorted Cartesian grids as defined in Equation (\ref{eqn:cart_distort1})-(\ref{eqn:cart_distort2}) with increasing $w_0$. Although these comparisons are less quantitative, they do suggest that the simulated distributions of plasma pressure are not significantly affected by the choice of the computational grid. Both smooth structures and shocks at $t=0.48$ using non-uniform, non-orthogonal grids are remarkably similar to those in Figure \ref{fig:ot2d}a using a uniform Cartesian grid. The effectiveness of the MHD solver is also illustrated in more quantitative comparisons of using line profiles. Figure \ref{fig:ot2d_cut} shows the comparisons of the simulated plasma pressure profiles (at $t=0.48$) along the $y$-direction with $x=0.5$. Although the simulated pressure exhibits some minor differences at the shock transition regions, possibly due to the inhomogeneity of the grid, very little difference occurs in the smooth regions. The Orszag-Tang vortex simulations demonstrate the capability of the GAMERA code handling MHD shock propagation and interaction, without the requirement of the orthogonality of the computation geometry. An animation of the Orszag-Tang vortex simulation using the four different grid geometries is included in the supplemental material.

\subsection{Spherical Blast Wave}

We use MHD blast wave simulations to test the robustness the numeric schemes in handling multi-dimensional, strong MHD shocks and rarefactions. In a two-dimensional simulation domain in the $x$-$y$ plane, The initial plasma has uniform density $\rho=1$, thermal pressure $P=0.1$ and zero velocity $\mathbf{u}=0$, with $\gamma=\frac{5}{3}$. Within the spatial region $r = \sqrt{x^2+y^2} < 0.1$, the thermal pressure $P$ is set to 10.0 (that is, 100 times greater than the ambient plasma pressure). The initial magnetic field is along the diagonal in the $x$-$y$ plane with $B_x = \frac{1}{\sqrt{2}}$, $B_y = \frac{1}{\sqrt{2}}$ and $B_z = 0$.

\begin{figure}[b!]
	\noindent\includegraphics[width=34pc]{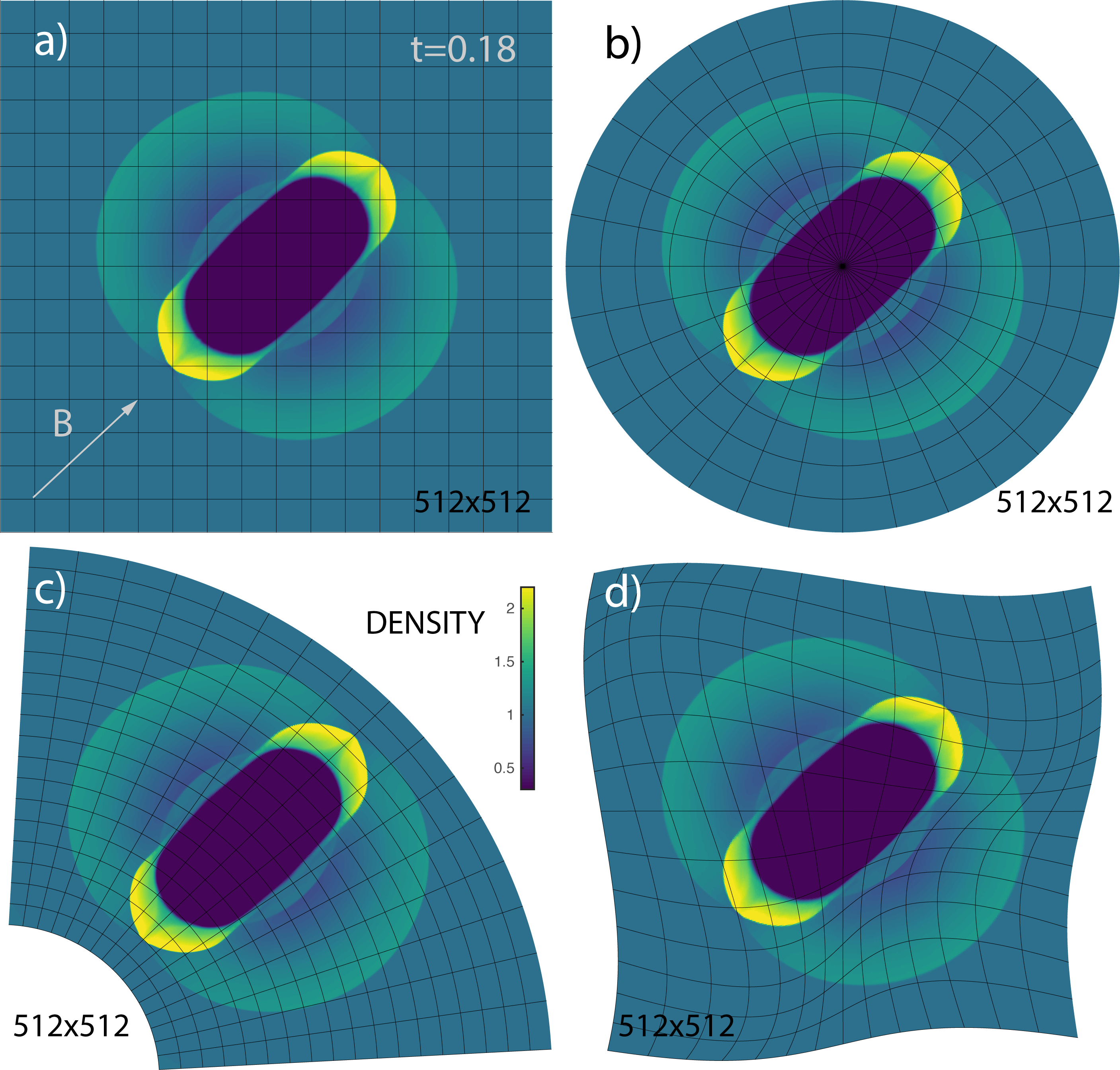}%
	\centering
	\caption{The spatial distribution of plasma pressure $P$ at $t=0.48$ in four MHD spherical blast wave simulations using different grids. Each simulation has $512\times 512$ cells.}\label{fig:bw}
\end{figure}

We use four different computational grids for the MHD blast wave simulation with the same initial conditions. These grids include a standard Cartesian grid (Figure \ref{fig:bw}a), a cylindrical polar grid (Figure \ref{fig:bw}b), a quarter of a ring (Figure \ref{fig:bw}c) and a distorted Cartesian grid (Figure \ref{fig:bw}d) as used in the Orszag-Tang simulations. Each grid has a resolution of $512\times 512$ cells with the center of the blast wave located at the origin. The Cartesian grid is used as a reference, while the non-Cartesian grids are used to test the propagations of strong MHD shocks and rarefaction in non-orthogonal, curvilinear geometries. The cylindrical polar grid is singular at the pole, which is treated using a conservative averaging-reconstruction technique developed by \citet{Zhang2018a} without changing the numerical schemes of the MHD solver. An animation of the MHD blast wave simulation in the four different grids is included in the supplemental material. As shown in the animation, at early simulation times (e.g., $t<0.05$), the out-going blast waves are approximately spherical and show no grid alignment effects due to non-orthogonal curvilinear geometries. The results shown in Figure \ref{fig:bw} are the spatial distributions of plasma density at $t=0.18$. Although the spherical blast wave simulations are not very quantitative, it shows that the spatial distributions of plasma density in the blast wave simulation at a later time are not sensitive to the choice of the curvilinear grid. In the simulation results shown in Figure \ref{fig:bw}, both smooth structures and shocks in the non-Cartesian geometries (Figure \ref{fig:bw}b-d) are remarkably similar to that in the reference Cartesian grid (Figure \ref{fig:bw}a) without noticeable differences. Note that the Richtmyer-Meshkov instability is suppressed by the strong magnetic field, and no fingers are evident in the interior of the field-aligned bubble as those in a hydrodynamic blast wave simulation.

\section{Summary}

The GAMERA code is a re-invention of the LFM MHD kernel with significant upgrades in both numerical schemes and software implementation. The improvements in numerical schemes include 1) $12^{th}$-order grid metric calculations using Gaussian quadrature; 2) high-order upwind reconstruction method; 3) the PDM limiter with an extremum-preserving option; 4) background magnetic field splitting for very low-$\beta$ conditions. The Ring-Average technique for spherical axis singularity described by \citet{Zhang2018a} is also implemented in the GAMERA code for high-resolution simulations in general geometry with axis singularities. The improvements in software implementation include 1) modern Fortran implementation with minimum external library dependence; 2) core computational routines that act on vector-length aligned blocks of memory; 3) extensive use of OpenMP threading within a socket allowing MPI ranks distributed across sockets; 4) High performance Computing (HPC) -friendly data structures, namely the use of large, contiguous arrays and contiguous memory access patterns. The GAMERA code is designed to be easily applicable to multi-dimensional MHD flow simulations in non-orthogonal curvilinear grids adapted to specific conditions. Current applications of the GAMERA code includes global simulations of planetary magnetospheres, the inner heliosphere and solar wind, and local simulations of basic plasma physics applications, e.g., current sheets.

This paper provides a comprehensive description of the numerical recipes used in the GAMERA code. Our hope in the development of this paper has been to give some context for the key questions and solutions that need to be addressed in order to design a large scale simulation code for MHD problems. The numerical recipes described in this paper are useful as a reference, but it also shows that, when compared with other codes, these recipes are not the same.  The nuances of disparate application areas will lead to different codes that excel in their tailored regimes.  Although the numerical considerations of the LFM, inherited by GAMERA, are tailored to heliospheric environments we have designed GAMERA to be more general.  We expect GAMERA to be a useful tool in any application where geometric flexibility is important.



\section*{Acknowledgement}

The research was supported by the National Center for Atmospheric Research (NCAR) and the Johns Hopkins University Applied Physics Laboratory Independent Research \& Development funds. We would like to acknowledge high-performance computing support from Cheyenne (doi:10.5065/D6RX99HX) provided by NCAR's Computational and Information Systems Laboratory, sponsored by the National Science Foundation (NSF). Michael Wiltberger was serving at the NSF during the production of the this paper. Any opinion, findings, or conclusions or recommendations expressed in this material are those of the authors and do not necessarily reflect the views of the NSF. 

\section{Appendix}

\section*{A. Normalization of the ideal MHD equations}

The mass and momentum equations in SI units follows
\begin{equation}
\frac{\partial\rho}{\partial t} + \nabla\cdot\left( \rho \mathbf{u}\right) =0,\label{eqn:mass_norm}
\end{equation}
\begin{equation}
\frac{\partial\rho \mathbf{u}}{\partial t} + \nabla\cdot\left( \rho\mathbf{u}\mathbf{u} + \bar{\mathbf{I}}P \right) - \frac{1}{\mu_0}\mathbf{j}\times\mathbf{B}=0, \label{eqn:momentum_norm}
\end{equation}
where $\mu_0 = 4\pi\times 10^{-7}\,\mathrm{T\cdot m/A}$. To normalize the MHD variables in the above equations, define
\begin{align}
\rho&=\rho_0\tilde{\rho}, \label{eqn:norm1} \\
\mathbf{u}&=u_0\tilde{\mathbf{u}},\label{eqn:norm2}\\
P&=P_0\tilde{P},\\
\mathbf{B}&=B_0\tilde{\mathbf{B}},\label{eqn:norm4}
\end{align}
where $\rho_0$, $u_0$, $P_0$ and $B_0$ are the normalization constants for mass, velocity, pressure and magnetic field, respectively. $\tilde{\rho}$, $\tilde{\mathbf{u}}$, $\tilde{P}$, $\tilde{\mathbf{B}}$ are the corresponding dimensionless variables evolved by the MHD solver. For time and spatial coordinates, let $t = t_0\tilde{t}$ and $\mathbf{x}=x_0\tilde{\mathbf{x}}$ with $\tilde{t}$ and $\tilde{\nabla}$ the dimensionless temporal and spatial differential operators:
\begin{align}
\frac{\partial}{\partial t} &= \frac{1}{t_0}\frac{\partial}{\partial \tilde{t}}, \\
\nabla & = \frac{1}{x_0}\tilde{\nabla}.
\end{align}

First substitute (\ref{eqn:norm1}) and (\ref{eqn:norm2}) into the mass equation (\ref{eqn:mass_norm}):
\begin{equation}
\frac{1}{t_0}\frac{\partial\rho_0\tilde{\rho}}{\partial \tilde{t}} + \frac{1}{x_0}\tilde{\nabla}\cdot\left( \rho_0\tilde{\rho} u_0\tilde{\mathbf{u}}\right) =0 \longrightarrow \frac{x_0}{t_0u_0}\frac{\partial\tilde{\rho}}{\partial \tilde{t}} + \tilde{\nabla}\cdot\left( \tilde{\rho} \tilde{\mathbf{u}}\right) =0. 
\end{equation}
If we choose the following normalization between time and velocity:
\begin{equation}
t_0 = \frac{x_0}{u_0}\label{eqn:norm_relation1}, 
\end{equation}
the dimensionless mass equation is written as:
\begin{equation}
\frac{\partial\tilde{\rho}}{\partial \tilde{t}} + \tilde{\nabla}\cdot\left( \tilde{\rho} \tilde{\mathbf{u}}\right) =0. \label{eqn:mass_eqn_norm}
\end{equation}
Next substitute (\ref{eqn:norm1}) - (\ref{eqn:norm4}) into the momentum equation (\ref{eqn:momentum_norm}) and given that $x_0 = u_0t_0$ from the normalized mass equation:
\begin{align}
&\frac{\rho_0u_0}{t_0}\frac{\partial\tilde{\rho}\tilde{\mathbf{u}}}{\partial\tilde{t}} + \frac{1}{x_0}\tilde{\nabla}\cdot\left( \rho_0u_0^2\tilde{\rho}\tilde{\mathbf{u}}\tilde{\mathbf{u}} + \overline{\mathbf{I}}P_0\tilde{P}\right) - \frac{B_0^2}{\mu_0x_0}\tilde{\nabla}\times\tilde{\mathbf{B}}\times\tilde{\mathbf{B}} = 0\\
\Longrightarrow & \frac{\partial\tilde{\rho}\tilde{\mathbf{u}}}{\partial\tilde{t}} + \tilde{\nabla}\cdot\left( \tilde{\rho}\tilde{\mathbf{u}}\tilde{\mathbf{u}} + \overline{\mathbf{I}}\tilde{P}\frac{P_0}{\rho_0u_0^2}\right) - \frac{B_0^2}{\mu_0\rho_0u_0^2}\tilde{\nabla}\times\tilde{\mathbf{B}}\times\tilde{\mathbf{B}} = 0.
\end{align}
The above equation gives the following normalization for the plasma pressure $P_0$ and the magnetic field $B_0$:
\begin{align}
P_0 & = \rho_0u_0^2\label{eqn:norm_relation2} \\
B_0^2 & = \mu_0\rho_0u_0^2. \label{eqn:norm_relation3}
\end{align}
Thus the dimensionless momentum equation is written as:
\begin{equation}
\frac{\partial\tilde{\rho}\tilde{\mathbf{u}}}{\partial\tilde{t}} + \tilde{\nabla}\cdot\left( \tilde{\rho}\tilde{\mathbf{u}}\tilde{\mathbf{u}} + \overline{\mathbf{I}}\tilde{P}\right) - \tilde{\nabla}\times\tilde{\mathbf{B}}\times\tilde{\mathbf{B}} = 0\label{eqn:momentum_eqn_norm}
\end{equation}

It is straightforward to show that with the normalization relations (\ref{eqn:norm_relation1}), (\ref{eqn:norm_relation2}) and (\ref{eqn:norm_relation3}), the dimensionless plasma energy equation and Faraday's law are written as:
\begin{equation}
\frac{\partial \tilde{E_P}}{\partial \tilde{t}} = -\tilde{\nabla}\cdot\left[ \tilde{\mathbf{u}} \left( \tilde{E_P} + \tilde{P} \right) \right] - \tilde{\mathbf{u}}\cdot\tilde{\nabla}\cdot\left( \frac{\tilde{B}^2}{2}\bar{\mathbf{I}} - \tilde{\mathbf{B}}\tilde{\mathbf{B}} \right)\label{eqn:energy_eqn_norm}
\end{equation}
\begin{equation}
\frac{\partial \tilde{\mathbf{B}}}{\partial \tilde{t}} = \tilde{\nabla}\times\tilde{\mathbf{u}}\times\tilde{\mathbf{B}}. \label{eqn:faraday_norm}
\end{equation}
where $\tilde{E}_P = \frac{1}{2}\tilde{\rho}\tilde{u}^2 + \frac{\tilde{P}}{\gamma-1}$ is the normalized plasma energy. To summarize, Equations (\ref{eqn:mass_eqn_norm}), (\ref{eqn:momentum_eqn_norm}), (\ref{eqn:energy_eqn_norm}) and (\ref{eqn:faraday_norm}) forms a set of dimensionless MHD equations with the following normalizations:
\begin{align}
t_0 & = \frac{x_0}{u_0}, \\
P_0 & = \rho_0u_0^2, \\
B_0 & = \sqrt{\mu_0\rho_0}u_0.
\end{align}

\section*{B. The Grid Metric Calculations}\label{Append:A}

As described in Section \ref{sec:fv_def}, the \textit{primary} grid is the cell corner coordinates $\mathbf{x}_{i\pm\frac{1}{2},j\pm\frac{1}{2},k\pm\frac{1}{2}}$. Thus the corresponding cell center positions $\mathbf{x}_{i,j,k}^C$ for cell $(i,j,k)$ as shown in Figure \ref{fig:FV-grid-def} are computed using the eight cell corners forming a hexahedron as:
\begin{equation}
\begin{split}
\mathbf{x}_{i,j,k}^C = & \frac{1}{8} \left( \mathbf{x}_{i-\frac{1}{2},j-\frac{1}{2},k-\frac{1}{2}}+\mathbf{x}_{i-\frac{1}{2},j+\frac{1}{2},k-\frac{1}{2}}+\mathbf{x}_{i-\frac{1}{2},j-\frac{1}{2},k+\frac{1}{2}}+\mathbf{x}_{i-\frac{1}{2},j+\frac{1}{2},k+\frac{1}{2}} \right. \\
& \left.+\mathbf{x}_{i+\frac{1}{2},j-\frac{1}{2},k-\frac{1}{2}}+\mathbf{x}_{i+\frac{1}{2},j-\frac{1}{2},k-\frac{1}{2}}+\mathbf{x}_{i+\frac{1}{2},j+\frac{1}{2},k-\frac{1}{2}}+\mathbf{x}_{i+\frac{1}{2},j+\frac{1}{2},k+\frac{1}{2}} \right)
\end{split}
\end{equation}

The calculated cell center location $\mathbf{x}_{i,j,k}^C$ should be located with in the control volume ($i,j,k$). In some cases, reasonably-shaped grid design could possibly put $\mathbf{x}_{i,j,k}^C$ close to a face when the cell is a concave, thus the numerical methods may not work well due to the extreme in the metric calculations. However, this situation has not occurred in any of the heliophysics applications developed using GAMERA. In the GAMERA code, the face center locations $\mathbf{x}^{\mu}$, $\mathbf{x}^{\nu}$ and $\mathbf{x}^{\zeta}$ are used directly in the calculation of magnetic field vectors. Thus they are calculated using high-order Gaussian quadrature, which gives better definitions of the ``barycenter'' of a cell interface especially when the cells are either very distorted or degenerated to tetrahedrons, e.g., spherical grid cells near the axis. For example, when using two-dimensional Gaussian quadratures, the $\mu$-face center locations $\mathbf{x}^{\mu}_{i+\frac{1}{2},j,k}$ are calculated as:
\begin{equation}
\mathbf{x}_{i+\frac{1}{2},j,k}^\mu = \frac{1}{A^\mu_{i+\frac{1}{2},j,k}}\int^{k+\frac{1}{2}}_{k-\frac{1}{2}}\int^{j+\frac{1}{2}}_{j-\frac{1}{2}}\mathbf{x}\left(\mu=i+\frac{1}{2},\nu,\zeta \right)d\nu d\zeta = \frac{1}{A^\mu_{i+\frac{1}{2},j,k}}\sum_{m=1}^{m=N}\sum_{n=1}^{n=N}w_nw_m\mathbf{x}_{n,m}^{\mu=i+1/2}|J_{n,m}^{\nu,\zeta}|\label{eqn:bary1_cent}
\end{equation}
where $A^\mu_{i+\frac{1}{2},j,k}$ is the $\mu$ face area at interface ($i+\frac{1}{2},j,k$) which is also calculated using Gaussian quadrature. $n$ and $m$ are the indices of Gaussian quadrature points in the $\nu$- and $\zeta$-direction, respectively. $w_n$ and $w_m$ are the corresponding weights at the Gaussian points. $\mathbf{x}_{n,m}$ is the corresponding spatial location of the 2-D Gaussian quadrature points with indices ($n,m$) on the $\nu$-$\zeta$ interface, and $|J_{n,m}^{\nu,\zeta}|$ is the corresponding Jacobian of local coordinate transformation. In the GAMERA code, the quadrature is done using the $12^{th}$-order two-dimensional Gaussian quadrature ($N=12$). The actual implementation of the $12^{th}$-order Gaussian quadrature can be found in the example Matlab source code. 

Similarly, the face center locations at $\nu$ and $\zeta$ interfaces are calculated as:
\begin{equation}
\mathbf{x}_{i,j+\frac{1}{2},k}^\nu = \frac{1}{A^\nu_{i,j+\frac{1}{2},k}}\int^{k+\frac{1}{2}}_{k-\frac{1}{2}}\int^{i+\frac{1}{2}}_{i-\frac{1}{2}}\mathbf{x}\left(\mu,\nu=j+\frac{1}{2},\zeta \right)d\zeta d\mu = \frac{1}{A^\nu_{i,j+\frac{1}{2},k}}\sum_{m=1}^{m=N}\sum_{n=1}^{n=N}w_nw_m\mathbf{x}_{n,m}^{\nu = j+1/2}|J_{n,m}^{\zeta,\mu}| \label{eqn:bary2_cent}
\end{equation}
\begin{equation}
\mathbf{x}_{i,j,k+\frac{1}{2}}^\zeta = \frac{1}{A^\zeta_{i,j,k+\frac{1}{2}}}\int^{i+\frac{1}{2}}_{i-\frac{1}{2}}\int^{j+\frac{1}{2}}_{j-\frac{1}{2}}\mathbf{x}\left(\mu,\nu,\zeta=k+\frac{1}{2} \right)d\mu d\nu = \frac{1}{A^\zeta_{i,j,k+\frac{1}{2}}}\sum_{m=1}^{m=N}\sum_{n=1}^{n=N}w_nw_m\mathbf{x}_{n,m}^{\zeta = k+1/2}|J_{n,m}^{\mu,\nu}|\label{eqn:bary3_cent}
\end{equation}
where the integrals are done on corresponding cell faces formed by four corner points (or three corner points when the cell faces are degenerated).  Numerical solutions exhibit fewer artifacts in grids with axis singularities or highly distorted aspect ratios when the ``barycenters'' of cell faces defined in Equation (\ref{eqn:bary1_cent}) - (\ref{eqn:bary3_cent}) are used.

The face areas $A^\mu_{i\pm\frac{1}{2},j,k}$, $A^\nu_{i,j\pm\frac{1}{2},k}$ and $A^\zeta_{i,j,k\pm\frac{1}{2}}$ are also calculated using the $12^{th}$-order two-dimensional Gaussian quadrature based on the four corner points forming the cell face:
\begin{equation}
A^\mu_{i+\frac{1}{2},j,k} = \int^{j+\frac{1}{2}}_{j-\frac{1}{2}}\int^{k+\frac{1}{2}}_{k-\frac{1}{2}}d\nu d\zeta= \sum_{m=1}^{m=N}\sum_{n=1}^{n=N}w_nw_m|J_{n,m}^{\nu,\zeta}|\label{eqn:bary1_face}
\end{equation}
\begin{equation}
A^\nu_{i,j+\frac{1}{2},k} = \int^{k+\frac{1}{2}}_{k-\frac{1}{2}}\int^{i+\frac{1}{2}}_{i-\frac{1}{2}}d\zeta d\mu= \sum_{m=1}^{m=N}\sum_{n=1}^{n=N}w_nw_m|J_{n,m}^{\zeta,\mu}|\label{eqn:bary2_face}
\end{equation}
\begin{equation}
A^\zeta_{i,j,k+\frac{1}{2}} = \int^{i+\frac{1}{2}}_{i-\frac{1}{2}}\int^{j+\frac{1}{2}}_{j-\frac{1}{2}}d\mu d\nu= \sum_{m=1}^{m=N}\sum_{n=1}^{n=N}w_nw_m|J_{n,m}^{\mu,\nu}|\label{eqn:baryc3_face}
\end{equation}

The cell volume $V_{i,j,k}$ is also calculated using a $12^{th}$-order three-dimensional Gaussian quadrature based on the hexahedron cell formed by eight cell corners shown in Figure \ref{fig:FV-grid-def}: 
\begin{equation}
V_{i,j,k} = \int_{i-\frac{1}{2}}^{i+\frac{1}{2}}\int_{j-\frac{1}{2}}^{j+\frac{1}{2}}\int_{k-\frac{1}{2}}^{k+\frac{1}{2}}d\mu d\nu d\zeta= \sum_{l=1}^{l=N}\sum_{m=1}^{m=N}\sum_{n=1}^{n=N}w_lw_nw_m|J_{l,n,m}^{\nu,\nu,\zeta}|
\end{equation}


\section*{C. The General Form of the PDM Operator}

\begin{figure}[htb!]
	\noindent\includegraphics[width=24pc]{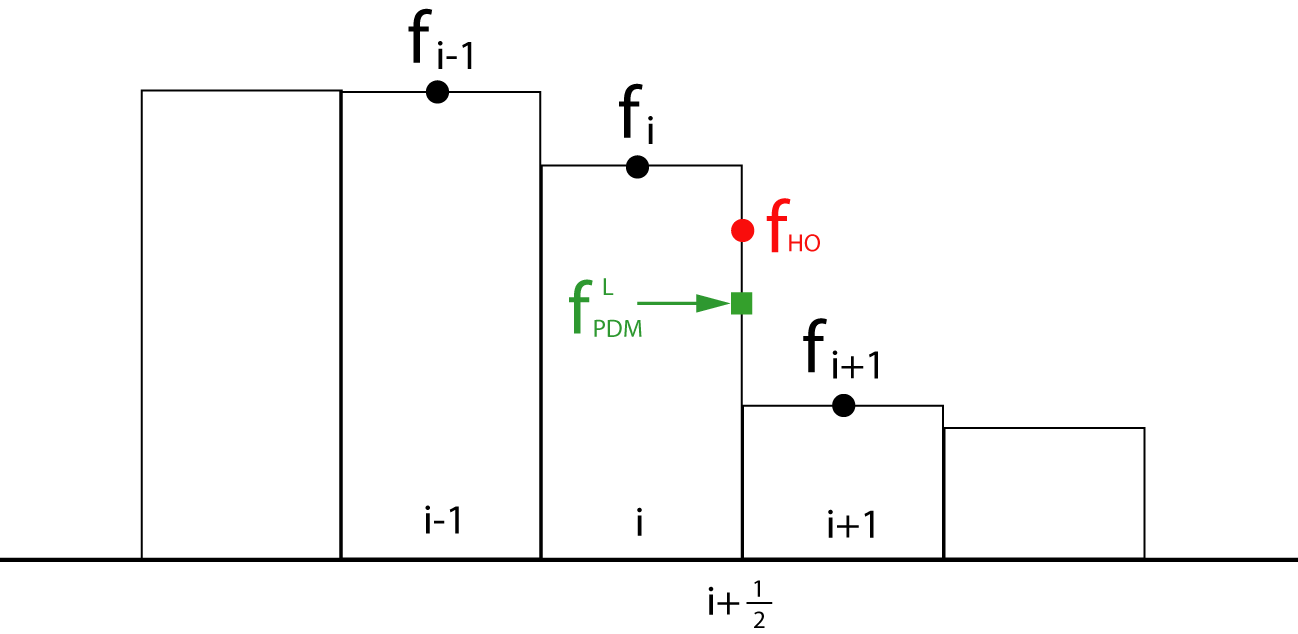}%
	\centering
	\caption{The calculation of left-state at interface $i+\frac{1}{2}$ using the PDM limiter.} \label{fig:pdm2}
\end{figure}

For non-linear problems, the form of the PDM operator $M^+$ using a $2^{nd}$-order centered reconstruction is written as \citep{Hain1987}
\begin{equation}
\begin{split}
M^+ = m^+ & - \frac{1}{2}\mathrm{sign}(v)\mathrm{sign}(\nabla^+)\mathrm{max}[0,|\nabla^+| \\
          & -\frac{A}{4}\cdot(1-\mathrm{sign}(v))\cdot|\mathrm{sign}(\nabla^+)+\mathrm{sign}(\nabla^{++})|\cdot|\nabla^{++}|  \\
          & -\frac{A}{4}\cdot(1+\mathrm{sign}(v))\cdot|\mathrm{sign}(\nabla^-)+\mathrm{sign}(\nabla^{+})|\cdot|\nabla^-|],
\end{split}\label{eqn:pdm_gen}
\end{equation}
where $m^+f = \frac{1}{2}(f_i+f_{i+1})$ corresponds to the $2^{nd}$-order reconstruction, $\nabla^{++}f = f_{i+2}-f_{+1}$, $\nabla^+f = f_{i+1}-f_i$, $\nabla^-f = f_{i}-f_{i-1}$, and $A$ is the parameter determines the amount of numerical diffusion. The direction of sweeping is $\mathrm{sign}(v)$, thus Equation (\ref{eqn:pdm_gen}) gives both the left- and right-state at interface $i+\frac{1}{2}$. If $A=0$, the above operator $M^+$ becomes the donor cell method which is one of the most diffusive first-order method. For linear problems, \citet{Hain1987} have shown that the monotonicity condition for the PDM parameter A is related to the Courant number $N_{CFL}$:
\begin{equation}
N_{CFL}<\frac{2}{2+A}.\label{eqn:pdm_CFL}
\end{equation}
In principle any value of $A>0$ could be used in the PDM limiter. Larger values of $A$ lead to more aggressive limiting as indicated by (\ref{eqn:pdm_gen}). However, the allowable Courant number decreases with increasing $A$ as shown in equation (\ref{eqn:pdm_CFL}).  In practice, we find that there is little improvement in resolving square wave profiles if $A > 4$.  Thus in the GAMERA solver, the default value for the PDM value $A$ is chosen to be 4.0, together with a fixed Courant number $N_{CFL}=0.3$ which satisfies (\ref{eqn:pdm_CFL}). The PDM limiter can be extended to use arbitrary high-order spatial reconstruction instead of using $m^+$ in Equation (\ref{eqn:pdm_gen}) as shown in Figure \ref{fig:pdm2}. For a high-order interface value $f^{HO}$ reconstructed at $i+\frac{1}{2}$, the left-state ($v>0$) at interface $i+\frac{1}{2}$ is calculated using the following equation:
\begin{equation}
f^L_{PDM} = f_{HO}^* - \mathrm{sign}\left(f_{i+1}-f_{i}\right)\cdot\mathrm{max}\left[0,\big |f_{HO}^*-f_i \big |-\frac{A}{4}\cdot\big |f_{i}-f_{i-1} \big |\cdot\frac{1}{2}\big |\mathrm{sign}(f_i-f_{i-1})+\mathrm{sign}(f_{i+1}-f_{i}) \big | \right].
\end{equation}

Here the $f_{HO}^*$ is a ``clipped'' version of the high-order reconstruction value $f_{HO}$:
\begin{equation}
f_{HO}^*=\mathrm{median}(f_i,f_{HO},f_{i+1}). 
\end{equation}
It is very straightforward to show that if $A=0$, the above equation becomes 
\begin{equation}
f^L_{PDM} = f_{i},
\end{equation}
which is the first-order Donor cell method. The numerical diffusion is reduced significantly as $A$ is increased from $0$ to $4.0$.

\section*{D. The Choice of Reconstruction Order}

The PDM limiter can be combined with an arbitrarily high order reconstruction scheme as discussed in Section \ref{sec:recon}. In the GAMERA code, the default choices of the reconstruction method is the $7^{th}$-order upwind reconstruction (the $8^{th}$-order centered reconstruction is the one used in the original LFM, which is also implemented in the GAMERA code for reference). The reason for choosing such high-order reconstruction schemes as the default is based on achieving low numerical diffusion with reasonable amount of computing resource, which is important for 3-D global-scale simulations of large space plasma systems such as planetary magnetospheres and the heliosphere. Here we show two sets of numerical solutions of the linear advection equation to justify the necessity of the high-order reconstruction implemented in the GAMERA code.

\subsection*{1-D Linear Advection of Four Shapes in Non-uniform Grid}

\begin{figure}[htb!]
	\noindent\includegraphics[width=39pc]{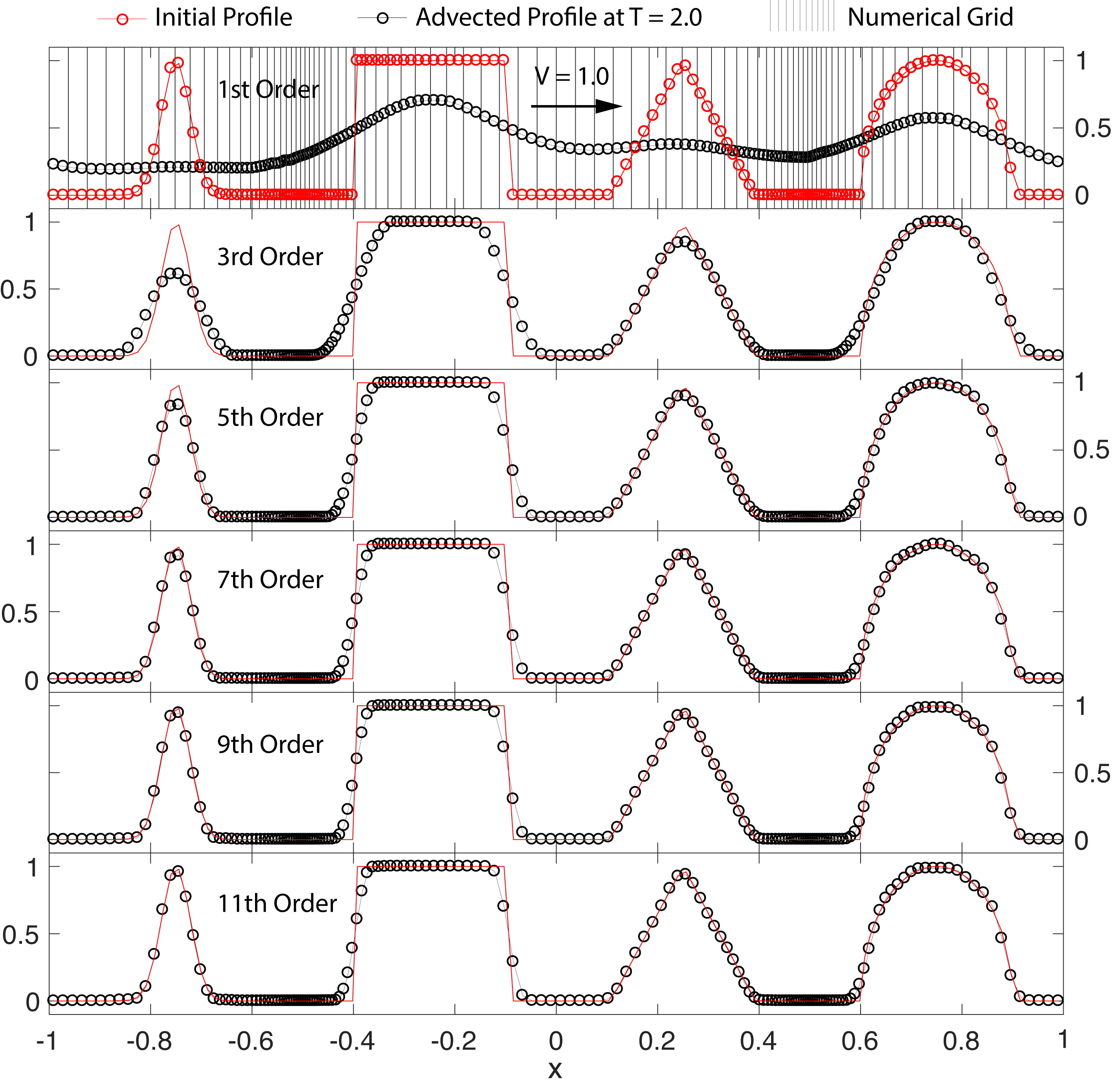}%
	\centering
	\caption{1-D linear advection of four shapes in a non-uniform grid using different orders of reconstruction. The vertical gray line shows the spatial distribution of the computational grid. The red profile is the initial condition at $t=0$. The black profiles are simulation results at $t=2.0$ with different orders of spatial reconstruction. }\label{fig:advect}
\end{figure}

The 1-D linear advection equation solved in this section follows
\begin{equation}
\frac{\partial f}{\partial t} + u\frac{\partial f}{\partial x} = 0, \label{eqn:A_adv_eq}
\end{equation}
where $u=1.0$ and $x$ is defined on $[-1, 1]$. The computational domain is non-uniform with
\begin{equation}
x_i=2(\frac{i}{N}+0.1\sin\frac{2\pi i}{N})-1,
\end{equation}
where $i=0,1,2,...N$, and $N$ is the total number of cell centers in the $x$-direction. Periodic boundary conditions are imposed at $x=1$ and $x=-1$, thus the initial profile of $f$ goes back to exactly the same location at $t=2$. The red profile in the top panel of Figure \ref{fig:advect} shows the initial profile of $f$ at $t=0$, with the gray vertical lines showing variation of the non-uniform grid geometry. Four distinctive shapes are used in the initial profile: a Gaussian peak, a square wave, a triangle and a half-circle. This linear advection test simulation has been used extensively in previous studies to test the quality of advection schemes in fluid simulations. 

As shown in the initial profile, smooth resolution is tested by a narrow Gaussian wave with a width of $2.4\Delta x$ centered at $x=-0.75$, while discontinuity resolution is tested by a square wave centered at $x=-0.25$. Discontinuity in the first derivative is tested using a triangular profile centered at $x=0.25$. Although there is no discontinuity in value, the half-circle profile centered at $x = 0.75$ is very challenging since it combines sudden and gradual changes in gradient with limited number of cells. In addition, there is a discontinuity in curvature at each side of the half-circle profile, which makes it more challenging to resolve the profile especially in non-uniform geometry. In the following simulations, the PDM parameter A described in Appendix C is chosen to be 4.0 and the Courant number is 0.3, which are the same as the default values used in the MHD solver. 

The black profiles in Figure \ref{fig:advect} are the simulated distributions of $f$ at $t=2.0$ using six different orders of upwind reconstruction (e.g., $1^{st}$, $3^{rd}$, ..., $11^{th}$-order). The $1^{st}$-order scheme, which is basically the Donor cell method, is over-diffusive - all the four profiles are smeared significantly as Error functions at $t=2.0$. As the order of reconstruction increases to $7^{th}$-order, the algorithm starts to resolve the full narrow Gaussian peak within 9 cells. Above 7th-order, the improvement on the Gaussian peak is small. The resolution of the square wave is also improved as the order of reconstruction increases. Above $5^{th}$-order, the resolution of the sharp transition of the right-side of the square wave remains at 4 cells. In other words, the improvement on resolving the square wave is small. Given that the total computing time (including both reconstruction and limiting) of the $7^{th}$-order scheme is only about $10\%$ more than the $5^{th}$-order scheme, we choose the $7^{th}$-order scheme as the default choice for spatial reconstruction in the GAMERA code. A more quantitative analysis on the effective numerical diffusion coefficient based on the advection of a square wave is in the next section.

\subsection*{The Effective Diffusion Coefficient}

When the one-dimensional linear advection equation (\ref{eqn:A_adv_eq}) is solved numerically, the effective equation solved by numerical schemes is written as
\begin{equation}
\frac{\partial f}{\partial t} + u\frac{\partial f}{\partial x} = D\frac{\partial^2 f}{\partial x^2},\label{eqn:A_adv_dif_eq}
\end{equation}
where $u$ is the advection speed and $D$ is an effective diffusion coefficient, which depends on both the numerical scheme and the grid resolution. In general, the above equation is a mixed advection-diffusion equation with an analytical solution written as
\begin{equation}
f\left(x,t\right) = \int_{-\infty}^{+\infty}f\left(\xi-ut\right)e^{-\frac{\left(\xi-x\right)^2}{Dt}}d\xi.\label{eqn:A_adv_solution}
\end{equation}
When the diffusion coefficient $D\equiv 0$, the solution becomes 
\begin{equation}
f\left(x,t\right) = f\left(x-ut,t=0\right).
\end{equation}

\begin{figure}[htb!]
	\noindent\includegraphics[width=39pc]{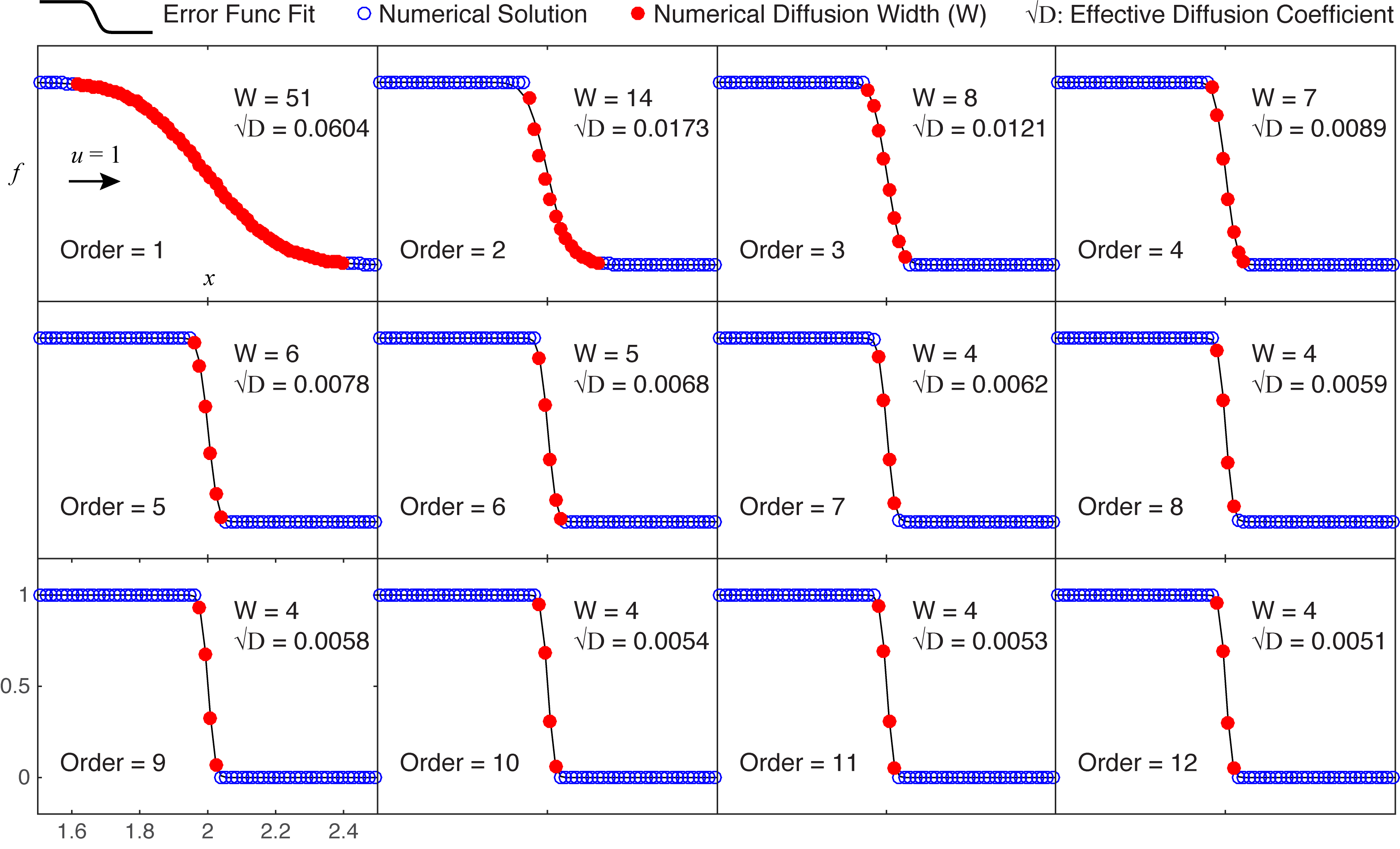}%
	\centering
	\caption{1-D linear advection of a step function using different orders of spatial reconstruction. Blue circles are the numerical solutions to the linear advection equatio while the black curves are least square fits of analytical solutions to the corresponding numerical solutions. The red dots show the numerical diffusion width which is defined in the transition region between 0.01 and 0.99. The effective diffusion coefficient $\sqrt{D}$ is listed in the top right corner of each panel. }\label{fig:rey}
\end{figure}

We use the numerical solution to a special case of the above linear advection-diffusion equation to approximately quantify the amount of effective numerical diffusion introduced by different reconstruction methods. If the linear advection-diffusion equation (\ref{eqn:A_adv_dif_eq}) is solved numerically in a semi-infinite domain $x\in[0, +\infty)$ with an initial condition $f(x,t=0)=0$ and a boundary condition $f\left(x=0,t\right)=1$, the analytical solution (\ref{eqn:A_adv_solution}) of the advection-diffusion equation with constant $u$ and $D$ is written as:
\begin{equation}
f(x,t)=\frac{1}{2}\mathrm{erfc}\left(\frac{x-ut}{\sqrt{D t}}\right),\label{eqn:A_erf}
\end{equation}
where $\mathrm{erfc(\cdot)}$ is the complimentary error function. After solving the linear advection equation (\ref{eqn:A_adv_eq}) numerically with $f(x,t=0)=0$ and $f(x=0,t)=1$, the numerical diffusion coefficient $D$ is approximated by fitting the numerical solution $f_i$ to the analytical solution (\ref{eqn:A_erf}). To simplify the calculations, we solve the linear advection equation (\ref{eqn:A_adv_eq}) in a uniform computational domain defined at $[0,4]$ with 256 cells. At $t=2$, the front of the step function is located at $x=2$, and a least square fit process is applied to the numerical solution $f_i$ to estimate the effective diffusion coefficient $D$. Figure \ref{fig:rey} shows the numerical solutions of the step function using reconstruction schemes from $1^{st}$-order to $12^{th}$-order, together with the effective diffusion coefficient $\sqrt{D}$ calculated for each profile. The numerical diffusion width $W$ listed in each panel is defined as the number of computational cells in the transition region of the step function ($0.01<f_i<0.99$). If the effective diffusion coefficient $D$ is exactly zero, the numerical diffusion width $W$ is expected to be 0.

The summary of the numerical solution to the advection equation (\ref{eqn:A_adv_eq}) is listed in Table \ref{tab:rey}. Reconstruction schemes from $1^{st}$-order to $12^{th}$-order are tested with the analytical solution to the advection-diffusion equation (\ref{eqn:A_adv_dif_eq}). The effective diffusion coefficient ($\sqrt{D}$) is listed in the second column, which decreases from 0.0604 to 0.0051 as the order of reconstruction increases from 1 to 12. The numerical diffusion width ($W$) is listed in the third column of Table \ref{tab:rey}. When the order of reconstruction is higher than $7^{th}$-order, the numerical diffusion width remains at 4, suggesting that the improvement on the slope of the square wave is small above $7^{th}$-order. This quantitative comparison is consistent with the linear advection test simulations shown in Figure \ref{fig:advect}. The third column shows the relative grid Reynolds number in the simulation normalized by the effective grid Reynolds number ($uL/D$) in the $1^{st}$-order solution. When using a $7^{th}$-order reconstruction scheme, the effective grid Reynolds number is enhanced by approximately two orders of magnitude compared to the $1^{st}$-order solution. Above $7^{th}$-order, there is still improvement to the relative grid Reynolds number but the improvement is small. If we use a simple measure ($R/Order$) to quantify the efficient of the reconstruction methods, it peaks at $7^{th}$-order as shown in the last column of Table 4.

\begin{table}[htb!]
\centering
\begin{tabular}{lcccc}
Order & $\sqrt{D}$ Fit & Width (cells)  & Relative $R$ ($\frac{uL}{D}$)  &  $R/Order$\\ \hline
1     & 0.0604           & 51                  & 1.0                                   &  1.0              \\
2     & 0.1730           & 14                  & 12.1                                  &  6.1  \\
3     & 0.1210           & 8                   & 24.9                                  &  8.3 \\ 
4     & 0.0089           & 7                   & 46.1                                  &  11.5 \\
5     & 0.0078           & 6                   & 59.9                                  &  11.9 \\
6     & 0.0068           & 5                   & 78.8                                  &  13.1 \\
7     & 0.0062           & 4                   & 95.0                                  &  13.6 \\
8     & 0.0059           & 4                   & 104.8                                 &  13.1\\
9     & 0.0058           & 4                   & 108.4                                 &  12.0\\
10    & 0.0054           & 4                   & 125.1                                 &  12.5 \\
11    & 0.0053           & 4                   & 129.8                                 &  12.9 \\
12    & 0.0051           & 4                   & 140.2                                 &  11.6 \\ \hline
\end{tabular}\label{tab:rey}
\caption{The numerical diffusion coefficient $\sqrt{D}$, diffusion width ($W$), relative Reynolds number $R$ normalized by the one derived from the first order method and $R/Order$ calculated using order of reconstruction from $1^{st}$-order to $12^{th}$-order. }
\end{table}

\bibliographystyle{unsrt}

{}


\end{spacing}
\end{document}